\documentclass[11pt]{article}
\usepackage{times}
\usepackage{latexsym}
\usepackage{comment}
\usepackage{geometry} 
\usepackage{tablefootnote}
\usepackage{multirow}
\usepackage{makecell}

\usepackage[flushleft]{threeparttable}

\usepackage{microtype}

\newcommand{\TODO}[2]{}

\makeatletter
\let\@svfnsymbol\@fnsymbol
\renewcommand*{\@fnsymbol}[1]{}
\makeatother
\newcommand{\toUOttawa}{$^{1}$}
\newcommand{\toMILA}{$^2$}
\newcommand{\toUdeM}{$^{3}$}
\newcommand{\toATI}{$^4$}
\newcommand{\toUOxford}{$^{5}$}
\newcommand{\toUPenn}{$^6$}
\newcommand{\toMcGill}{$^7$}
\newcommand{\toBLG}{$^{8}$}
\newcommand{\toDecLab}{$^{9}$}
\newcommand{\toHEC}{$^{10}$}
\newcommand{\toMaxPlanck}{$^{11}$}
\newcommand{\toLibeo}{$^{12}$}
\newcommand{\toUToronto}{$^{13}$}

\author{
\footnote{
\toUOttawa University of Ottawa,
\toMILA Mila,
\toUdeM Universit\'e de Montr\'eal,
\toATI The Alan Turing Institute,
\toUOxford University of Oxford,
\toUPenn University of Pennsylvania,
\toMcGill McGill University,
\toBLG Borden Ladner Gervais LLP,
\toDecLab The Decision Lab,
\toHEC HEC Montr\'eal,
\toMaxPlanck Max Planck Institute,
\toLibeo Lib\'eo,
\toUToronto University of Toronto.
}
\footnote{Corresponding author general: \texttt{richard.janda@mcgill.ca}}
\footnote{Corresponding author for public health: \texttt{abhinav.sharma@mcgill.ca}}
\footnote{Corresponding author for privacy: \texttt{ywyu@math.toronto.edu}}
\footnote{Corresponding author for machine learning: \texttt{yoshua.bengio@mila.quebec}}
\footnote{Corresponding author for user perspective: \texttt{brooke@thedecisionlab.com}}
\footnote{Corresponding author for technical implementation: \texttt{jean-francois.rousseau@libeo.com}}
  Hannah Alsdurf\toUOttawa, \and
  Edmond Belliveau, \and
  Yoshua Bengio\toMILA$^,$\toUdeM,\and
  Tristan Deleu\toMILA$^,$\toUdeM,\and
  Prateek Gupta\toMILA$^,$\toATI$^,$\toUOxford,\and
  Daphne Ippolito\toUPenn,\and
  Richard Janda\toMcGill,\and
  Max Jarvie\toBLG,\and
  Tyler Kolody\toMcGill,\and
  Sekoul Krastev\toDecLab,\and
  Tegan Maharaj\toMILA$^,$\toUdeM,\and
  Robert Obryk,\and
  Dan Pilat\toDecLab,\and
  Val\'erie Pisano\toMILA,\and
  Benjamin Prud'homme\toMILA,\and
  Meng Qu,\toMILA$^,$\toHEC\and
  Nasim Rahaman\toMILA$^,$\toMaxPlanck,\and
  Irina Rish\toMILA$^,$\toUdeM,\and
  Jean-Fran\c{c}ois Rousseau\toLibeo,\and
  Victor Schmidt\toMILA$^,$\toUdeM,\and
  Abhinav Sharma\toMcGill,\and
  Brooke Struck\toDecLab ,\and
  Jian Tang\toMILA$^,$\toHEC,\and
  Martin Weiss\toMILA$^,$\toUdeM,\and
  Yun William Yu\toUToronto
}

\date{}

\usepackage{amsmath,amssymb,amsfonts}
\usepackage{hyperref}

\usepackage{algorithmic}
\usepackage{graphicx}
\usepackage{textcomp}
\usepackage{xcolor}
\usepackage{array}
\newcolumntype{L}[1]{>{\raggedright\let\newline\\\arraybackslash\hspace{0pt}}m{#1}}
\usepackage{soul,color}
\soulregister\cite7

\begin{document}

\title{COVI White Paper - Version 1.1}

\maketitle

\vspace{-2.7em}

\begin{abstract}
The SARS-CoV-2 (Covid-19) pandemic has resulted in significant strain on health care and public health institutions around the world. Contact tracing is an essential tool for public health officials and local communities to change the course of the Covid-19 pandemic. Standard manual contact tracing of people infected with Covid-19, while the current gold standard, has significant challenges that limit the ability of public health authorities to minimize community infections.
Personalized peer-to-peer contact tracing through the use of mobile applications has the potential to shift the paradigm of Covid-19 community spread.
Although some countries have deployed centralized tracking systems through either GPS or Bluetooth, more privacy-protecting decentralized systems offer much of the same benefit without concentrating data in the hands of a state authority or in for-profit corporations.

Additionally, machine learning methods can be used to circumvent some of the limitations of standard digital tracing by incorporating many clues (including medical conditions, self-reported symptoms, and numerous encounters with people at different risk levels, for different durations and distances) and their uncertainty into a more graded and precise estimation of infection and contagion risk. The estimated risk can be used to provide early risk awareness, personalized recommendations and relevant information to the user and connect them to health services. Finally, the non-identifying data about these risks can inform detailed epidemiological models trained jointly with the machine learning predictor, and these models can provide statistical evidence for the interaction and importance of different factors involved in the transmission of the disease. They can also be used  to monitor, evaluate and optimize different health policy and confinement/deconfinement scenarios according to medical and economic productivity indicators.

However, such a strategy based on mobile apps and machine learning should proactively mitigate potential ethical and privacy risks, which could have substantial impacts on society (not only impacts on health but also impacts such as stigmatization and abuse of personal data). Here, we present an overview of the rationale, design, ethical considerations and privacy strategy of `COVI,' a Covid-19 public peer-to-peer contact tracing and risk awareness mobile application developed in Canada.

\textbf{Addendum 2020-07-14:} The government of Canada has declined to endorse COVI and will be promoting a different app for decentralized contact tracing. In the interest of preventing fragmentation of the app landscape, COVI will therefore not be deployed to end users. We are currently still in the process of finalizing the project, and plan to release our code and models for academic consumption and to make them accessible to other States should they wish to deploy an app based on or inspired by said code and models.

\end{abstract}

\newpage

\tableofcontents

\newpage
\makeatletter
\let\@fnsymbol\@svfnsymbol
\makeatother

\section{Overview}

\subsection{Introduction}
The SARS-CoV-2 (Covid-19) outbreak is the most widespread pandemic in a century and is currently the largest global crisis since the Second World War \cite{gates2020responding}. The pandemic has created a global health emergency that has drastically impacted all aspects of modern life and has strained global healthcare systems, economies, and political institutions \cite{fernandes2020economic,anderson2020will,barometer2019january}. Our rapidly evolving understanding of Covid-19 has challenged evidence-based decisions regarding strategies used to contain and prevent the spread of the virus \cite{niehus2020quantifying,irfan2020case}.

Manual contact tracing, which attempts to identify and isolate individuals at high risk of acquiring Covid-19, is the primary strategy used by public health authorities to track and reduce the spread of the virus \cite{flint2003assessment, ferretti2020quantifying}.
In recent months, automatic contact tracing solutions have been proposed to address several of the challenges with manual contact tracing \cite{tang2020contact, bay2020bluetrace, chan2020pact, rivest2020pact, applegoogle2020, pepp, trieu2020epione}.
They aim to reduce the cost and workload on healthcare professionals, the recall bias in remembering prior contacts, the inability to identify contacts from the general public (for example a supermarket worker), and the time delay between identifying whether a person was infected and manually alerting their contacts about a high-risk interaction with the person.
Furthermore, while manual tracing can be very effective at the early stages of a potential epidemic, its efficacy declines as infection becomes more widespread and the number of contact tracers needed grows \cite{npr_contact_tracing, ferretti2020quantifying}.
Finally, it requires individuals to have a certain level of comfort in disclosing potentially sensitive information about their social circle to government officials \cite{flanagan1988equality}. 

Privacy is a serious concern for both manual contact tracing and automatic tracing solutions \cite{levine1988contact, cho2020contact, raskar2020apps, chan2020pact}.
Contact tracing as it is typically conceived requires that individuals provide some amount of personal information to state authorities \cite{burgess1963contact}. 
With manual contact tracing, this takes the form of diagnosed patients trying to recall all of their physical encounters in the past two weeks and report them to a public health official, who then attempts to contact each individual listed and asks them to undertake a similar recall exercise.
Automatic contact tracing, which relies on either Bluetooth or GPS from users' phones to detect when they have been in contact with each other, can also be implemented in a centralized way.
The disclosure of personal information to state authorities takes the form of a centralized database that tracks contact encounters among identifiable individuals \cite{pepp, bay2020bluetrace}.

While technological solutions can amplify the impact of contact tracing, if implemented incorrectly, they may also pose significant risks to citizens \cite{millar2020well, raskar2020apps}, including loss of civil liberties, erosion of privacy, and government or private surveillance \cite{aclu2020principles}.
There has recently been a robust debate on the difference between centralized and decentralized digital contact tracing, ranging from whether such a distinction is truly meaningful \cite{inria2020proximity} to a focused critique of the privacy implications of centralized systems \cite{meyer2020controversy, greenberg2020clever}.
These concerns may significantly limit the efficacy of such applications as citizens who do not trust an app would be unlikely to use it or use it appropriately.
Democratic societies must therefore employ a privacy-protecting approach to digital contact tracing to enhance public trust in such applications \cite{millar2020well, raskar2020apps, sharma2018using}.
The need for decentralized contact tracing has prompted a surge in proposals for privacy-preserving automatic contact tracing strategies \cite{applegoogle2020, chan2020pact, rivest2020pact, trieu2020epione, troncoso2020decentralized}.
Decentralized approaches automatically notify recent contacts of their risk without entrusting any identifiable contact information to a centralized governmental authority.

Beyond the most basic feature of contact tracing---notifying individuals when they have been in contact with an infected person---an effective public health app can provide a wide range of capabilities to assist its users in making actionable decisions \cite{chen2017use} and to assist in the development of better epidemiological models and better public policies \cite{dandekar2020safe}.
While most automated contact tracing proposals can only provide a binary notification (i.e., an individual either was exposed to someone who tested positive for Covid-19 or not), a more realistic multi-level risk gradation, where risk is based on additional factors such as symptoms, comorbidities, and occupation, allows for nuanced suggestions and an augmented understanding of individual and collective risk.
While this additional sensitive information should by default be kept securely on each user's phone, giving app users the option to volunteer their data pseudonymously for the construction of epidemiological and risk assessment models empowers them with the ability to contribute meaningfully to the fight against Covid-19 and to the quality of the predictor used in their phone.

In this white paper, we describe the design of COVI, a privacy-protecting mobile application based on Bluetooth proximity detection, developed for Canada. The COVI application aims to achieve the following goals: 1) reduce and control the number of infections by empowering citizens, in the broadest sense of the term,  to protect themselves and others by following targeted recommendations based on their infection risk; 2) extract crucial information to inform and guide a data-driven approach to public health policy regarding pandemic confinement and deconfinement planning; and 3) establish strong privacy protections using a decentralized strategy to keep users' personal information away from other individuals, companies and governmental organizations. This white paper introduces the overall philosophy and approach behind COVI, and demonstrates how---by combining research from public health, epidemiology, privacy, machine learning (ML), ethics and psychology---COVI aims to mitigate risks while empowering citizens to make evidence-based decisions about their personal response to the Covid-19 crisis.

\subsection{Primary goals of COVI}

\subsubsection{Reduce the spread of Covid-19}
\label{sec:goal1}

The primary aim of COVI is to reduce the spread of Covid-19, in order to: 1) reduce the morbidity and mortality associated with Covid-19 infection; and 2) reduce the burden of Covid-19 on healthcare systems. To achieve this aim, the primary strategy of COVI is to inform individuals of their infection risk so they can act responsibly to protect themselves and others and limit the spread of the virus.  

COVI leverages probabilistic risk levels (as detailed below), rather than binary results of confirmed infection (or not), to assign Covid-19 infection risk levels to app users. Using probabilistic risk levels in this way formalizes and quantifies the long-standing use in public health practice of case definitions with qualitative levels of certainty (e.g.,  possible, probable, and confirmed). The probabilistic risk is modified and updated based on two main categories of information---users' individual profile and interaction profile. Individual profiles arise from user-entered information including demographic information, baseline comorbidities, occupation and the presence of new symptoms. The interaction profile arises when two or more users converge in physical proximity. The probability of transmission depends on details such as distance apart, time spent in an interaction, the prior contacts of an individual, and the use of masks or other physical separation devices. The combination of this information generates a machine learning (ML) derived personalized probability
distribution of the user`s likeliness of being infected, when the
infection may have occurred, and the expected contagiousness on different days after infection.

COVI then empowers citizens to take a progressive, proactive approach to managing infection risk for themselves and others. The specific individual risk level will not be displayed as the response to receiving a specific risk number can be heterogeneous and elicit counter productive behaviour \cite{zikmund200428}. Instead, when a user's risk level increases, evidence-based messaging \cite{kang2012development} previously approved by public health authorities \cite{gov_canada_covid_guidance} will be provided through the app. This messaging will recommend various actions that can be taken by the user to reduce the risk of Covid-19 transmission, in coordination with local public health.      

While the app uses contact information to identify and notify individuals at high risk of transmission, it cannot (and is not intended to) fully replace manual contact tracing, especially in difficult cases where professional judgment is needed. Instead, COVI aims to complement  manual contact tracing by helping individuals to make informed choices that will decrease the risk of transmission even in the context of constrained public health resources. When COVI computes a high risk for an individual, that person may be guided towards local public health services for testing followed by manual contact tracing, with consent.

\subsubsection{Inform a data-driven approach to Covid-19 public health policy}
\label{sec:goal2}

In addition to the direct benefit of automated contact tracing, providing earlier awareness of infection risk by propagating Covid-19 risk across users of the application, COVI aims to generate information to guide local and global public health actions to control Covid-19. 

Of global relevance is information about the fundamental factors that influence disease transmission and drive decisions about containment and de-containment policies. The application will gather data on demographics and comorbidities which influence the risk of Covid-19 infection and the risk of adverse outcomes following infection. Individuals can also enter symptoms in real time, thereby creating a symptom evolution profile. These details are often challenging to acquire in retrospect due to recall bias \cite{hunger2013official}. These demographic and health factors are used within ML and epidemiological models (see below) to gain a more granular understanding of viral transmission dynamics. For example, it has been modeled that infectiousness can start approximately 2.5 days prior to symptoms onset with peak virulence in the day prior to symptom onset \cite{he2020temporal, lauer2020incubation} Viral shedding appears to plateau for 5-7 days, followed by a decline. While these details are key in understanding general viral profile in an infected individual, the ability to understand how this information translates into real-world viral transmission is limited \cite{leung2020first}. By matching individual level symptoms onset, digital contact tracing, the individual profile, and details of the contact profile (as described in Section~\ref{sec:goal1}) with self-identified and officially verified health data of Covid-19 diagnoses, COVI will provide critical information to enhance our understanding of real-world viral transmission and of the dynamics of the spread of the virus in the population of app users. 

Of local relevance, the epidemiological ML models trained on local data can be used to evaluate and optimize public health policies. The models can simulate the future effect of policies on the spread of the disease based on the response of citizens to recommendations directly delivered through the app as a function of the user profile and risk level of the individual. As the pandemic progresses, effective de-confinement strategies gain greater importance to aid in transitioning a population from strict to more lenient social distancing. With an understanding of how different user profiles influence infection risk, more targeted approaches to de-confinement can be proposed. The ML models and their output can be provided in aggregate de-identified formats to local public health agencies. Analysis of these data can provide a more precise understanding of the progression of the pandemic, which is critical to the creation of effective public health policy \cite{wang2020response}.This may inform region-specific strategies to loosen social distancing. COVI can also provide real-time and geographically localized feedback of viral transmission. For example, data about factors associated with higher numbers of contacts or environments with higher transmission risk could inform public health messaging. Similarly, monitoring the distribution of users across probabilistic risk levels, geography and other factors could enable rapid detection of increasing transmission risk in relation to de-confinement policies.  These aggregated monitoring data can also be used to evaluation the effect of policy changes, enabling data-driven adaptation of de-confinement strategies. 

\subsubsection{Protect privacy and maintain public trust}
\label{sec:goal3}

Despite the advantages of digital contact tracing, it raises legitimate concerns about individual privacy and civil liberties in democratic societies. In order to achieve the goals provided in Sections~\ref{sec:goal1}
and~\ref{sec:goal2}, strong privacy protection is paramount to maintain public trust in the technology and mitigate risks to human rights and democracy. 

COVI is designed to minimize the collection, use and disclosure of personal information and maximize the opportunities for users to give consent, while fulfilling its core purposes. There are two principal reasons for COVI's privacy-centric approach. First, we regard privacy as a form of public good, having value for the proper functioning of democratic political systems \cite{regan1995privacy, solove2008understanding} 
and civil society generally \cite{solove2008understanding, kundera1984being}. Massive adoption of contact tracing systems that do not sufficiently integrate privacy into the system design can pose a genuine danger to these institutions. Second, individuals all value their privacy at least to some degree, and because contact tracing depends so strongly on the network effect---there needs to be a critical mass of users with the app installed for it to be most effective---maximizing privacy protection therefore becomes a defining criterion for user adoption \cite{keith2010privacy, cho2020contact}.

It has been widely documented that automatic tracing systems risk offending the core privacy values of both those citizens who are suspicious of state intrusion as well as those who fear the potential humiliation or even reprisals brought on by having personal information exposed to third parties \cite{levine1988contact, whitman2003two, bayer2000surveillance, raskar2020apps}. Because automatic tracing systems can have far greater reach than manual tracing efforts, the potential effect on individual privacy is commensurately extensive. 
Despite the significant impact on privacy, however, the ability to collect data about the spread of the epidemic at a much faster rate than is possible with manual tracing may prove a key weapon to combat it. This leaves the governments of democratic states apparently impaled on the horns of a serious dilemma: to traverse the pandemic, must those governments choose between the preservation of life or fundamental human
rights principles?

Fortunately, it is a false dilemma. While any contact tracing system involves trade-offs, some of those trade-offs will be more palatable to democratic societies than others. COVI's contact tracing methodology is not an extension of central state authority efforts to locate and inform individuals. The COVI system also is designed to avoid as much as possible the possibility of third parties knowing an individual's risk level or infection status, while still exchanging enough information to positively affect individual behaviour and enable those other users to protect themselves and others. 

The opportunity cost entailed by the COVI approach is borne principally by the state, which has less information in its hands to exert direct control over the pandemic. However, this lack of direct control need not translate into harm to the community. Those using COVI in lieu of more invasive contact tracing methods can still benefit from the primary aim of contact tracing, which is to isolate and test those at higher risk of having an infectious status, provided that the recommendations made by the application are followed by the great majority of participants. State actors, moreover, can still benefit from the aggregate data that the system collects, which will aid in the formation of policy. In consequence, the trade-off presented by COVI results in minimal negative impact on the civil society of democratic states.

Furthermore, COVI seeks consent for the collection, use and disclosure only of that information that is needed for application function or to optimize its efficacy. In order to promote individual empowerment within the COVI consent framework, we also provide options where feasible. For further discussion of the consent framework, please see Section \ref{section:privacy}.

COVI's approach to maintaining privacy involves a multi-faceted strategy that makes privacy central to the application's functionality. COVI's primary approach to protect privacy is to decentralize the communication of risk between users \cite{tcn, covidwatch, applegoogle2020}. Furthermore, sensitive information about the digital contact trail (their networks of contacts) is by default only stored on their phone. This is done through cryptographic protocols to communicate securely between phones and between the server hosting the ML and epidemiological model. Finally, pseudonymized and aggregated data about users' individual and interaction profiles compiled with their consent is managed by an independent, not-for-profit organization whose sole mission is to protect the privacy, dignity and health of users for the time of the Covid-19 epidemic. This secured data will be used to train the ML predictive and epidemiological models. The data collected can never be used for commercial purposes, nor sold to private companies and will all be deleted as soon as the pandemic is over. Furthermore, information stored on user's devices will be purged on a rolling basis typically every 30 days (See Sec.~\ref{sss:data-retention-policy}). It cannot be used for surveillance or to enforce quarantine, and the government does not have access to the data beyond the aggregate level data shared with health authorities to inform decision making. 

Empowering individuals and protecting privacy are natural complements to each other, and both are important features for proactive initiatives affecting individual health \cite{patil2014big}. Information confers power; citizens provide power to government, for instance, by supplying the government with information. By retaining control of their information through robust privacy protections, citizens themselves hold the power—as well as the responsibility—to take action to address the crisis.

The application serves to organize information in such a way as to support individuals in taking informed decisions, so that they can use their power effectively and fulfill the responsibilities that they take on to influence the outcome of the crisis. Agency always remains squarely in the hands of individuals who choose to install COVI: citizens decide how much of their data to share, and how to respond to the recommendations it provides.

However, it bears mentioning that COVI's use of a more sophisticated risk message passing protocol, as opposed to the more traditional binary contact tracing, presents additional privacy risks. We believe these risks to be acceptable in the face of the extraordinary public health challenge presented by the pandemic and the advantage brought by machine learning in terms of reducing the reproduction number of the virus (the number of new persons
infected per infected person), see Section~\ref{sec:ML-results} so we have developed a protocol designed to minimize the privacy risks they introduce. From the outset of the project, we have worked with the Office of the Privacy Commissioner of Canada on a review of our protocol and have applied the principles of the Joint Statement by Federal, Provincial and Territorial Privacy Commissioners of May 7, 2020 entitled ``Supporting public health, building public trust: Privacy principles for contact tracing and similar apps'' \cite{privacystatement2020}. Furthermore we have have sought to respect the foundational principles of ``privacy by design'' \cite{privacybydesign}. For detailed discussion of the protocol and its attack surface, please see Section \ref{ss:protocol}.

\subsubsection{Protect human rights}
\label{sec:goal4}

The right to privacy is protected by the \textit{Canadian Charter of Rights and Freedoms} as well as by the \textit{Quebec Charter of Human Rights and Freedoms}. However, technologies – including the COVI app – can have an impact on other rights and freedoms. COVI has taken a human-rights by design approach to building and governing the app \cite{penney2018advancing}. We have also put thought to ensure that the app not only ``passively'' respects human rights (e.g.~ensuring that the algorithm not reproduce discrimination or biases), but we have also put in place a set of measures to ensure that Canadians’ rights and freedoms be actively protected at all times. 

By way of example, the COVI Project has taken the following steps:
\begin{itemize}
    \item COVI will be made available in multiple other languages than French and English, including Indigenous languages.
    \item  COVI will be collecting – voluntarily and based on users’ consent – additional data to help foster our understanding of whether and how the pandemic and the associated public health measures impact different communities or populations.
    \item The governance of COVI as an independent not-for-profit is structured to ensure that members of marginalized communities as well as Indigenous peoples are involved in making all important decisions related to the app and its associated data. Moreover, one of the core mandates informing COVI governance will be to ensure inclusion, diversity and equity at all times.
    \item COVI and its governance are independent from governments and will not allow any of its data to be used for surveillance, punitive or allied purposes.
\end{itemize}

\subsection{App overview}

\subsubsection{User interface}
The COVI interface for users works as follows. Upon download, users are provided with an overview of how the app works and the privacy implications of using COVI (See Section \ref{section:privacy}). After the user's age is verified, they are prompted to fill out a short demographics and health pre-conditions questionnaire to initialize the application. By default, all this data stays on the phone. Once the onboarding is complete, the user arrives on the home screen. There are four primary elements to the screen:
\begin{itemize}
    \item The tailored recommendations feature that helps users make real-time decisions daily about their activities based on their personal level of risk (out-of-app actions).
    \item  The ``action cards'' feature that prompts users to input additional/updated information to further tailor their risk profile (in-app actions).
    \item A survey and data visualization feature to allow users to express what is important to them and see how the crisis is unfolding.
    \item A ``share'' button for the user to help promote adoption of COVI among their friends, family, colleagues, etc.
\end{itemize}

Beyond these on-screen elements, COVI also supplies notifications to the user, either to update their in-app information (low-priority actions) or when urgent recommendations are updated (high-priority actions). When users are not actively using the app, it runs in the background, exchanging risk levels (in a cryptographically protected way) with other app users that they encounter. If a user gets tested for Covid-19, they will, in earlier versions, be able to self-report the results of the test. In upcoming versions, users will be able to fetch their test results directly within the COVI application.
When they input (or receive) a positive result, they will be asked for further consent to have it shared (through their elevated risk level) confidentially with recent contacts.

\subsubsection{Background Processes}
The user-facing functionality of the COVI app is powered by background processes which predict the user's risk level, record contacts, and retrieve the risk level information for recent contacts.
At its core, the COVI app is built around contact tracing and exposure notification.
Like many other automated contact tracing proposals \cite{tcn, bay2020bluetrace, chan2020pact, rivest2020pact, applegoogle2020}, phones make use of Bluetooth information to determine contacts.
We are currently testing several different Bluetooth frameworks---including TCN \cite{tcn}, Google-Apple Exposure Notifications \cite{applegoogle2020}, and a new system developed by the NHS \cite{nhs_app}---for determining contact events (See \ref{ss:protocol}).

Unlike many other digital conact tracing efforts, instead of simply tracing binary exposures to diagnosed positive Covid-19 cases, COVI uses Machine Learning (ML) to locally compute scalar risk levels which estimate when a user may have been infected and what their contagiousness might be at different days in the recent past. Although diagnosed Covid-19 cases are maximum risk, individuals who haven't been diagnosed can still present non-zero risk, e.g.~they might be asymptomatic but contagious. When these risk levels are substantially modified for a particular day in the past, they are then sent to all contacts of that day, enabling the network of COVI apps to recompute everyone's risk levels in a decentralized way, improving the overall accuracy of predictions. While risk levels play a major role in the background, they are never presented explicitly to the user.

Note that although we use Bluetooth to identify proximity with other users, we will use coarse geographic information through an augmented GeoIP database as an indirect feature for the risk predictor and for epidemiological modeling---we will not be directly using GPS due to privacy concerns.
For volunteers consenting to contribute their data for research, coarse-grained location information will also be sent to the COVI ML server and aggregated across users for the purposes of creating heat-maps, without associating them to traces of individual users. It suffices that a smaller fraction of the population opts in to yield sufficient training data for the predictor.

\subsection{Use of machine learning}

The information generated from raw automatic contact tracing is not very actionable to users.
Should someone who has had a single short contact with an infected person take self-quarantine as seriously as someone who has had multiple lengthy contacts (such as at a workplace)?
If a person has reported many of the symptoms of Covid-19 but has not yet been able to get tested, should their contacts still be alerted?
In manual contact tracing, a health official uses their professional judgment when making recommendations to contacts, which is time consuming and requires substantial expertise. 
COVI aims to take advantage of ML to optimize and automate the integration of clues regarding the possibility that a person is infected, and use the resulting graded risk levels to drive appropriate recommendations and 
signals sent to other users so they can update their own risk assessment.
Users of COVI can opt into sending their data (see Section \ref{ss:opt-in}) to COVI's secure ML server, where it is used to train two distinct but complementary models, the risk predictor and the epidemiological simulator.

\subsubsection{Risk predictor}
Instead of presenting users with raw data on contact events, COVI internally computes a set of risk levels for the past
two weeks. The computed personal risk levels are based on a combination of user-reported symptoms, demographic information, and information about contact events, including estimated
contagiousness (risk levels) of those encountered. They are then used for two purposes. First, the current risk level is fed into the personalized recommendations the app makes to users (see below). More precisely, an ML model predicts the
probability that a person has been infected, and how contagious that person was in the recent past and today. The estimated contagiousness in past days is crucial to inform the app of other users encountered in the recent past so that they can recompute their own risk levels. 
For example, imagine that Alice and Bob spent a lot of time together 3 days ago, and that because of newly available information, Bob's device estimates that he is probably infected and that he was likely very
contagious in the last 4 days. Bob's phone would then send a message to Alice's device about his updated expected contagiousness 3 days ago. Alice's device would then recompute her risk levels. By having that updated risk information at a time when she could
herself become contagious but before developing symptoms, COVI enables
early awareness of possible contagiousness. COVI would highlight messages on Alice's app suggesting that she increase isolation and minimize contacts. If Alice reacts like most people who realize they may be infected, she will act responsibly and considerably reduce the silent spread of the virus which would have otherwise occurred. From a probabilistic perspective, the risk predictor takes observed data
from the last two weeks and predicts the probability distribution
of past unobserved variables (like having been infected during a particular encounter, or the degree of contagiousness on different days in the past).

\subsubsection{Epidemiological simulator}

The volunteered pseudonymized data can also be used to fit an individual-level epidemiological model which captures the stochastic flow of events forward in time, through asynchronous events corresponding to movement of people, encounters between people, medical events (like becoming infected, having a particular viral load or some relevant symptoms) and behaviours (like wearing a mask at work, spending more or less time in different categories of locations like shops, offices, hospitals or parks). The epidemiological model includes prior knowledge about the relevant aspects of people's lives (like displacements and behaviours such as wearing a mask) and is structured around many conditional probabilities
for the above events which change the state of the system. These conditional probabilities are parameterized and these parameters can be estimated, with methods described in Section~\ref{sec:ML}, taking advantage of the risk predictor to sample the unobserved variables such as being infected and the degree of contagiousness.
These epidemiological models can then be incorporated in a simulator that
can be used by public health officials in several ways: to geographically map out the development of the disease (e.g., areas where people are getting infected faster), to understand the choices
of citizens (e.g., where are people better or worse at following recommendations) and to better define the factors which matter for contagion and how they interact. These epidemiological
models can also be used to simulate the evolution of the outbreaks under different hypothetical scenarios, and to optimize public policy with respect to objectives such as minimizing the number of hospitalizations due to the disease or
keeping the daily reproduction number $R_t$ below 1.

See Sections~\ref{sec:epimodel} and~\ref{sec:ML} for more details on both the risk predictor and the epidemiological model and how they can be trained together with
methods such as amortized variational inference.

\subsection{User experience}
While the technological implementation presented here represents a viable tool in reducing Covid-19 infections, the success of voluntary digital tracing is highly correlated with citizen uptake, participation in data sharing, and sustained use of the application.
In the same vein, because of the inherent risks associated with citizens sharing private data and receiving public health information, COVI was designed to align with  with the interests of users. Psychological science has shown us that the best manner of doing this is not through coercion but by eliciting the evolving preferences of users.
A large emphasis of COVI's design will thus be on creating measurement mechanisms (in-application surveys, focus groups, etc.) that allow us to better understand users and adjust settings and informational environments in a way that corresponds with their expressed (not assumed) best interests.
Evidence-based approaches from various sub-fields of science are used to achieve these goals in a number of ways, giving rise to the following foundational principles that guide COVI's road-map.
\begin{itemize}
    \item \textbf{User preferences drive end-to-end experience}. As government recommendations about appropriate personal responses to the crisis evolve and become more gradated and situation-dependent, it is important to understand users' evolving risk preferences. COVI uses a variety of measurement tools---at a population level, during the on-boarding process, and throughout the application's life-cycle---in order to elicit these preferences. Importantly, decades of research have shown that it is insufficient to ask people what they prefer---preferences must be elicited, validated and updated often \cite{gregg2006easier, friese2006implicit}.
    
    At the same time, we understand that initial engagement, regular interaction and sustained use are critical to COVI's impact on population level health outcomes \cite{ferretti2020quantifying}. Thus, the effective support of user preferences must be complemented by an engaging informational and visual design. We accomplish this by leveraging ergonomic checklists, user experience best practices and constant usability audits. In addition, emphasis is placed on creating engagement measurement mechanisms that allow for constant variant testing and iteration. This combination of closely tracked user preference and effective user experience result generates a dynamic interface that adapts to each user, allowing their ongoing interactions to be both empowering and engaging.

    \item \textbf{User comprehension is prioritized and verified rather than assumed}. Disclosure psychology tells us that there is a world of difference between ``technically disclosing'' and ``effectively communicating'' \cite{veltriivchenko2017}. In an effort to bring full transparency to COVI, the app (1) continuously and prominently displays a link to the privacy protections COVI utilizes, as well as to the privacy agreement itself and (2) employs user testing to ensure that these statements are not just de-facto present but are also reflected in users' awareness.

   \item \textbf{User empowerment to protect themselves and others is maximized}. Effective communication with the public regarding the evolving situation and appropriate course of action during a crisis is critical---especially when the public's reaction is the key driver of recovery. Information must be communicated clearly and recommendations must come from a place of authority. However, as with all communication, the meaning is in the response. With this in mind, we have combined an approach based on (1) established evidence from the field of crisis communication \cite{chang2019}, (2) an ongoing collection of primary data examining Canadians' reactions to  message variations and (3) user data examining the link between messaging and likelihood of reducing risky behaviour. We use these tools to ensure that the application empowers users to act in a way they deem appropriate.

    \item \textbf{User psycho-social well-being is promoted}. Given the sensitive nature of the communication contained within COVI, it is important to monitor users' reactions and ensure that we are not creating undue strain. While a complete lack of stress is an inappropriate reaction to a crisis, creating non-actionable urgency can cause mistrust and fatigue that impedes engagement with recommendations. In addition, such psychological strain can have more serious effects---from raising levels of anxiety to increasing instances of domestic violence \cite{gachter2011relationship}. Therefore, special attention must be paid to ensure that the data and recommendations contained within COVI are carefully crafted to deliver need-to-know information that limit the psychological strain.

    \item \textbf{User inclusivity acknowledges the diversity of their needs}. The Covid-19 crisis presents differential pressures and risks for different segments of the population \cite{devakumar2020racism}. Unfortunately, marginalized groups \cite{gostin2020responding} are both the most likely to be affected and the least likely to be able to access and use a tool such as COVI. For this reason, COVI uses accessibility best practices from the very beginning and leverages access, diversity and inclusion frameworks to identify possible gaps as early on in the product life cycle as possible in order to then address them as quickly as possible.
    
    Gender dynamics in Canadians' response to Covid-19 is an issue that is of particular interest in the context of inclusivity \cite{wenham2020covid}. By including gender analysis frameworks in our ongoing user behaviour tests, we are continually gaining a better understanding of the potential risks associated with gender dynamics (e.g.~the risk that tracing technology be used as a way to limit freedom of movement of a partner) as well as the opportunities (e.g.~developing models of household penetration of COVI while factoring in gender issues).
\end{itemize}

See Section~\ref{sec:behaviour} for more details on the psychological and user interface aspects of the app.

\subsection{Comparison with other approaches}
There are many different approaches to contact tracing \cite{rivest2020pact, bay2020bluetrace, troncoso2020decentralized, applegoogle2020, raskar2020apps, novid, chan2020pact, white2020slowing}.
Although a full taxonomy of contact tracing methods is beyond the scope of this white paper, we do think it is important to discuss some of the major design choices we made in building COVI as they relate to choices made elsewhere.

\subsubsection{Manual vs. automatic}
One important consideration is the degree to which a human has to be in the loop.
Recall that classical fully manual contact tracing involves a human contact tracer asking the patient to recall all of their contacts and locations in the last two weeks.
The MIT PrivateKit SafePaths/SafePlaces approach \cite{raskar2020apps} and the Singapore's TraceTogether \cite{bay2020bluetrace} app are augmentations to manual contact tracing, providing extra information to make manual contact tracing easier, but still retaining human judgment and touch.
This involves a significant amount of work on the part of public health authorities, but also enables careful professional judgment of the severity of contact.

On the other hand, fully automatic approaches \cite{novid, troncoso2020decentralized} have the benefit of requiring much less work on the part of the public health authorities, but also may lack the benefits of professional judgment.
Fully automatic approaches may also be more vulnerable to malicious parties attacking the system, because there is no built-in safeguard of human judgment at every step.

Due to how overwhelmed Canadian public health authorities are, we chose to have COVI fall much closer to the fully automatic end of the spectrum, while retaining a point of contact with public health authorities by providing high-risk or infected users with recommendations to report for testing. In this way, COVI can be complementary to manual tracing while also having the potential of making a significantly positive impact on its own, with no need for a human in the loop at the time where it matters, i.e., when early warning signals are being propagated through the network of contacts. Finally, the removal of a human in the loop reduces the risk of privacy breach and misuse of user personal data by a government authority.

\subsubsection{Types of risk messages}
The vast majority of contact tracing apps send binary exposure notifications, or at most a two-level notification where symptomatic diagnoses and clinical tests are differentiated. Like the NHS contact tracing app \cite{bbcnews_nhs}, COVI is seeking to send multi-level risk messages. The newly proposed Google-Apple API and the TCN Coalition protocol also includes some support for non-binary risk messages \cite{applegoogle2020, tcn}.

Although binary contact tracing reveals less information about users (notably, users who are not diagnosed reveal no information at all), sending multi-level risk messages allows for early and accurate customized recommendations to individuals.
We believe that the additional value COVI offers in more accurate personal risk levels and recommendations is worth the trade-off made in asking users to send more information to their contacts. This additional information translates into more precise customized recommendations, and can have a significant impact on the ability of COVI to empower users with the knowledge they need to protect themselves and others, especially in the pre-symptomatic phase of the disease.
The use of ML to integrate complex clues which would otherwise require human intuition mitigates the absence of direct human intervention into a fully automatic contact tracing app.
By sending multi-level risk levels and enabling accurate risk predictions, COVI can automatically give more relevant recommendations and perform triage on potential exposure events.
Of course, users will receive recommendations to contact a health professional when appropriate, and by augmenting binary tracing with risk messages, COVI addresses the shortcomings of automated tracing. 

\subsubsection{Centralization of data}
The types of attacks and adversaries one seeks to thwart play a central role in the design of contact tracing apps.
The volume of potential sensitive information, ranging from social contacts to medical histories to location histories, on a large subset of a citizenry rightly raises questions about abuse \cite{cho2020contact, chan2020pact, rivest2020pact}.
Thus, the degree to which people trust central authorities should be central in the design of a tracing app.

One simplistic way to design a contact tracing app is to upload full trajectory/contact information of all users to a central authority, who performs the matches.
The Israeli government seems to have proposed such an approach \cite{bbcnews_israel}.
Obviously, this is not at all private with regards to government authorities, but it does have the advantages of allowing the government detailed data on which to make public health decisions, in addition to allowing them to exercise professional judgment for manual contact tracing.
Furthermore, because all of the data is directly held by the central authority, in the absence of a data breach (though those are worryingly common \cite{liu2015data}), everyone's data is protected from other individuals.

However, many residents of other countries are less willing to hand over all of their data to a single central authority.
Especially if app installation is voluntary, it then becomes necessary to design apps that have greater privacy guarantees.
Many apps have thus instead taken a partially centralized approach in the binary tracing setting, where only diagnosed users upload their contact/trace data to a central server.
The central server can then use that data to notify users that they may have been exposed.
Classical manual contact tracing falls into this category, as well as the augmented manual contact tracing app TraceTogether \cite{bay2020bluetrace}.
This partially centralized approach gives the central authority data on all diagnosed users and their contacts/traces, and is a compromise that seems to be gaining traction in some circles, such as the Pan-European Privacy-Preserving Proximity Tracing approach \cite{pepp}.
Unfortunately, this partially centralized approach is incompatible with COVI's sending of non-binary risk messages, because non-diagnosed users are sending around risk messages alongside infected users. Any central authority would therefore gain data on nearly all users of the app.

Even further along that spectrum is the group of fully decentralized approaches, which attempt to prevent any one authority from having full contact/trace information.
Although most of these approaches involve sending data via a central server, the server does not get unencrypted/identifiable data in decentralized approaches.
COVI, and many of the other recent proposals, including DP-3T \cite{troncoso2020decentralized}, covid-watch \cite{white2020slowing}, TCN Coalition \cite{tcn}, MIT PACT \cite{rivest2020pact}, Washington PACT \cite{chan2020pact}, and the new Google-Apple API \cite{applegoogle2020} attempt to meet this standard using a variety of different technologies.
Fully decentralized approaches for binary contact tracing are arguably easier to protect, though there are still inherent privacy limitations (Section \ref{ss:inherent}).

Due to technical limitations related to the bandwidth of sending around many risk level messages, COVI Canada currently uses a different strategy (Section \ref{ss:protocol}) while also evaluating the feasibility of switching to either the Google-Apple API or TCN in the interests of interoperability and using open standards.

\section{Privacy and Consent Details}
\label{section:privacy}

As discussed above in Section~\ref{sec:goal3}, COVI is designed to collect, use and disclose as little personal information as possible and present as many opportunities for users to consent while still performing its intended purposes. By doing so, COVI complies with what Canadian private sector privacy laws require with respect to the principles of data minimization and consent. In several respects, however, COVI aims to go well beyond the minimum effort needed to satisfy the requirements of such privacy laws, which usually permit a flexible, pragmatic balance to be struck between individual rights and the needs of organizations that use personal information to provide their services. In other words, COVI does not aim to take advantage of what the law allows; instead, COVI seeks to provide its services while achieving privacy in the robust sense of maximizing individual control over information about oneself \cite{warrenbrandeis1890privacy, prosser1960privacy}.

In Canada, the collection, use and disclosure of personal information in the private sector is governed by the \textit{Personal Information Protection and Electronic Documents Act}, , SC 2000, c 5, and all provincial laws deemed substantially similar thereto. While there are differences between the federal and provincial statutes, these laws generally seek to ensure that organizations that collect, use and disclose personal information, among other things, (i) are held accountable for personal information under their control, (ii) are transparent about their privacy practices and the purposes for which personal information is collected, (iii) respect the principle of data minimization, (iv) apply appropriate security safeguards and (v) seek meaningful consent in a form appropriate to the circumstances and the nature of the personal information. 

When considering COVI's privacy model, it is important to bear in mind that privacy laws -- Canadian laws included -- are typically framed with the assumption that some level of trust will need to be vested in an organization which will act as custodian or controller of personal information. These laws aim to ensure that this trust is well founded by providing rules governing considerations such as those raised in the previous paragraph, and by endowing regulators with the power to investigate complaints, issue public reports and in some circumstances levy monetary penalties. 

With respect to its core functions, the COVI privacy model has been designed to eliminate, to the extent possible, the need for this kind of trust. As reflected in the discussion below in Section \ref{ss:protocol}, the steps taken to encrypt and obfuscate both the content and the routing of data that traverses the messaging system, and the deliberate decentralization or federation of control over various elements of that system, collectively seek to eliminate the need to vest significant trust in the organizations that act as service providers in the system, or in other users (within limits). While not entirely achievable in fact, the aim of the technical and organizational framework is to render the information circulating within the system meaningless to anyone but the intended recipient and frustrate third parties seeking to undermine that aim through snooping, coercion or bribery. Privacy laws as interpreted by regulators do not typically require organizations to go to these extremes to respect requirements related to security safeguards and data minimization.

That being said, because trust must be vested by users in the system as a whole, COVI also seeks to maximize accountability and transparency, and follows an express consent model to ensure meaningful consent. In support of accountability, COVI and its information ecosystem will be placed under the supervision of a not-for-profit independent entity constituted for this purpose, which will act as a fiduciary administrator of the COVI system and take responsibility for its operation as well as ongoing optimization of the privacy model. With respect to transparency, in addition to the customary privacy policy, COVI will publish accessible infographics through its website, this whitepaper, and release the source code for inspection under an open source licensing model. In support of obtaining meaningful consent, COVI seeks consent separately for different personal information elements and makes consent optional to the extent practically possible, as explained further in Section \ref{ss:consent}.

\subsection{Consent}
\label{ss:consent}
COVI seeks consent for the collection, use or disclosure of different personal information elements at different moments of the user experience. Although the form of consent requested is always express, it is presented as a condition of service (i.e., required) or optional depending on the nature of the information at issue and the purposes for which it will be used. In what follows we describe the consent sought at each of these moments. 

\subsubsection{Consent for use of data for core functions}
\label{sss:consent_core}
During the installation and onboarding process, COVI users are asked to expressly consent to permit COVI to collect, use and disclose the minimum amount of information needed to perform contact tracing, calculate risk of infection and exchange risk messages. In the initial implementation, the consent collected will be as follows:

\begin{itemize}
    \item  Consent will be obtained from application users for the \textit{collection, use, and disclosure} of the following information, as a condition of use of the service, by means of consent language and a privacy policy:
\begin{itemize}
\item GeoIP-based geolocation history
(only blurred positions
are kept, at the level of no finer-grained resolution than Canada Post's forward sortation areas, for the purpose of predicting location-based risk and modeling the spatial evolution of the disease)
\item Random contact IDs generated by the application
\item The user's current risk levels
\end{itemize}
The application will not function properly without this information and as such the user's consent is a condition for use of COVI. It is, however, important to note that with respect to the \textit{disclosure} of this information, the privacy protocol seeks to reduce its informational content to the greatest degree possible before it leaves the user's device; risk levels, for example, are sent to contacts without third parties (including government) having access to this information or the ability to connect it to any individual (see \ref{ss:protocol}).
    \item Consent will also be obtained from application users for the \textit{collection and use} of the following information, as a condition of use of the service, by means of consent language and a privacy policy:
\begin{itemize}
\item Age (user-reported)
\item Sex (user-reported)
\item Health conditions (user-reported)
\item Active symptoms (user-reported)
\item Ongoing relevant behaviour (user-reported)
\item Coarse geographical location (measured by GeoIP database)
\item Analytics information (use of application features that does not reveal any sensitive information about the user)
\end{itemize}
All of these data, apart from the analytics information, will be fed into the application’s risk assessment function (along with current risk level). Risk assessment will be undertaken locally on the application installed on the user’s device. None of this information will leave the device unless the user opts to allow it to be sent to COVI Canada for training the ML model and assist (in aggregated form) in epidemiological research by government or other third parties (see \ref{sss:consent_ml}).

The analytics information, which allows COVI Canada to track non-sensitive events such as whether users have completed installation and onboarding of the application, will be sent to COVI Canada in a pseudonymized form. We consider this collection as necessary for the function of the COVI application as it is critical that the application gain wide adoption in order to maximize its efficacy in respect of its core purpose. If a significant percentage of users are failing to complete onboarding, thereby preventing the application from functioning to propagate risks, knowing this will allow us to change messaging strategy, UI or UX in order to encourage completion of the onboarding process. While the information so collected leaves the device, this is not a disclosure, as the information is provided only to COVI Canada, which is the organization accountable for the application and its information ecosystem. The information is also manifestly non-sensitive in nature and is not associated with any information that would allow a third party to trace it back to the originating device. We distinguish this analytics information from a separate class of potentially sensitive analytics information that is only collected and used if the user has opted into sending information to COVI Canada to train the ML model, as we explain further below in subsection \ref{sss:consent_ml}.
 
\end{itemize}

\subsubsection{Consent for use of official positive test result}
\label{sss:consent_official_result}
Upon receipt of an official diagnosis,  COVI users may optionally consent to have COVI utilize an official positive test result for updating the user's risk level, which can then be used to improve the estimation of other users' risk level on their devices. The entry of the official diagnosis requires an authentication step involving interaction with the databases of health authorities that contain official diagnostic information; given the particular legal requirements in each jurisdiction a uniform approach may not be possible. In any case, following this authentication step the user will be expressly asked to consent to the \textit{collection, use and disclosure} of the official positive test result. If consent is granted, the application will onboard the official result and use it to calculate a new risk level. The updated risk level will then be communicated to recent contacts, to be factored into each contact's risk assessment. Although the official result itself is not shared, because the effect of the positive result on the risk level calculation will be to drive it towards its maximum possible output value, we treat the updated risk level as a \textit{de facto} indicator of positive infection status and therefore the communication of this information as a disclosure of the result.
Those contacted will not be told which of their contacts tested positive. Each contact's copy of the COVI app will calculate that contact's new risk level, and the end-users will only be given a set of customized recommendations based on that new internal risk level assessment.

\subsubsection{Consent for use of data for research purposes}
\label{sss:consent_ml} 
Finally, upon installation or at any time thereafter, users may grant consent on an opt-in basis (and may subsequently opt-out, at any time) to send certain information at scheduled intervals to the COVI ML server for the purposes of training ML and epidemiological models and sharing aggregated data with the government and other third parties. Users will be asked to consent to the \textit{collection and use} of the following information, which will be sent to COVI researchers in pseudonymized form:

\begin{itemize}
\item Age (user-reported)
\item Sex (user-reported)
\item Health conditions (user-reported)
\item Active symptoms (user-reported)
\item Ongoing relevant behaviour (user-reported)
\item Coarse geographical location (measured by GeoIP)
\item Certified positive infection status (if entered, pursuant to the consent provided under
\item Analytics information (use of application features that may reveal sensitive information about the user)
\end{itemize}

This data, apart from the analytics information, will be used for improving the risk prediction and epidemiological models.  It will also form the basis for generating aggregated, population-level data to be shared with government actors and other third parties, solely for purposes relating to efforts to understand or combat Covid-19. The analytics information will allow COVI Canada to assess such matters as the efficacy of recommendations with respect to lowering risk levels. 

All pseudonymized data will remain on COVI servers, and all processing of such data into aggregate form will take place before any aggregated data is provided to government or other third parties.

It is important to note in relation to the consent sought here that we regard the training of the ML model of fundamental importance to the core purpose of the application, even though we provide users with the power to opt-in or out. Although the ML model is initially trained with synthetic data to a degree that provides a moderately effective risk predictor, the model requires training on real data in order to yield a risk predictor with the level of accuracy needed to reduce the spread of Covid-19. 
Typically, privacy laws permit the collection of information necessary to serve a core purpose of a service as a \textit{condition of service} rather than an option. By making the collection and use of this \textit{necessary} information \textit{optional}, the COVI consent model goes well beyond what privacy laws require. Although arguably counterintuitive, our reasoning is easily explained. While the information collected for the purpose of training the machine learning model is necessary, it is only necessary that a certain quantity of such information be collected. For any given user, it is not necessary that information be collected from that user. As such, in recognition of the “privacy preserving” nature of the application, the application allows individuals to opt-in to allowing COVI Canada to access and use this information notwithstanding its necessity. 
The potentially sensitive analytics data obtained here is brought under this opt-in consent for similar reasons. As necessary as it is to provide the best possible recommendations, we do not need to track the correlations between recommendations made and risk levels over time for each and every user in order to assess the efficacy of the recommendations made.

Naturally, we hope that many Canadians will choose to volunteer their data so that we can build better risk prediction and better recommendations into the app, and create better epidemiological models to guide public policy. Whether any given individual decides to do so or not, the core functions of the app will nonetheless function for everyone. We accept those who do not participate in the hope that we will have a sufficient critical mass of volunteers providing data for training the ML model.

The data collected will be kept in the independent entity mentioned earlier in this section, and will be destroyed once it is no longer needed. The user may revoke their consent at any time, upon which their data will be deleted from the server. If they do not revoke consent, their data will still be automatically expired after a period of at most 90 days. Following some fixed period after which there have been no new cases, any data remaining in the unaggregated dataset will be deleted.

\subsection[Inherent privacy limitations of decentralized automatic contact tracing$^1$]{Inherent privacy limitations of decentralized automated contact tracing\footnote{This section was published in revised form in JAMIA. Ref: \url{https://doi.org/10.1093/jamia/ocaa153}} }

\label{ss:inherent}
Although there are many technological and cryptographic means of protecting information in transit and at rest, an automated contact tracing system is by its very nature a tracking system, albeit one with limited scope.
Because the system has to inform exposed users that they were exposed to someone who has been diagnosed with Covid-19, the system leaks information about the diagnosed users' identities.
There are endemic privacy risks that cannot be removed by technological means.
We believe it is of paramount importance to acknowledge and analyze these inherent risks, allowing both end-users and the government to make informed decisions on the amount of privacy loss they are willing to tolerate for the purposes of fighting the pandemic.

Before getting into details on our proposal, we will first analyze some of the systemic risks to decentralized automated contact tracing by considering an abstract system with the following desirable properties:
\begin{enumerate}
  \item The contact tracing is mediated through a smartphone app, such that when two phones are within 2 meters of each other, a \textit{contact} is recorded.
  \item When a user (Bob) is diagnosed with Covid-19 (or has an increase to risk level), all of their contacts (who we'll refer to as Alices) for the past 14 days are notified of the following fact: on day X, Alice was in close proximity with an infected individual.
\end{enumerate}
Even if Alice herself isn't directly notified of the day by the app---e.g.~the app only tells her that she should self-quarantine---this is equivalent from a security perspective so long as the phone is notified of the day, since a malicious app could extract the information.
Thus, we treat them the same in the privacy analysis.

There are two primary differences between this decentralized automatic contact tracing model and a more traditional manual or centralized contact tracing models:
\begin{enumerate}
    \item With automatic decentralized contact tracing, if multiple diagnosed users are in contact with Alice, she will receive an exposure notification for each individual. In normal manual contact tracing, Alice may only receive a single notification, despite being exposed to multiple individuals.
    \item Because it is decentralized, it is difficult to prevent an adversary from acquiring multiple identities in a Sybil-style attack, whereas in a more traditional model, there may be mitigation possibilities. Since the automatic contact tracing is smartphone mediated, an adversary with multiple smartphones may be able to acquire multiple identities.
\end{enumerate}
These differences allow for a set of attacks on Bob's privacy, where Bob is a user sending out exposure notifications/risk messages to his contacts.
For simplicity of discussion, we will describe the attacks below in the simplest binary exposure notification setting for Covid-19, but most of the attacks apply to any party that sends messages to anyone who has been in close proximity to them.
This is true whether the message contains a transmission risk value for Covid-19 or a list of self-reported symptoms.
We believe it is important to acknowledge these risks as a baseline before diving into discussion of the COVI architecture.

\subsubsection{Attacks on medical privacy}
One inherent privacy leakage of contact tracing is that it is derived from Bob's location history.
An attacker who has sufficient information about Bob's location history can perform a linkage attack to learn Bob's medical status.
Luckily, Bob's location history does not itself have to be broadcast, but even so, sending information to Bob's contacts implicitly reveals where Bob was.
Thus, businesses that have access to even parts of Bob's location history can gain access to his diagnosis status.
One example of such a business is the operators of a hotel; we claim that the hotel can determine the diagnosis status of any of their guests who they know to be using the app.

Let's begin with the simplest version of the attack.
The hotel places a phone in every hotel room running the contact tracing app.
If a guest Bob stays in Room 100 on June 1, and is later diagnosed with Covid-19, then the phone in his room will say that there was an infected individual in that room on June 1.
Because the hotel knows the guest register, they are trivially able to determine that Bob was diagnosed with Covid-19, breaching his medical privacy.

Of course, this super simple version of the attack can be partially thwarted by not allowing the hotel 1000 phones.
If you validate every single copy of the app so that only real people can possess them, then you prevent the simplest version of the attack, because the hotel cannot acquire 1000 identities.
There are of course other privacy issues that are raised by validation of app installation, but those may be surmountable through other means.

However, although the simple attack of having a copy of the app in every room can be blocked by user registration, that does not block a slightly more sophisticated version of the attack.
Suppose the hotel has 1000 rooms, and only 10 phones running the app, which is trivially achieved---e.g.~they have 10 employees running the app.
Then, at night when all the guests are in bed, each employee walks past half of their doors, only turning on the phone at the correct doors.

This is effectively a 10-bit code for each room, identifying it by which set of employees walked past their rooms. If employees 1, 3, and 5 walked past Bob's room, then his code would be 1010100000. Since a 10-bit code has $2^{10} = 1024$ possibilities, every room can get its own unique code with the right pattern of employees walking past.
Later, if employees 1, 3, and 5 get messages stating that on June 1st, they were in contact with an infected user, and none of the other employees get that message, then the hotel immediately knows that Bob was the one diagnosed.

While this may seem logistically challenging to coordinate, it is trivial to simulate by having a device in each room simulate employees walking by in specific patterns.
All a hotel needs is access to 10 real accounts, and with that, it is straight-forward to turn devices and identities on and off to identify every guest room.
These devices are no longer running the app as normal, but they are simulating the behavior of a real person walking past rooms in a weird pattern, and so this attack cannot be easily detected/stopped by the contact tracing system.

Another mathematically equivalent attack is the vigilante, or triangulation attack, where an attacker seeks to `out' an individual they have encountered as infected.
One motivation might be that the vigilante (say, Mallory) wants revenge on Bob for having exposed her to Covid-19.
This is in many ways mathematically equivalent to the hotel attack, but the difference is that Mallory does not know Bob's location for a fixed period of time, as the hotel does.

However, Mallory does know her own location history and when she encountered other people while going about the world.
If Mallory can narrow down the time of exposure to Covid-19 within a 5 minute period, she can reasonably guess when and where she crossed paths with an infected individual.
If during that time window, the only person she was in close proximity with was Bob, then she learns (1) that Bob was diagnosed and (2) that Bob was the source of her exposure, both of which are key leaks of information.

Again, Mallory could turn on a separate phone for every 5 minute period of the day, but there is also a logarithmic version of the attack.
Bcause there are 1440 minutes in a day, there are only 288 5-minute periods.
Using 9 phones, Mallory can similarly assign a 9-bit binary code to each 5-minute period, and depending on which phones get the exposure notification, Mallory will know when she was exposed.
As with the hotel attack, with a small amount of technical expertise and access to 9 identities, Mallory can write an app on a single phone that pretends to be 9 phones with the app installed.

In the above, we showed that it only requires logarithmically many identities for an attacker to reveal information about their contacts or guests.
In practice, many proposed decentralized contact tracing protocols do not require even that many, because they do not require strong user validation when users are attempting to determine their own exposure status.
For example, in several of the decentralized proposals we will consider later, all of the contact matching happens locally on the phone.
This is extremely powerful for protecting the user privacy of users who don't send messages---since they do not transmit any information off their phones---but also means that there is no straight-forward way to prevent an attacker from processing the contact matching multiple times.

\subsubsection{Attacks on user location history}
The previous subsection dealt with leakage of medical data, namely the diagnosis status of a user.
However, there is also leakage of the movement patterns of users.
In particular, any user that transmits information about their infection status (i.e.~Bob) also implicitly transmits information about their previous locations.
This is of course necessary for Alice to make a contact.
For example, in Bluetooth based systems, Alice records Bob broadcasting a Bluetooth advertisement; these advertisements are often random or pseudorandom, so they cannot be matched together without Bob's cooperation.
However, once Bob sends notifications to all of his previous contacts, information on his locations is leaked to at the very least Alice.

For the sake of simplicity in discussion, we will consider the case where Covid-19 exposure events are uncommon.
This may seem like an odd assumption in the midst of a pandemic, but it is a reasonable assumption for contact tracing because if most users are experiencing exposure events, then there is little signal in informing users that they have been exposed.
Unfortunately, in this case, the timing of exposure notifications also leaks a lot of information.

Let us again consider Mallory, who wants to reveal information about Bob.
We saw above that with a logarithmic number of identities, Mallory can reveal the time and place she encountered Bob.
If that time and place is in a public setting (e.g.~on public transit), Mallory may not be able to exactly identify who Bob is.
But now let's suppose that Mallory repeatedly encounters Bob on many occasions.
Normally, Mallory does not necessarily know that her many encounters with Bob are with the same person.
However, once Bob is diagnosed with Covid-19, Mallory receives a notification for every single one of her encounters with Bob, for which she knows time and place information.
Since exposure events are uncommon, Mallory is able to infer that all of her exposure notifications were likely for the same person.
Thus, Mallory is able to build a partial record of Bob's movements.

Note that this attack is not related to medical diagnoses, but is enabled simply because Bob is sending a notification to every time and place he's been, which is the point of contact tracing.
In some ways, this attack is limited in scope because Mallory could have simply remembered her encounters with Bobs in other ways (e.g.~by recording her vision and then post-processing using facial recognition technology, or with older smartphones, WiFi MAC addresses were also traceable).
However, it is still a leakage of the protocol.

The danger of the premeditated vigilante timing attack can be amplified if a large institution is the adversary; let's call her Grace.
Suppose Grace places devices running the contact tracing protocol and nothing else around at many locations around a city.
Using the premeditated vigilante timing attack, Grace is then able to correlate together location histories of many diagnosed individuals.

There is additional noise in the signal that Grace receives, because with enough devices, the assumption of relatively few exposure events is no longer sufficient to group together the location histories---multiple users will be reporting diagnosis events at roughly the same time.
However, location histories are contiguous in space, and if Grace has sufficiently many devices placed around, reconstructing trajectories is very possible.
Obvious candidates for Grace include a governmental actor or a large corporation, because they will have the means to deploy devices across a large geographical area.
Of course, this information leakage should be kept in context, as a large governmental actor already has access to many other sources of tracking information, such as cell tower pings or CCTV feeds. Hence what really prevents these attacks is the legal framework and social norms in the country, along with the political strength of public opinion and media. Transparency in the management of COVI is thus of primary importance to alleviate such attacks.

\subsubsection{Mitigations}
Unfortunated, both of the attacks given above work for every automated contact tracing protocol with the given properties.
Although there are technological solutions that can be make it harder or more annoying for an attacker to carry out, none of them can actually stop the attacks.
This is of course because identifying contacts with Covid-19 is the entire point of contact tracing.
Any real-world decentralized contact tracing system will furthermore have additional risks due to the design of the system.

A privacy maximalist would reasonably consider these attacks to be a reason to not use any decentralized automated contact tracing system.
However, even privacy pragmatists may be concerned about these privacy trade-offs, so we believe it is important to directly acknowledge, so that the users and the government can balance the value proposition of contact tracing for public health with the amount of data that is being exposed.
Furthermore, while technological solutions may be limited, we believe that there are legal and economic protections that can be put into place.

\subsection{Private risk messages protocol choices}
\label{ss:protocol}

Although as discussed in the previous section, some risks are endemic to contact tracing and cannot be removed, that of course does not absolve us of the responsibility to provide as much privacy protection as we can, while still achieving the aims of the system.
In order to enable the privacy guarantees for core functionality, we have to build a private messaging system \cite{van15, tyagi2017stadium, corrigan15} that ensures that no information about an individual's risk-level or contact history is revealed to the authorities or to other users, other than what is absolutely necessary in order to exchange those risk messages.
With the help of a number of external auditors and reviewers, we are currently evaluating three different systems for the private messaging system, each with their own benefits and limitations:
\begin{enumerate}
    \item Google-Apple Exposure Notification (GAEN) Framework \cite{applegoogle2020}
    \item TCN Coalition protocol \cite{tcn}
    \item NHS Bluetooth protocol \cite{nhs_app} + mix-nets for message exchange \cite{van15}
\end{enumerate}

All of the three systems have their pros and cons.
In this white paper, we do not go into full system-level design details, but instead evaluate at a high level the ways each of these messaging systems could be applied in our design.
Our initial deployment uses option (3) NHS Bluetooth + mix-nets, because there are technical and practical limitations to GAEN and TCN.

For the remainder of this section, we will refer to the different actors by name. Several of these characters we already met in the previous section, but here's a quick recap.
\paragraph{Dramatis Personae}
\begin{itemize}
    \item Alice, a user of the app. She encounters Bob on day $d$.
    \item Bob, a user of the app. He encounters Alice on day $d$. Later, with new information, Bob's quantized estimated contagiousness (communicated through the risk level) for day $d$ changes, and he wants to privately communicate that change to Alice.
    \item Grace, the government (or other central authority). She runs the central mailbox server containing all the reports.
    \item Eve, a passive eavesdropper, who tries to obtain information by overhearing communication, but doesn't do anything active.
    \item Mallory, an actively malicious actor, who tries to break the system, and will try to send false information to the servers and other parties. Any malicious user or entity can be or become Mallory.
\end{itemize}

\subsubsection{Google-Apple Exposure Notifications}
The Google-Apple Exposure Notification (GAEN) \cite{applegoogle2020} API has the significant advantage of being directly supported by the smartphone manufacturers, who have lower-level access to the Bluetooth stack than any other solution can.
Without going into too much detail, Alice and Bob's phones are constantly broadcasting Rolling Proximity Identifiers (RPIs) via Bluetooth.
These RPIs are derived from a daily Diagnosis Key, allowing later regeneration.

When Alice encounters Bob, the Alice's phone stores the RPI she hears from Bob.
Later, if Bob wishes to send a risk message to Alice, Bob publishes Diagnosis Keys to Grace, along with an attached quantized transmission risk message.
Alice periodically downloads all the Diagnosis Keys in a specified geographical area from Grace, and then locally regenerates the RPIs; whenever she regenerates an RPI she heard, she knows that the message is meant for her, informing her that she was exposed to Bob.

However, to prevent misuse of their system, GAEN also has strict limitations on participating apps, which make certain types of data collection and use cases difficult.
They do not currently permit advance consent for sharing Diagnosis Keys, which we require for fast risk propagation of risk along the network.
Additionally, they do not allow phones to access location services; although apps can ask users their location manually, this restricts the types of epidemiological data that can be sent to a public health authority.
Luckily, GAEN is a work-in-progress, and we are in discussions with both Google and Apple on workarounds for these limitations.

\subsubsection{TCN Coalition}
An alternative to using the officially supported Google-Apple framework is the TCN coalition protocol \cite{tcn}.
The lack of support from Google and Apple imposes some technical limitations on the capabilities of the framework.
Notably, iOS-iOS background Bluetooth communication is severely restricted.
However, the TCN approach does not have the consent and location limitations of GAEN.

The TCN protocol is so-named for having users share Temporary Contact Numbers (TCNs, roughly equivalent to GAEN's RPIs) with each other over Bluetooth if the phones' owners have been in proximity with each other \cite{tcn}.
For example, Alice has four of Bob's TCNs, and each time quantum is 5 minutes, then they have spent approximately 20 minutes within the distance boundary established by the protocol.
Phones can publish a `report' to a central server, associating a set of temporary contact numbers with a risk level payload (using the TCN report \textit{memo} field).
TCN reports allow the receiver to regenerate a set of TCNs that Bob broadcast over an app-specific range of time (e.g.~6 hours).

As with GAEN, to establish a match, users will download all new reports from a geographical area.
TCN reports allow the user to regenerate the TCNs that were originally broadcast, and users can locally check those TCNs against their internal log of TCNs received.

At regular intervals, users check for matching reports, and update their own risk levels based on the information they've received.
When a user's risk level changes substantially, the app posts their risk information to all relevant mailboxes associated with their contacts from the last two weeks.
Users may also update their risk levels based on other information, such as self-reported symptoms, or official Covid-19 diagnoses.
Their contacts can then regularly check the risk levels associated with the TCNs that the phone has logged.

\subsubsection{NHS Bluetooth + mix-nets}
One of the major drawbacks of both GAEN and TCN is the need to download all of the reports/Diagnosis Keys in a large geographical area for Alice to determine a match locally on her phone.
This approach has the advantage of revealing very little information about Alice when she retrieves messages, but if there are many risk messages being sent, also incurs a significant bandwidth requirement.
Furthermore, the TCN Bluetooth stack has not been as extensively tested as some other systems.
Thus, our initial deployment uses a third option.

The National Health Service of the UK has also designed and deployed a Bluetooth contact tracing app \cite{nhs_app}
and the NHS code has been further validated by field tests.
Additionally, instead of simply broadcasting a BLE advertisement, they actually create a communication channel between pairs of phones, which allows longer messages than a BLE advertisement.
The NHS Bluetooth messages between phones contain encrypted IDs to support a more centralized contact tracing system than the one we intend for COVI.
However, their libraries and code support sending Bluetooth messages in support of a more decentralized system (by simply changing what messages are sent).

When Alice and Bob are in close proximity, we use Diffie-Hellman secret sharing \cite{merkle1978secure} over Bluetooth to generate two shared secret contact tokens, one for messages from Alice to Bob, and the other for messages from Bob to Alice.
The use of Diffie-Hellman secret sharing prevents Eve or Mallory from being able to falsify messages from a user, though of course Mallory can act as a user herself.
A contact token can then be used to derive a `mailbox address' and encryption key using a one-way hash function.

Then, when Bob posts his risk status to Grace, he sends encrypted risk messages to the correct address.
Afterwards, Alice can check only mailbox addresses where she expects to receive her messages.
These messages contain exactly the same transmission risk information that would be attached to GAEN Diagnosis Keys or TCN reports.

However, active checking by Alice runs the risk of social graph attacks: Grace can see that Bob sent a message which Alice checked, and from that infer that Alice and Bob were in close proximity.
Thus, direct retrieval of messages (instead of downloading a large batch) necessitates additional layers of communication secrecy, which we achieve by using mix-nets (Section \ref{sss:mixnets}.

\subsubsection{Hiding sending/retrieval patterns}
\label{sss:mixnets}
Unfortunately, the pattern of mailbox retrievals can reveal sensitive metadata even if the addresses and messages themselves are encrypted \cite{greschbach12}.
For example, if Alice's IP address is seen checking a message Bob's IP address sent, then Grace knows that there was a contact between the two of them, allowing her to infer the social graph.

Thus, we either have to hide retrieval patterns or sending patterns.
There are a number of different ways to hide retrieval patterns.
The default mechanism in GAEN and TCN is for the database to be geographically sharded, and then Alice downloads the entire set of new messages in her geographic region.
GAEN and TCN do not protect Bob's message sending from the authorities, which reveals a Bluetooth trace, but because Alice downloads the entire database, there's no way to link individuals together in a social graph attack.

Other proposals \cite{berke2020assessing, trieu2020epione} make use of Private Set Intersection Cardinality (PSI-CA) protocols instead, which allow Alice to query the server for intersections, without revealing her set and without learning Grace's database. Unfortunately, these protocols tend to be very computationally and/or bandwidth expensive, and so are often not feasible in practice for a deployment on the scale of an entire country.
Furthermore, these protocols do not easily work with a message payload in addition to intersection detection.

Although GAEN and TCN hide retrieval patterns by downloading a sharded database, this comes at a significant bandwidth cost.
However, recall that to defend against social graph attacks, we need to \textit{either} hide sending patterns or retrieval patterns.
In the NHS + mix-net approach, we use a mix-net  \cite{chaum81} to hide sending patterns instead.

We onion-encrypt each of Bob's risk messages \cite{reed1998anonymous}, which are then decrypted in layers by the different mix-net servers, which also shuffle Bob's messages with those of other individuals before forwarding them on the next server in the network, and finally depositing the messages to Grace.

\paragraph{Mix-net design}
Each of the mixing servers $1,\ldots,N$ publishes a public key $p_1,\ldots, p_N$. For the purposes of this discussion, we consider Grace to control the last mixing server $N$. For each message $(x, m)$, which includes both an address $x$ and an encrypted message $m$, Bob sends $p_1(p_2(\ldots p_{N-1}(p_N((x, m)))\ldots))$ to the first mix server, where $p_i$ is encryption via public key $p_i$. The first mix server removes the first level of encryption, getting $p_2(\ldots p_{N-1}(p_N((x, m)))\ldots)$. The first mix server waits until it has received encrypted reports from multiple Bobs, groups them all together and shuffles them, mixing together the messages from different Bobs, and then forwards them as a batch to the second mix server. The second mix server does the same thing. At the end of the protocol, the final mix server (controlled by Grace) is left with a series messages of the form $(x, m)$, which have been decoupled from Bob.

Alice can then directly check all messages to an address $x$; although Grace learns the set of addresses Alice is checking, those addresses are not obviously linked to Bob, preventing Grace from linking Alice to Bob.
So long as one of the mix servers was honest and there is not an active attack with malformed data (i.e.~the remainder are honest but curious), Bob's messages have been shuffled in with other people's, and so are decoupled from his identity.

\paragraph{Attack (tracking message)} If the first mix server colludes with Grace in an active attack, they can discard all messages except for Bob's, replacing them with garbage messages. In this manner, they can determine which mailboxes Bob is talking to. In order to properly address this attack, each server in the mix-net must introduce appropriate noise messages to hide Bob's identity \cite{van15}. For our initial implementation, we do not include this noise.

However, we note that it is possible to detect this attack by sending ourselves messages through the mix-network. If they are discarded, then we will know that this attack is happening, and can take appropriate actions then.
Furthermore, this type of canary can be implemented by any user; thus, independent third parties can verify that the first mix-net is not performing a tracking message attack.

\subsection{Opt-in data for ML training and aggregation for governments}
\label{ss:opt-in}

Users will have the choice to opt in to sending pseudonymized data to COVI ML servers for use in further refining the ML model that determines `risk levels' based on past contacts, symptoms and demographic information.
If a user consents to this, the following information is sent at regular intervals (roughly daily) to the COVI server: a pseudonymized data packet and heat-map information.

\subsection{Pseudonymized data packet}
\begin{itemize}
    \item Age (in approximate bands), sex, pre-existing conditions
    \item User-reported symptoms
    \item Certified diagnosis status
    \item Number and duration of contacts, along with the risk levels of those contacts---this does not include the risk messages themselves, to prevent possible social graph attacks, only the risk levels themselves, along with the metadata of date of contact.
    \item Location types visited and activity. This will NOT include actual location information. Instead, the phone will locally group locations by type (e.g.~residence, grocery store, street, etc.), and will only send the types of locations visited.
    \item Pseudonymous identifier---this is necessary to allow the user to later revoke consent and delete their data from the COVI ML server.
\end{itemize}
Using the classification laid out by El Emam \& Malin \cite{el2014concepts}, we note that this packet includes no direct identifiers, but several quasi-identifiers, including age, sex, pre-existing conditions, and certified diagnosis status.
Age is coarse-grained into 10-year bands to reduce the reidentification risk.
The number and duration of contacts is not considered an identifier, as it is not temporally stable.
Similarly, we do not consider the types of locations visited an identifier, as we do not include specific locations in the data packet.
The pseudonymous identifier is a randomly generated string, needed only so that we are able to comply with deletion requests by the user.
The information contained in Packet 1 thus corresponds to de-identified information comparable to those found in a clinical trial database.

The pseudonymized data packet will be packaged in a data file by the phone, compressed, then encrypted with the public key of COVI organisation before being sent to the COVI ML server.
This data is of course \textit{highly} sensitive (similar to de-identified medical records or clinical trial information), running a re-identification risk should a malicious party get access.
The sensitivity of the data is of course why the data requires an active opt-in consent by the end user (Section \ref{ss:opt-in}).
The primary means of protecting the pseudonymized data are legal, through our data trust, rather than technical.
The technical pseudonymization techniques discussed here are only meant to reduce the risk of disclosure, but cannot and should not be relied upon to remove it. Standard security protocols for data protection such as SOC2 certification will be used to protect these files.

\subsection{Separated geographical information}
\label{sec:separate-zones}
Heat-maps of local risk levels are essential for public health officials to track and react to local outbreaks.
To avoid identifying a user by their geotrace, we do not send a user's geotrace directly to the server, and instead only send separate packets containing localized information about contact events and risk levels through a mix-net.
These packets will again be sent only with the additional consent of willingness to volunteer data for modelling and statistical purposes (the same consent as willingness to volunteer data for training the ML estimator).

Each of these contact event packets will include rough location information at the spatial resolution level of a Canada Post forward sortation area (a \textit{zone}; see Section \ref{section:geolocation}), which have median roughly 20,000 residents.
There will also include a few other latent variables like mobility and risk-averseness.

For modelling and outbreak tracking, several different types of data packets are needed.
These will all be sent separately through a mix-net to the ML server so that they cannot be easily connected to a single user.
Furthermore, as soon as the ML server receives the packet, it immediately aggregates it with other packets corresponding to the same zones/days, discarding the original packet.

\paragraph{Heat-map packet}
This information will allow public health officials to map locations where high-risk users are frequenting, without revealing who those users are.
\begin{itemize}
    \item Zone traversed during the day
    \item Day that zone was traversed
    \item Personal risk level on that day
    \item Mobility/risk averseness latent variable (4-bits)
    \item Old risk level if a previous packet was sent, but the phone now has better information on the risk level for that day (e.g.~after diagnosis).
\end{itemize}
Note that the heat-map data does not include any direct or quasi-identifiers, as none of this information is stable temporally.
We do note though that sometimes the zone traversed will correspond to someone's residence; this is why it's important to not send exact location, but only a forward sortation area with at least 100 persons.
For privacy reasons, we ensure that COVI never gets exact location using GPS, but rather instead only gets a rough location through a GeoIP database.
The mobility/risk averseness latent variable is a summary statistic corresponding roughly to how much the person goes out and meets people, as well as their hygiene habits, like mask-wearing and hand-washing. While hygiene habits are relatively stable, the phone will compute a single log of how risky a person's movements are, and use that to compute a quantized riskiness score for each day, which helps inform transmission risk.
We believe that this riskiness score is not an identifier because it will change as the person's movement habits change over time.

\paragraph{Flow-map packet}
This information will allow public health officials and epidemiologists to map the flow of transmission risk across residences.
\begin{itemize}
    \item Home residence zone.
    \item Day of contact event.
    \item Zone of contact event.
    \item Risk level received from another contact.
    \item Old risk level received from another contact (if a previous packet was sent, but was updated).
\end{itemize}
This packet does include the quasi-identifier of home residence zone, which may identify a user to within 100 possible persons.
The other three pieces of information are not identifiers.
This data packet is more revealing than the basic heat-map packet, but of course will still be immediately aggregated upon receipt by the ML server.

This data will be sent via a sending mix-net \ref{sss:mixnets}, to shuffle the data packets in with those of other users, and to hide the IP address of the sender.
In addition to the immediate aggregation on the ML server as soon as it is received, this procedure will mitigate against geotracing attacks, even on this opt-in data.

\subsubsection{Geolocation discretization}
\label{section:geolocation}
For privacy reasons, we believe that it is important that the COVI app not record exact locations.
As such, although the app may ask for Location Services permissions in order to use Bluetooth (as the permissions are combined), the COVI app will not directly record GPS.
Instead, the COVI app will make use of GeoIP services.
The app will send an anonymous query with no information other than external IP address to a GeoIP server.
The server will then resolve the IP address to a coarse location (depending on the resolution of the database, this may range from as fine as a Canadian Forward Sortation Area\cite{census2016} to as coarse as a city-level resolution, though we will ensure that no information finer than a Forward Sortation Area will be returned).
The server will delete all query logs within a day, keeping them only long enough to defend against Denial-of-Service attacks.
Because this process does not make use of GPS, we hope that it will be compatible with the Google-Apple Exposure Notification API, as their policy is that one should not combine that with Location Services.

Forward sortation areas (`FSAs' or `zones') are relatively large stable geographic regions with a median population of 20,000 residents.
As of 2016, there are a total of 1641 FSAs. The largest FSAs have >100,000 residents, and only 25 have fewer than 100 residents.
These regions vary greatly in size. In remote regions, extreme cases can stretch well over tens of kilometers.
To further preserve privacy, we will only send data about zones that contain at least 100 persons, and lump all the other low population-count FSAs together into a single code.

\subsubsection{Aggregation procedure for government use}
\label{sss:aggregation}
As described above, aggregated data will be shared with health authorities for public health purposes.
The COVI ML server plays the role of an aggregation intermediary, both for the immediately aggregated geolocation data, and for the demographic/symptom data in the pseudonymized data packet.
These data will include:
\begin{itemize}
    \item A daily aggregate heat map of infection hot spots and flow of transmission risk. To reduce the risk of reidentification, locations will be bucketed as described above (in forward sortation areas at the finest scale). 
    \item Epidemiological models. Trained risk models can be jointly fitted with epidemiological models and they are of course also a form of data aggregation, and one that our approach is uniquely beneficial for. These models will inform public policy on what kinds of contacts and symptoms are most at-risk for the spread of infection, as well as
    how different policy decisions could unfold according
    to simulations run with these models.
    \item Aggregate demographic information on symptoms and infection status. While extraordinarily sensitive, data on the relationship between symptoms and demographics is invaluable in providing accurate information to the public, and taking the appropriate policy measures to best control the spread. 
\end{itemize}
We will strive to meet the bar of 100-anonymity (in the sense of $k$-anonymity \cite{sweeney2002k}) for all aggregate data.
We will make sure that the
learning algorithms used 
are robust to model inversion attacks \cite{fredrikson2015model}.

Additionally, over the course of the pandemic, we expect that public health authorities may ask us for other types of aggregate information.
Provided that we can answer them while meeting the bar of 100-anonymity (or something comparable), we will compute those answers from the pseudonymous data and release that information.

\subsubsection{Data Storage Policy}

The pseudonymized data and geographical zone risk packets necessary for training predictive statistical and epidemiological models will be stored in a secured server with restricted access to selected AI researchers who will train these models.
This machine will not be managed by the government; we are in the process of setting up COVI Canada, a not-for-profit organization focused on managing these data according to the highest standards of good governance and with the sole mandate to protect Canadians' health, well-being, dignity and privacy.
All data in encrypted before leaving the user phone with the COVI Canada public key, stored encrypted and never decrypted before being used by an AI researcher, except for the geographical zone risk packets which will be immediately aggregated and then destroyed.
We expect to get new batches of data daily and to retrain risk prediction models at that rate.

\subsubsection{Data Retention Policy}
\label{sss:data-retention-policy}
All non-aggregated pseudonymous data will be automatically expunged after a period of no longer than 90 days.
Users additionally have the option to revoke their consent at any point using the app, which will cause their records to be deleted from the COVI server.
Because of the existence of rotating off-site backups, revocations may take up to 60 days to fully propagate (though in most cases, should be done within 30 days).
We note that for technical reasons, consent revocation can only happen through the app, because the server needs to know the pseudonymized identifier to determine which records to delete.
As such, a user whose phone is factory reset will be unable to revoke their consent; for them, however, the data will still expire within 90 days.

The aggregated data and risk models will be retained indefinitely for research and reproducibility purposes.
Of course, since the aggregated data and risk models are being widely distributed---in the former, to government health authorities, in the latter, to user's phones for local risk prediction---it should be assumed that this data is available to malicious users.
It is for this reason that the aggregated data and risk models must be suitably de-identified, as described above.
We expect that we should be able to give strong $k$-anonymity privacy \cite{sweeney2002k} guarantees for such an aggregated dataset, though details for this remain to be determined, and are dependent on the exact aggregation strategy.

Additionally, we may use the characteristics of the pseudonymous data to generate a synthetic dataset with similar features to the original raw data.
This synthetic dataset may similarly be retained indefinitely, but should not be vulnerable to linkage attacks, and may even be made fully public for use by other researchers.

\subsection{Residual Risks and mitigations}
\label{sec:residual}

Unfortunately, hackers, scammers, and other nefarious agents are a part of life, preying on, among other things, users' trust in institutions. Once COVI rolls out with governmental support, it will become part of the ecosystem of attack vectors, and we need to keep that in mind as we design our messaging and protocols. In this section, we will discuss both remaining technical attacks on the protocol which we do not fully protect against and social engineering attacks on the users themselves. Below, we will list out some of the attacks we envision, as well as potential means of mitigation (note that there is of course some overlap with Inherent Privacy Limitations in Section \ref{ss:inherent}, as the risks of COVI are a superset of the risks of decentralized contact tracing in general).

\subsubsection{Vigilante attack}
\label{section:vigilante}
The vigilante, or triangulation attack is one where an attacker seeks to `out' an individual as infected. Our privacy model is built such that triangulation is (1) impossible to do retroactively (i.e.., after Alice's risk level increases, she then tries to track down the source), and (2) technically and logistically annoying to do prospectively.

Once an app determines a contact event, we will have the app forget the original messages that were received over Bluetooth; the app will thus no longer have information on the exact time of contact. By deliberately forgetting as much fine-grained time information as possible, we hope to reduce triangulation risk while retaining the ability to do contact tracing and risk awareness propagation.

\paragraph{Single-party premeditated attack}
\label{sss:single-party-premeditated}
Unfortunately, someone who has built a cracked version of the app in advance could have it record all of the contacts and associated risk messages, along with exact time and location of contact. This attack is a particularly egregious data breach when performed by a party (e.g.~a hotel) that knows the whereabouts and identity of a particular user, because that party can then expose the medical status of a user.

As discussed earlier, this style of attack is possible with any contact tracing app, no matter the safeguards (Section \ref{ss:inherent}). Our technical protocol does not attempt to prevent this premeditated attack, though we should note for completeness that an attacker only needs a cracked copy of the COVI app in advance, rather than additional physical phones. As it is not possible to prevent this attack technologically, we must explore legal and economic solutions instead.

\paragraph{Multi-party retroactive attack}
The multi-party triangulation problem, where multiple individuals band together to try to figure out who infected them, is even more difficult to prevent. However, note that this is not specific to a contact tracing app. If several people all get sick after a group encounter, with or without an app, they may be able to coordinate together to figure out that it was someone in that encounter who infected them. We try to ensure our app does not exacerbate this problem by not exposing contact event details to the users themselves, rather instead just providing recommendations which depend on an updated risk level. For example, we do not provide the risk level in clear (the app only gives recommendations which generally also depend on other factors like location or medical conditions) and we do not tell the user what contact event may have caused their recommendations to change, if any.

\subsubsection{Rogue authority attacks}
While it may be possible for individuals to perform some of the attacks below, we believe that the biggest risk here comes from a rogue authority corrupting the mailbox server and ML server.

\paragraph{Social graph attack}
In order to prevent social graph attacks by the mailbox server/authorities, we hide either Bob's sending patterns with a mix-net or Alice's retrieval patterns by downloading all messages in a geographic area. 
In the mix-net setting, note that while we protect the social graph by hiding Bob's sending patterns through a mix-net, Alice retrieves directly, so the mailbox server can see how many social interactions she has had, though not with whom.
This can be defended against by having Alice further retrieve messages through a mix-net or through anonymizing proxies, though we do not provide any guarantees on this front.

Additionally, some residual risk remains in targeted social graph attacks on subgroups where sufficient numbers of individuals choose to opt-in to sending pseudonymized data to the COVI ML servers.
The rogue authority can inject specific unique risk levels by having a device broadcast uncommon risk levels to contacts.
Those contacts then upload those unique risk levels to the ML server for the flow-map, revealing their home forward sortation areas because those messages can be tagged by the ML server.
Those unique risk levels then also can be used to tag the pseudonymous data packet in the same way, which may allow for membership inference attacks.

We mitigate this vulnerability in two ways: (1) the risk levels are quantized to 4-bits (16 levels), and (2) the risk predictor makes full use of the entire range of 16 levels.
Because the ML risk predictor regularly outputs scores in the entire range, there will not be unique risk levels that can be used for tagging.
That the ML risk predictor honestly uses the entire range can be independently verified by 3rd parties, as the risk predictor is public information.

\paragraph{Location history attack}
A rogue authority (Grace) with control over the mailbox server and who has deployed contact tracing devices around a city can learn a significant amount of information on the location histories of users. 
This is an inherent limitation of decentralized contact tracing (Section \ref{ss:inherent}).
Although social graph attacks are defended against by hiding either Bob's sending or Alice's retrieval patterns, whichever one of the two is not hidden may have implicit location leaked to the mailbox server.

In Bob's case, if sending patterns are not hidden (the default in the Google-Apple Exposure Notification framework \cite{applegoogle2020} and TCN \cite{tcn}), then the set of Bluetooth advertisements he sent are in theory known to the mailbox server.
By deploying a set of Bluetooth listening devices at many locations, the server thus learns a partial location trace for Bob.

Similarly, in Alice's case, if retrieval patterns are not hidden (the default in the NHS + mix-net approach), then the set of mailboxes Alice checks are sent to the mailbox server.
If the mailbox server deploys active Bluetooth devices at many locations that perform the Bluetooth handshake with Alice, then the server thus learns a partial location trace for Alice.
In a binary post-diagnosis exposure notification system, this attack is worse, because only some users send notifications (as Bob) but every user checks (as Alice).
In our risk-message passing system, many users have non-baseline risk, so the distinction between Alice and Bob is primarily one of role, as most users will play both roles.

A potential partial mitigation for these location history attacks is to hide both sending and access patterns.
For example, if a sending mix-net is used for Bob in the GAEN and TCN proposals, that would provide some amount of anonymity (though partial traces of 6-24 hours may still be available to Grace).
Alternately, a retrieval mix-net could be used for Alice in the NHS + mix-net approach, though retrieval mix-nets are more difficult to scale than send-only mix-nets

We are actively exploring both of these mitigations, but have not yet come to any conclusions and cannot make promises as to their feasibility/scalability.
Furthermore, we note that even with these protections, a rogue authority that deploys Bluetooth devices around a city can masquerade as real contacts, receiving risk messages.
These risk messages themselves are correlated, and may reveal partial identities of message senders, so hiding the IP address of the sender through a mix-net is only partial protection anyway.

Additionally, we note that although contact tracing can be used as a mechanism for location tracking, there are many other options already available to adversaries with governmental resources, such as using CCTV and facial recognition or cell tower pings.
It is due to these difficulties that we have categorized the location history attack under residual risks.
As these are attacks by the central authority, we hope that there may be legal and governmental oversight defenses.

\subsubsection{Privacy leakage to businesses}
Another source of residual risk is large institutions other than the mailbox server wishing to find out more information about a targeted subset of people who they physically have on their premises:
for example, an employer wishing to spy on their employees.
There are already many examples of employers doing hyperlocal location tracking with e.g.~motion detectors or WiFi triangulation, and it is important
to consider how COVI might present another means of tracking.
Alternately, a less nefarious version of this may be a grocery store wanting to know how high risk their customers are, so some of this information is may be beneficial to `leak.'

\paragraph{Medical status leakage}
Any party, including a business, can set up a device pretending to be a COVI install, which will then receive the risk messages days later from anyone who walks past the device.
Again, this is an inherent leakage of decentralized automatic contact tracing (Section \ref{ss:inherent}).
Luckily, the time delay of up to a day between
the encounter and the transmission of the risk message makes it impossible to associate a risk level in real-time to a particular person walking past the device.

For the legitimate use cases, the COVI team should consider providing a local monitoring app that does not reveal exact times, but only reveals the aggregated risk statistics of a location.
This information is similar to the aggregate heat maps that we plan to provide public health officials, and they should only reveal the same aggregate information with k-anonymity privacy protections.

While we cannot technologically prevent the illegitimate tracking, there may be legal protections that can be put into place.
For example, in the terms of service, we ask users to not hack the system; this will not prevent a malicious attacker, but may at least discourage businesses from sending/receiving false messages on the system, especially if we can provide them with an avenue for legitimate local monitoring.

\paragraph{Hyperlocal location leakage}
Unfortunately, although medical and risk information is only available through the app protocol, which we control, there is some amount of information that is leaked simply by way of broadcasting the Bluetooth messages.
By setting up Bluetooth receivers, a business can triangulate to within 2 meters the position at all times of every person on their premises.
While some of this is already possible with motion detectors and WiFi triangulation, Bluetooth likely permits higher resolution versions of the same thing.
Note that because our Bluetooth messages change every 5-15 minutes, such a system does not know exactly who is each person that they detect.
However, much can be inferred, from, say, the amount of time a particular desk has a person present
(although it is also trivial for an employer to
know that a particular person is present at their
desk by other means).

Such a system can also be used to track movements of individuals around a public place (e.g.~a grocery store).
The system would not know who it was who entered, because the Bluetooth messages are varying randomly, but would allow the grocery store to determine how long people stopped in front of particular displays.
We do not believe this information to be significantly different from what can already be tracked via security cameras, motion detectors and WiFi triangulation, but it is an additional data leakage.

\subsubsection{Phishing attacks}
Phishing attacks are a fact of Internet life. They are characterized by a scammer masquerading as a trusted entity, and using that trust to convince a mark to do something. COVI, as a new entity with governmental support, will have to deal with scammers masquerading as it.

\paragraph{Accessing fake COVI URL on roll-out}
Once the COVI publicity campaign rolls out, residents of Canada everywhere will be encouraged to download COVI to assist in contact tracing efforts. Since COVI is a phone-based app, an obvious initial attack vector is for Mallory to blast untargeted text messages to as many numbers as possible, hoping to get Alice before she downloads COVI. That text message would claim to be from government, encouraging them to visit \url{http://fake-covi-url.ca} to download the app. Once Alice visits this URL, a malicious payload can be downloaded onto their phones, or perhaps the URL asks them for personal information (e.g.~SIN, health number, etc.). The attack can be customized based on the real information that COVI asks for (e.g.~demographic information), so that if Alice asks a trusted tech-savvy friend for advice, they may not realize that Alice is on a fake COVI page.

For this reason, it is important to be very careful what information we ask the user to disclose. Any personal information we train them to disclose is information that they may be less hesitant to disclose to a malicious website.

Furthermore, a message can also ask the user to virally forward it around, preying on Alice's sense of civic duty. This message makes more sense as a social media post. e.g.~``The Canadian government asks us all to download COVI to help fight Covid-19. Go to \url{http://fake-covi-url.ca} now, and forward this to all your friends so we can beat back Covid-19 together!'' The social media post can be structured with a real link to a news source or COVI press release, to give it an air of legitimacy, with the only fake payload the malformed URL.
As soon as such as scheme is uncovered, it is thus
important to warn people about it.

\paragraph{Installing a fake app}
Notably, the URL above does not have to actually lead to an app download. In fact, the URL can even redirect to the real app after getting user information, to help the attack go unnoticed. Alternately, though, as an ongoing attack, Mallory can instead direct Alice and Bob to a fake app download, which then completely compromises the users' phones. This attack will hopefully be  prevented by the Apple and Google app stores, but it's possible something like that could slip through the net. An attacker can modify an existing app to sound like COVI, and people could be sent text messages from the ``Health Authorities'' with a link to that fake app.

\paragraph{Fake COVI diagnosis text message}
When the diagnosis notification server is set up, COVI users will be trained to expect a text message with a code/URL to enable their apps to send an infection notice. If Mallory knows that Bob has recently been tested, she can send him a text message claiming to have his test results at a specified URL. The URL can be any malicious payload, as described above. This attack is even more nefarious, because Bob, having recently been in contact with the health authorities, may be more amenable to disclosing a health number or SIN, because those are numbers he sometimes does give out to the health system. The malicious website can plausibly ask to need Bob's name, address, SIN, and health number, which can later be re-purposed for identity theft.

Mallory does not actually need to know Bob was recently tested, because she can of course just broadcast out the text message generally. However, luckily, this attack does not seem amenable to viral social media spread.

\paragraph{Revealing protected app information}
Sometimes, phishing attacks are designed to get individuals to perform an action to reveal data stored on the phone. However, because COVI pre-supplies the COVI encryption key and URL, as well as the mix-net coordination server, the user does not have the ability to reveal their location or Bluetooth trace to a 3rd party. This is in contrast to apps which give Alice and Bob the ability to send their geotraces to a health provider (e.g.~via email).

\subsubsection{Spread of misinformation through risk levels}
Phishing attacks generally provide Mallory with useful information. However, if Mallory just ``wants to watch the world burn,'' she may also just try to incite panic in users and/or their contacts. COVI allows users to self-report symptoms, which factor into the risk levels. If Mallory can convince a sufficiently large number of users to submit false information, she may be able to break the contact tracing system and incite panic in the population.
This attack is mostly feasible only until the Public Health authorities are linked to COVI to send official confirmed diagnosis with one-time code. Afterwards, no unofficial test results can go through the network; while self-reported symptoms will still factor into the risk level, they will play a much smaller role and cannot influence risk levels as much as a claimed diagnosis.

\paragraph{Fake diagnosis results}
This is a variant on the fake diagnosis text message above, but should be performed using a phone call for maximum effect. Mallory calls Bob, claiming to have test results in the standard fashion used by public health authorities. She tells Bob he is infected, and needs to inform his contacts through COVI. Although COVI does not allow sending a confirmed diagnosis without a one-time code from the real Public Health authorities, Mallory can still tell Bob that he should mark all the symptoms, raising his risk level. Thus, even without a code, Mallory can increase the risk levels of Bob's contacts. Done at scale, this may incite panic, and certainly will decrease the utility of COVI for contact tracing. The primary mitigation for this attack is simply having self-reported symptoms play a much smaller role in risk estimation once the official confirmed diagnosis integrations with provincial health authorities are complete.

\paragraph{Incentive to falsely report}
Should COVI become government sanctioned, there is some possibility that employers may use COVI as a proof of illness. Even if that's not the case, Mallory can convince people that they'll be able to get something by self-reporting as infected. This probably makes the most sense as a viral social media post, a malicious `life hack' tip for individuals.

The message content will tell readers that if they self-report as infected, they'll get better medical care or time off of work. The self-reports are magnified by the contact graph risk messages. Overall, this will incite panic and take up medical resources. These messages also have the side-effect that they may increase distrust in COVI, because if people believe that other people are lying, then they themselves will trust COVI's recommendations less.

Although there is no way to fully defend against false reports before integration with provincial health authorities, we can build in suitable incentives in the confirmation prompts for self-reports.
Because of the negative impacts of user self-reports, we also recommend that COVI risk status/recommendations should not be used as verification of illness by 3rd parties, removing these types of incentives.

\subsubsection{Appealing to desire for vengeance}
We have designed the privacy protocol in its final incarnation to make it difficult for a non-technical user Alice to determine who exposed her after the fact.
It is impossible to prevent Alice from determining that Bob exposed her if Alice is premeditated (see Section \ref{section:vigilante}), because COVI is designed to forget exact time/location information of contact before any risk messages are sent.
This hopefully makes it more difficult for Alice to later determine that it was Bob who exposed her, though it is of course imperfect.

However, a non-technical user who desires vengeance may not understand the privacy protections of the app. Mallory can advertise to Alice a service where Alice pays/downloads an app and Mallory will claim to be able to figure out Bob's identity. For the purposes of this attack, it doesn't actually matter whether or not Mallory is able to do that. Mallory can still either get Alice to download a malicious app or get bitcoin out of Alice to perform that service.

\subsubsection{Disinformation campaigns}
Although we do not expect active disinformation campaigns in the early stages of the app launch, we cannot discount the possibility of such attacks, given their prevalence in modern online life, sometimes from foreign state actors \cite{garrett2017echo}.
Many disinformation campaigns will take the forms of social engineering attacks described above (and can be guarded against as such), but there are a few specific to the motivations and scale of such attacks.

\paragraph{Fake high risk levels}
One easy means to destroy the usefulness of the app would be to create a large number of false high-risk reports---e.g.~consider 6 million apps reporting high infection risk in Montreal, Toronto, and Vancouver.
Luckily, Bluetooth provides proof of presence; furthermore, all three of the approaches were are considering are designed to prevent replay attacks making it infeasible for an attacker to replay existing messages with false updates. Thus, this attack requires the attacker to have a physical device present broadcasting Bluetooth signals.

Although we expect the physical device requirement to be the primary deterrent, we can further somewhat limit the impact of this kind of attack by limiting the number of messages per day per IP and per network block.
This can help prevent a single phone from pretending to be many phones simultaneously (though it is of course imperfect protection).
Additionally, if the servers start detecting an anomalous number of messages from an IP block, or from IP addresses of non-Canadian origin, that can be filtered as a possible disinformation campaign.
Furthermore, these same abuse protection mechanisms can help guard against Denial-of-Service attacks on the infrastructure.

Of course, Mallory can craft a malicious version of COVI that just sends max undiagnosed risk levels to everyone she encounters while walking about, possibly multiple times, and then finding sufficiently many IP addresses to broadcast the appropriate fake messages. This is impossible to prevent, because that's similar to the behaviour of a real COVI app with an infected user. She can only do so much damage in this way, because she can only be in one place at a time. However, if she can convince a large number of people, spread across Canada to do this, it'll again break the usefulness of the risk predictors.

A partial mitigation of this attack is for an ML model to count multiple simultaneous high-risk contacts as equivalent to a single one, because a malicious app can fake 1000 infected patients at a particular location. In our epidemiological simulations, we plan to explore whether this adjustment will decrease prediction accuracy, and if not, this mitigation may be implemented.

\paragraph{Data leaks}
A well-funded adversary may also target the privacy of the system as a way to prevent its adoption.
Most of the contact tracing data is guarded behind the private messaging systems we are considering (Section \ref{ss:protocol}).
However, one easy target is the COVI ML data collection server, which will contain pseudonymized data on hundreds of thousands of Canadians.
Although we do not associate the location data with individual users,
a data breach on the COVI ML server would still be on the same scale as a data breach on a major hospital systems' records (though without any full identifiers), and thus, the server should be treated with similar caution.
Such a breach would not only expose personal information on hundreds of thousands of people, but also hurt public trust in the privacy on the system as a whole.

For this reason, the raw pseudonymized data should not be kept in such form for longer than absolutely necessary to train the models.
As described, some amount of data retention is needed to have sufficient data to train accurate risk models, but the raw data should be (and will be) automatically expired on a regular basis, leaving only the aggregate data.
In particular, any location-associated data will be immediately aggregated and deleted, as we do not need the individual location-associated data and it is of particular sensitivity.
See Data Retention Policy \ref{sss:aggregation} for more details.

\section{Epidemiological Model Details}
\label{sec:epimodel}
The volunteered pseudonymized data will be used to fit individual-level epidemiological models capturing the stochastic flow of events forward in time.
This will include such events as movement of people, encounters between people, medical events, and behaviours.
These models can then be incorporated into a simulator that can be used by public health officials to geographically map out the development of the disease, understand the choice of citizens, and better define the factors which matter for contagion.

\subsection{Structure of the epidemiological simulator} \label{ss:sim}

The simulator is a stochastic agent-based model, implemented in Simpy. A population of humans is created in a city, and each human moves around the city according to mobility patterns generated by an EHR (Electronic Health Record) model. A portion of the humans have the disease, and as they move around the city (spending time in places like home, work, transportation, stores, hospitals, long-term care facilities, etc.) they may infect each other, have symptoms, become hospitalized, etc. We track the spread of the disease through several metrics, ($R$, attack rate, etc.), and we tune parameters of the simulator to match these metrics to real data and to the output of a compartment (SEIR \cite{li1995global}) model fit to COVID-19 data from Wuhan and adapted to Canadian demographics \cite{lauer2020incubation}. The simulator outputs sequences of encounters with disease transmission information, and we use this output to create a dataset for machine learning models to predict individual risk from observed variables like symptoms, pre-existing conditions, and the places the individual has visited. In turn, this risk predictor can be used to tune the parameters of the simulator to data which would be collected on a mobile app.

\subsubsection{Implementation details}
The City has a graph structure of Locations---including households, stores, parks, hospitals (including ICUs), and nursing homes---and various modes of Transit---including walking, biking, metro/bus, rideshare/taxi, and car. Each location and mode of transit has a capacity and disease transmission properties associated with it.

Each Human has individual characteristics (age, sex, pre-existing medical conditions, carefulness, whether they have the app, how often they wear a mask, where they work, etc.), which are sampled according to demographic information for Canada.
Humans have epidemiological properties including viral load, infectiousness, and symptom progression during the disease, which depends on these individual characteristics (see below for details).

Many Events happen as humans move around and time passes:
\begin{enumerate}
    \item An Encounter happens when two humans are close enough in space and time. Currently space has roughly 2m resolution, but there is significant variance due to differences in Bluetooth implementation on different devices. To mitigate that, we use the signal strength rather than the distance as input and allow the trained predictor to best take into account the inherent uncertainty in distance. Currently time is in 15 minute windows. If one of the humans is infectious, they will infect the other human with a probability proportional to their infectiousness.
    \item If humans log their Symptoms, or get a positive or negative Test result, a new risk level is calculated. We currently only model one test type: lab tests with 0\% false positives and 10\% FNR; others are planned.
    \item Humans may become ill with a cold or flu (currently a random subset of 1\% of the population, weighted by age and some other characteristics). This generates a more realistic distribution of symptoms (Covid-19 is not the only thing which causes symptoms). Seasonal allergies are also planned.
\end{enumerate}
We have implemented various degrees of social distancing and other interventions and are currently exploring their effects on disease propagation.

The Infection timestamp is the exact time a Human was infected. Humans may be infected either by a location, e.g.~if a very infectious human had been there not too long ago, or more likely by an encounter with another human. We track the source of exposure for each infected Human.

Viral load is modeled as a piecewise-linear function with three pieces: increase, plateau, and decrease.
The increase starts after a number of incubation days (a Gaussian centred on 2.5 days), and proceeds up to the plateau value over a number of days sampled from a Gaussian centred on 2.5 days.
Currently the plateau is sampled just according to age, but we plan to make this depend on other individual characteristics (e.g.~pre-existing medical conditions, behavioural changes), as well as the initial viral load during the encounter. The plateau lasts for a number of days sampled from a Gaussian centred on 5 days
The decrease lasts for a number of days sampled from a Gaussian centred on 5 days \cite{lauer2020incubation, chang2020modelling}.

Infectiousness is proportional to viral load, but depends on characteristics such as being asymptomatic, immune-compromised, wearing a mask, coughing, etc.

Symptom progression depends on the viral load \cite{he2020temporal,ferretti2020transmission}. For each of the 3 stages, symptoms are sampled according to their average prevalence in Covid-19 patients. Symptoms start a number of days after infection that is sampled from a Gaussian centred around 2.5 days.

There is an approximately 40\% chance that someone will be Asymptomatic, (which lowers their infectiousness to 10\% of what it would have been otherwise), 15\% that they will get Really Sick (requiring hospitalization), and 30\% of those Really Sick will get Extremely Sick (requiring ICU). There is a 0.2\% chance the person will never recover. These values are sampled individually, and depend on things like age, pre-existing medical conditions, etc.  The simulator could model re-infection, but currently does not due to uncertainty in the literature about how frequent this is.

Whether someone wears a Mask is currently sampled according to how careful someone is. Masking is 98\% effective for hospital workers, and 32\% effective for others \cite{jefferson2020mask}.
We validate several population-level metrics about the simulated data, including:
\begin{enumerate}
    \item $R$ broken down by household, hospitals, and other locations,
    \item Encounter transmission rates,
    \item Secondary attack rate (\#tested positive / \#symtomatic),
    \item Fraction of symptomatic cases by age, and
    \item Qualitatively matching an SEIR curve.
\end{enumerate}

\section{Machine Learning Details}
\label{sec:ML}

The pseudonymized data volunteered to the COVI ML server will be used to train machine learning (ML) models which predict contagiousness risks and fit an epidemiological model. The models will be trained offline so as to not overload phones: training on
phones would require substantial compute power because of the
iterative and lengthy process of training. Another reason
for training offline is that we will need to try different models to determine the optimal configuration of the learning algorithms, only then sending the algorithms and parameters for the selected predictive model. For the chosen learning algorithm currently in the phones, the parameters should be re-estimated regularly
(up to a daily frequency if necessary)
and their updated values then sent to the phones.

To aid in this process and ensure predictions are well-calibrated even in early days of app adoption, we pre-train the ML models on simulated data generated from
an \textit{a priori} version of the epidemiological model,
described above. That epidemiological model is also
a simulator which can create histories of individual
contacts, behaviours and viral transmission. When we begin to collect real data from the app, we will ``close the loop`` and also tune parameters of the simulator to match the data as it is collected.  Including the use of the app and its risk predictions inside the simulation allows us to accurately model the impact of various interventions, since the whole
point of COVI is to empower citizens with information leading to targeted rather than uniform confinement
in order to contain the virus; see Section \ref{ss:sim} for details of the epidemiological simulator.

\subsection{Encounters between users}

When two phones with the app meet each other, they will exchange
(with a delay of up to a day) information about each other's risk
(more precisely how contagious their app estimates them to be at the
time of the encounter). Later, as additional information accumulates
on each phone, those risk estimates regarding the day when they
met may be revised. If the revision is sufficiently important (because
the risk level changed from one discrete level to another), an updated
message is sent to the other phone. For example, if a user starts having symptoms related to Covid-19, this user's phone would increase the probable contagiousness in the preceding days and send an update message to all the phones of the people this user met in the past 14 days. This enables each user to obtain an updated personalized  risk level and propagate this updated risk across prior contacts. If the change in risk is significant enough (which is likely if new symptoms in a contact emerge or if a user just tested positive to the disease), an updated risk will then be propagated through update messages sent to the user's past contacts. The purpose of the risk estimator is to predict a user's current and past contagiousness. The former influences the personalized recommendations so individuals can better protect those around, while the latter is sent to prior contacts, so they can become aware of their risk of being contagious and act accordingly.

\subsection{Privacy considerations}
For a full description of technical privacy, see Section \ref{section:privacy}.
Here, we give a brief summary of the points most salient to protecting privacy of individuals on the ML server used to train models and gather epidemiological data.

\subsubsection{Message passing}
When a phone sends a message to another phone via the secure messaging protocol,
the receiver will not know from which phone (neither phone number nor IP address)
the message comes from. To provide additional protection against stigmatization,
these messages are sent with a random delay of up to a day. In this way it is
not possible for an honest user Alice to know that her increase in risk level is due to
an encounter with a particular person Bob, unless she only had
a single encounter that day (See Section \ref{sec:residual} for attacks by malicious parties).
To further improve user privacy, risk levels are quantized to 4 bits of precision before being exchanged.
We note that this is a comparable amount of information to the 3-bit transmission risk level used in the Google-Apple protocol \cite{applegoogle2020} and to the self-reported symptoms memo that the coEpi project attaches to TCN reports \cite{tcn}.

\subsubsection{Data sent to ML server}
As described in Section \ref{ss:opt-in}, users may opt into contributing further data for research purposes. 
For opt-in users, two types of packets are sent to the ML server, both via a mixnet or proxy so that the ML server cannot know from which person the data originates.
The first type is a pseudonymous data packet containing health information as well as details about recent contacts.
The second type is a separate geolocation packet that is de-associated with the user information in the pseudonymous data packet.

The geolocation packets consist of a location ID corresponding to a forward sorting area (zones) containing several hundred to tens of thousands of people (Section \ref{section:geolocation}), attached to some metadata (including risk score and home forward sortation area for the user).
This data is used for building epidemiological heat maps (Section \ref{section:geolocation}), which can help local public
health authorities locate areas where the disease is concentrating or spreading faster, or where most dangerous contacts are happening.
As discussed, we do not use GPS to acquire this data, but instead only use a GeoIP database, which relies only on IP address.
In the pseudonymized data packets sent to the ML server, the zones are not sent at all; rather instead, the risk factor derived locally on the phone from the zone is included.
This is later used as input to the ML predictor.

The opt-in volunteered de-identified per-person records are kept in a secure server under security protections
appropriate for a pseudonymized medical record database.
As noted in Section \ref{sss:aggregation}, aggregated data is shared
by COVI with public health officials.
This aggregate information includes both these hotspot maps
and the parameters of the epidemiological model described in Section \ref{sec:epimodel}.

\subsection{Viewing the simulator as a generative model}

Let us try to abstract out some of the most important random variables
involved in the epidemiological simulator. First of all, the simulator samples
these variables in the order they would happen in time. Each event may
happen at a particular time or in a particular time interval (such as a day).

In addition to random variables attached with a particular time, there are
time-independent variables such as the answers to the questions the user may
provide when they install the app. This includes for example the age, biological
sex, and pre-existing medical conditions which could have an impact on the course
of the disease, but also questions on their household (how many people live in their
lodging), their work (e.g., if they work with Covid-19 patients) and their
behaviour (e.g., do they wear a mask). These {\em static} variables may
be revised after installation but they are considered a static property of the person.
In the simulator, these variables are sampled from a prior distribution $P({\rm static})$ based
initially on known statistics, and when data are available from the app, from the
population-level averages of these answers. Another very basic kind of variable
consists in the displacements of everyone in the population. These displacements are
not available for modeling directly but can be abstracted from other sources of
mobility data in order to build a model of how and with what frequency
people typically move around, from
their home to their work, to hospitals or shops, etc. Let us lump these
displacements under the {\em mobility} variable and call $P({\rm mobility})$
their distribution. Sampling from {\em mobility} gives hypothetical trajectories
of hypothetical people spending time in different places in the simulation.
Below, we note that mobility of a person may depend on {\em awareness} of the
person (either because of her symptoms or because her phone warns her to take
more precautions). Governmental public health policy can also influence mobility (e.g., by
allowing or not different types of locations to operate normally). Hence
the actual mobility model has the form 
$P({\rm mobility} | \text{\rm awareness},\;{\rm public\; health\; policy})$.
When they spend a few minutes near each other, this will trigger a {\em contact}.
Thus the simulator has a process for capturing the conditional distribution 
$P({\rm contacts} | {\rm mobility})$,
which essentially amounts to finding out when two people spend 5 minutes or more 
near each other. The specific attributes of the contacts may include how long and at what
distances the two people were. Now some of these contacts will be recorded by people who have
a phone with the COVI app and will constitute some of the observations available for training
(minus the precise position and exact time of the encounter, for privacy protection).

The most important latent variables modeled by the simulator are the infection status
and contagiousness
of each person, on each day of the simulation. A person can be in four states:
susceptible (not yet infected), exposed (infected but not yet infectious), infectious
(i.e., contagious) or recovered (including the unfortunate case of being dead).
While the person is infectious, a continuous variable which we call contagiousness
carries information about the ability to transfer the virus to someone else
(which may be because of the viral load and because of behaviour like coughing, etc).
The simulator models the temporal evolution of these latent variables as 
$P({\rm infection} | {\rm contacts},\;{\rm static})$,
i.e., conditional on the contact events and on the medical conditions of the person.

The time-dependent variables observed on the phone come in two main categories: 
on one hand, the symptoms and test results,
which may be entered on any day, and on the other hand
the observations associated with {\em contacts}, e.g., the noisily corrupted duration and
distance characteristic of the contact 
(along with an estimation of spatial precision because
it may differ depending on which kind of sensor
is used and measurement conditions), and the risk level sent by the other phone some time after the
contact itself, making the whole thing an event-based
asynchronous form of Dynamic Bayesian Network. Let us call {\em medical observations} those entered symptoms and test results.
The simulator thus has a model $P({\rm medical\;observations}|{\rm infection})$ which
can be sampled from for each person separately.

There are also several feedback loops in the probabilistic graphical model that we are
sketching, going through the risk predictor itself. First of all, the outputs of the
risk predictor are used to send {\rm messages} to other phones, which are part of
the {\rm contact} information discussed above and observed on the receiving phone.
Second, the simulation can take into account the influence of the app on behaviour,
by producing {\em awareness} (of potentially being contagious) in the person, 
which creates another feedback loop with $P({\rm mobility}|{\rm awareness},\;{\rm public\; health\; policy})$.

To summarize, exogeneous variables of the
system are ${\rm public\;health\;policy}$
and ${\rm static}$ variables about each
person. From these and initial states
of the simulation (some proportion of
susceptible, exposed, infectious, or
recovered individuals, as well as the
contagiousness of the infected ones),
the simulation can run forward in time
for several weeks or months over a hypothetical population with some mobility and density characteristics
(which may have to do with whether this is
an urban or rural environment, for example).
Because of the temporal obfuscation done in the phones,
it is enough to consider the time steps of the simulation as the successive days $t$.
We can view the whole system as an asynchronous dynamic
Bayes net, in the sense that the same types
of variables (and dependencies between them)
are instantiated for every time step of the
simulation. During the simulation, the following abstract variables are updated
each day according to this order and the above
logic:
\begin{align}
 P({\rm mobility}(t)|{\rm awareness}(t-1),\;{\rm public\; health\; policy}) \nonumber \\
 P({\rm contacts}(t) | {\rm mobility}(t), {\rm messages}(t-1)) \nonumber \\
 P({\rm infection}(t) | {\rm contacts}(t),\;{\rm static})\nonumber \\
 P({\rm medical\;observations}(t)|{\rm infection}(t))\nonumber \\
 P({\rm risk\; levels}(t)|{\rm phone\; data}(t))\nonumber \\
 P({\rm messages}(t) | {\rm risk\; levels}(t))\nonumber \\
 P({\rm awareness}(t)|{\rm risk\; levels}(t))
\end{align}
where ${\rm phone\;data}(t)$ is a shorthand for the sequence of observations available
inside the phone in order to compute the risk levels. That includes the 
static information, the messages received, the observed part of the
contacts and the observed medical observations (when 
they are observed).

Of course, disease spread, even in the midst of a pandemic, is hyper-localized.
Single individuals can cause an entire cluster of infection events \cite{washpost_korea}.
Although the simulated epidemiological models by themselves can predict such sporadic events, they cannot tell officials where those events might be.
For this reason, COVI will be providing public health authorities with an aggregated heat map of risk levels and infection events (See \ref{sss:aggregation} for more details).

\subsection{Observed and latent variables, inference predictor and generative simulator}

The pseudonymized data collected on the phones
will serve to train the risk predictor but also
go into training the above epidemiological simulator,
which can be seen as a generative model from which
one can sample new synthetic histories of contacts and
contagions, including into the future. This simulator would
be useful so that public health officials and epidemiologists can better understand the spread of the disease. The specific building blocks of the
simulator are conditional distributions (such
as the above, or decomposed into finer
levels of detail) which characterize
events like the transmission of the virus from one person
to another, the appearance of symptoms, the results
of a test, or how people change their behaviour
due to the messages displayed by the app.
The structure and parameters of these conditional
distributions, after being fitted to the observed data,
will also contain precious information for the 
epidemiological understanding of the disease (such as on the importance
and interactions of specific factors like distance, duration, or wearing a mask,
during an encounter with someone else) and 
about how people react to recommendations.

Let us call the risk predictor $Q$ and the epidemiological
simulator $P$ by analogy with variational methods
and variational auto-encoders~\cite{Kingma+Welling-ICLR2014} in particular, 
a correspondence developed below.
To understand how the risk predictor and the
epidemiological simulator interact and help each other,
it is important to understand that the simulator can
sample values for two kinds of random variables:
the ones (let us call them $X={\rm phone\; data}$) 
which can be observed on individual phones
(like the occurrence of contacts, the reported symptoms,
or the test results) and the ones which cannot be directly
observed, called the
latent variables (like the actual viral load, 
contagiousness or infection status of a person).
Let us denote the latent variables
not observed in one way or another
by $Z$. The simulator $P$ is
actually a generative model $P(X,Z)$ for the joint distribution
of $X$ and $Z$ over time and across users.

For much of the discussion below, we will focus on the
data observed within each phone, since this is the only data available
to the risk predictor running on that phone. The latent variables
$Z$ for Alice's phone 
contains important quantities about the underlying state of Alice,
such as whether she is infected, since when, and what has been her
contagiousness since she got infected. These variables are not directly
observed but characterize the risk she poses to others, and a quantized
form of her contagiousness (the risk level) will be sent by Alice to
the phones of the people (like Bob) she met in the past, as 
described above. The risk predictor $Q$ is actually an estimator
of the conditional probability $Q(Z|X)$ and in general it should
be viewed as an approximation of $P(Z|X)$.

For non-trivial generative models like $P(X,Z)$, typically parameterized
as $P(X,Z)=P(X|Z)P(Z)$, it is not computationally tractable to compute
$P(Z|X)$ exactly. In other words, whereas simulating samples and trajectories
from $P$ is easy given the parameters and structure of $P$, inverting 
that process to recover the latent $Z$ given the observed $X$ is difficult.
Several machine learning approaches exist to perform this approximation,
also called inference, and whatever the solution chosen, we will call
the resulting predictor $Q(Z|X)$. We will assume that one can sample
from $Q(Z|X)$, and that if pairs $(X,Z)$ were observed, they could be
used to train $Q(Z|X)$ by some optimization procedure, i.e.,
a training method.

\subsubsection{Training the risk predictor and the simulator together}

We thus need to learn both a generative model $P$ (here a part of the epidemiological
simulator) and an approximate
method $Q$ (the risk predictor)
which figures out the non-observed variables $Z$ relevant for
a particular user, given the data $X$ for that user. If we knew
the true $P$ we could just sample $(X,Z)$ pairs from it and train $Q$
by supervised learning (typically by regularized maximum likelihood). As a first step
in building a predictor, we have in fact constructed an epidemiological
simulator $P$ based on medical and mobility statistics, and generated
a large set of trajectories (e.g.~30 days over a population of 30,000
people in a town), thus leading to many $(X,Z)$ pairs. We can then
use these pairs to train a first risk predictor which is faithful to
the simulator and this is the predictor shipped in the first version
of the app, before any phone-level data is collected.

When data are collected, what we obtain is a collection of observed $X$ records (one per phone with the app, for each day of data collection).
We can then use our pretrained $Q(Z|X)$ predictor to obtain samples
of corresponding $Z$'s, and a possible strategy for training $P$ from this
is then to use these inferred and sampled $Z$'s along with the corresponding observed $X$'s
as a training set of $(X,Z)$ pairs to update $P$. This would lead
to a different $P$, one more faithful to the data distribution
over $X$, and we would then need to retrain $Q$ to be consistent
with the new $P$ (if $Q$ is trained rather than a fixed
procedure). Iterating this procedure is essentially
the wake-sleep algorithm~\cite{hinton1995wake}. Unfortunately, it is not guaranteed to optimize a well-defined objective function. A modern variant of this idea was introduced in {\em amortized variational methods} such as
the variational auto-encoder~\cite{Kingma+Welling-ICLR2014}. An upper bound on the log-likelihood $\log P(X)$
can be written which involves both $P$ and $Q$ and that can be optimized
by gradient-based methods. This is for now our method of choice for
training both $P$ and $Q$ together. We also propose to use samples from $P$
(which may go outside of the range observed in $X$) to enrich the training
of $Q$ (in a fashion similar to the wake-sleep algorithm). This should help
to address the challenge that the data distribution is not stationary, 
as society quickly evolves to face the pandemic. Keeping $Q$ aligned
with epidemiological knowledge not just around the data points $X$ but
more broadly should help bring more robustness to the system.

\subsubsection{Amortized Variational Inference}

The log-likelihood of the observed data $X$ can be lower-bounded
by this variational objective, also known as the ELBO or expected
lower bound:
\begin{align}
    {\cal L} = E_{Q(Z|X)}\left[ \log \frac{P(X,Z)}{Q(Z|X)}\right] \leq \log P(X)
\end{align}
where equality is achieved when $Q(Z|X)=P(Z|X)$.
Hence, jointly maximizing ${\cal L}$ over both $P$ and $Q$ has been found to be very successful in modeling observed data $X$
when we suspect that it is better explained by invoking
latent variables $Z$, which is exactly our situation.
Classical variational methods optimized $Q$ separately for each given $X$, whereas
modern amortized inference methods like the variational auto-encoder~\cite{Kingma+Welling-ICLR2014}
parameterize $Q(Z|X)$, e.g., as a neural
network, and allow a faster inference at the time when the data are given.
Instead the iterative optimization is done offline when training $Q$,
while just computing the output of the neural network $Q(Z|X)$ is very quick and can be performed on a phone, which makes amortized
variational inference appealing for training the app's risk
predictor.

The ELBO can be maximized by usual gradient-based methods. The typical
flow of computation proceeds as follows. Given the data available on one
phone $X$ up to some date, one samples latent variables $Z$ from 
the risk predictor $Q(Z|X)$. Then one computes the log of the joint probability
$\log P(X,Z)$ as well as $\log Q(Z|X)$. Finally, one can back-propagate 
the ELBO into $P$ to update its parameters and into $Q$ to update its
parameters. The only complication comes from the fact that $Z$
includes both continuous variables (the contagiousness values for
each day) and discrete variables (the contagion events). Back-propagating
through the continuous variables is easy and efficient through the
reparameterization trick~\cite{bengio2013estimating,Kingma+Welling-ICLR2014}. Propagating
gradient information through the sampling of discrete variables is 
slightly more involved, with the simplest technique involving a
Monte-Carlo approximation of the gradient obtained by sampling
(once or a few times) the discrete variables and using the log-likelihoods
of $P$ and $Q$ as reinforcement signals (similarly to what is done
in the REINFORCE algorithm \cite{williams1992simple} and its variants). More sophisticated methods
such as the Gumbell softmax method~\cite{jang2016categorical} lead to more efficient estimators
at the price of a bit of bias in the gradient.

Note that $Q$ is just an approximation for information stored in $P$, so its
parameters are in a sense not really free as they have to be consistent with
both $P$ and the observed data. In our case, $P$ is going to be fairly compact
and have rather few degrees of freedom, so the complexity in $Q$ arises from
the need to compile in the neural network not just the knowledge about $P$
but how to perform the kind of inversion or inference which is needed
to go from data to latent variables (whereas $P$ goes from latent variables
to data).

Note that many questions remain open to improve on this
early design. The tension between accuracy of the risk prediction 
and privacy are forcing the development
of novel machine learning methods. For example, not
knowing the full contact graph (for privacy reasons but
because simulations would hardly scale if we had to represent
the contact graph of all Canadians) is forcing us to imagine
novel ways of performing inference, possibly at different
spatial scales. Our current implementation sidesteps this question
by only doing inference at the level of individuals but more
work is needed to draw larger scale conclusions about the
spread of the virus through a network of individuals.

And of course, other relaxation methods (to share risk information
through the network of phones and converge to a consensus)
or inference
methods might be more efficient than the ones proposed here.
Thanks to the fact that the viral transmission probability is tiny
for any particular contact, when a user's risk is bumped up
significantly (e.g.~because of a positive test result), the
information diffuses through the network and dies out exponentially
fast (quickly reaching zero effect because of the quantization
of risk levels).
Future work on the machine learning aspect will clearly
have to include comparisons between different approaches.
The metrics used should also be application-dependent.
In the short term we are using the model's log-likelihood,
precision and recall of infected individuals via their
risk level.

However, ultimately, the best way to evaluate different methods
is in terms of how well they manage to curb the spread
of the virus for a given amount of average mobility.
How should that be measured? In silico, a measure of this
can be obtained by running a simulation with the predictor
in the loop of behaviour and evaluating how this impacts
mobility and the spread of the virus, and how the trade-off
between the number of hours out of isolation (e.g.~at work)
and keeping the reproduction number can be best achieved.
Another ongoing question is how to scale simulations
to large populations? Current simulations have been run on up
to 30,000 individuals. How do we scale to the size of a
country like Canada? Although local effects are the most important
to understand viral propagation and growth, deconfinement will
open up questions about the spread of the virus between regions
and countries and we need appropriate computational tools at that scale.

\subsubsection{Risk message privacy}
Since users are sending risk messages to all of their contacts, there is data leakage in the risk messages themselves.
This is true regardless of what protocol/mechanism is used for the contact tracing (for the Google-Apple API, this is a 3-bit risk message \cite{applegoogle2020}, for the coEpi proposal, this is a self-reported symptom list \cite{tcn}), and is endemic to sending risk messages.
As an extreme case for the sake of illustration, if the risk message included the SIN of the user, then there would not be privacy at all in the protocol, because users would effectively be broadcasting their IDs with a delay through the private messaging system.
Even in more realistic scenarios, the risk messages themselves may carry correlatable information.
For example, if Alice and Bob run into each other every day, and most people are not infected, then Alice receiving the same risk level update from Bob for a set of contact events may allow her to infer that all of Bob's messages came from him.
While we hide this information from users of the app, a dishonest user with a hacked app could potentially record the actual messages.

For this reason, the space of possible risk messages should be sufficiently small that each user has some plausible deniability for having been the sender of the risk message.
To this end, we quantize the risk levels to 4 bits, having risk levels between 1 and 16.
With only 16 risk levels, Bob has plausible deniability as to being the sender of a risk message.

We have thus reduced the amount of information available for inference and training of the machine learning predictor in order to increase privacy,
in the following ways:
\begin{itemize}
    \item we have eliminated access to the exact location and trajectory of each person (instead of using GPS location, we use only a coarser statistic at the level of a forwarding sortation area, as computed through GeoIP)
    \item we have eliminated access to the exact time of the encounters (to protect the privacy of the other person), keeping only its day
    \item we have eliminated the information about the global structure of the contact graph (who met with whom): the only thing left is the view from each phone about the non-identifying contacts  and their risk levels,
    \item we have eliminated the exact knowledge needed to allow unambiguous matching of different encounters as belonging to the same contact (instead we have noisy information derived by the approximate time of receiving update messages)
    \item we are not able to model the joint distribution of geographical zones and per-user medical questionnaires, by splitting the data sent to the ML server into two kinds of packets going into different files whose entries cannot be matched anymore (one indexed by zones and the other by pseudonymous identifier).
\end{itemize}
Let us review what is lost with each of these and how we mitigate these issues. Geographical location is presumably important because
some areas have a larger base probability of infection in the population and for example density, cultural aspects etc which influence the probability
of dangerous contacts. We mitigate this by using demographic information from Statistics Canada as proxys for the actual geographical
location, but this may actually help to generalize better in zones for which less data are collected.

The exact time of encounters
could tell us about the circumstances of the contagion (e.g at work during the day vs during the night sharing a bed) but is too sensitive
an information when we want to avoid stigmatization (identifying who it was we met who had a high risk). Similarly, the contact graph is
crucial from a privacy perspective (most people do not want anyone tracking who they met), and although it would be interesting to see
how this impacts accuracy of predictions, we have chosen not to consider methods (like loopy belief propagation) which would require
the contact graph in order to perform inference and learning.

The ability to match different contact events as belonging to the same person
(while not necessarily knowing who that person was) is important to correctly estimate probabilities of infection. To see this, consider
two scenarios: over the course of a few days, Alice has $N=100$ encounters of 15 minutes each with Bob (they happen to live together), versus Alice has $N=100$
encounters of 15 minutes each, but this time each encounter is with a different person. In the first case, to first approximation, 
Alice's probability of becoming infected may increase to approach that of Bob's. In the second case, the probability that Alice has
been infected is very high because it is enough that any one of the 100 people were contagious for her to have become infected.
This is why it is very important to make the difference between repeated and not-repeated contacts. Although the communication protocol
makes it impossible to make an exact association between the different contacts, a probabilistic one can be made, and we have designed
a clustering algorithm based on the risk levels and the time of arrivals of update messages to cluster the different contacts into
blocks corresponding putatively to the same person. We use these noisy labels as extra input for the predictor.
The last point has already been discussed in Section~\ref{sec:separate-zones}.

\begin{figure}[!ht]
\centering
\vspace*{-3mm}
\includegraphics[width=\textwidth]{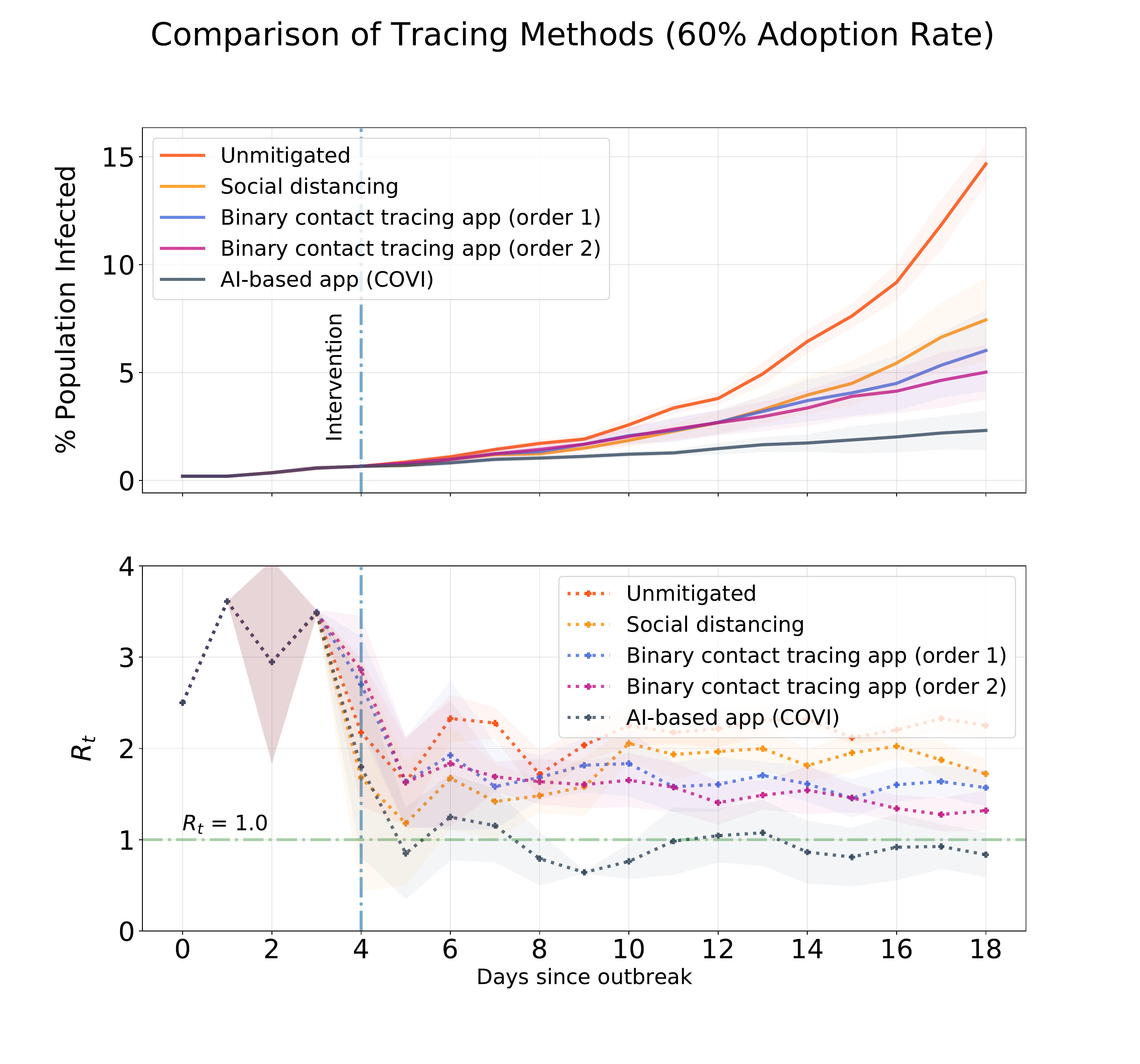}
\vspace*{-10mm}
\caption{Comparison between four different scenarios: unmitigated (pre-lockdown behavior), social distancing
(same mobility policy for all agents), binary digital tracing (standard method used in digital tracing apps without AI),
and AI-based app implementing a simple version of COVI's ML predictor (based
on a small Transformer). Top: evolution of accumulated number of
cases with respect to number of days elapsed. The intervention day (4) is when the different mobility policies are
put in place. Bottom: estimated reproduction number ($R_t$) as simulation progresses. We see a substantial gain in $R_t$
by using tracing of one form or another, but a much larger gain with ML-based risk prediction.
}
\label{fig:results}
\vspace*{-3mm}
\end{figure}

\subsection{Preliminary results on impact of machine learning}
\label{sec:ML-results}
We have trained a machine learning predictor in a supervised way using data generated from the simulator, using a simple contact tracing heuristic to generate the messages. To obtain preliminary results measuring the impact of using machine learning to predict contagiousness and obtain risk level messages, we have then used that predictor inside the simulator to influence the behavior of the agents according to four levels of recommendations
associated with different thresholds of risk levels. This has allowed us to simulate (with different random seeds than those used for generating
training data), how the ML predictor would impact the reproduction number 
$R_t$ of the disease and the growth of infections in a small pilot population of 1000
(it is indexed by time $t$ because it can evolve over time, depending on the
recommendations received by citizens from public health authorities and their app).
The simulation was performed with the assumption that 60\% of the population was using the app and users
at the strongest level of recommendation (quarantine level) got tested. 

Note that the number of inputs of the predictor is variable as it depends
on the number of contacts, so we have used a Transformer 
deep learning architecture~\cite{transformer},
which can also capture dependencies across the whole history
of the user (14 days and the list of all contacts) without
suffering from the vanishing gradient difficulty of recurrent neural networks~\cite{ke2018sparse}.
The contagiousness predicted by the transformer for each day in the past 14 days is converted
to a 16-level (4 bit) message sent to the contacts of the corresponding day. The conversion
from real-valued output to risk level was done by picking thresholds making the 16 bins of
approximately equal frequency. The risk level for today was converted into a recommendation
level. We used 4 recommendation levels in this simulation: 
\begin{enumerate}
    \item Recommendation level 1 (risk levels 0 and 1): Encouragement to wash hands, activating the hygiene factor to reduce infection.
    \item Recommendation level 2 (risk levels 2 and 3): Like Level 1 plus wearing a mask and standing 2m apart. This keeps the hygiene factor from above. It also activates wearing a mask outside the household. The efficacy of wearing a mask is set differently based on if someone is a healthcare worker or not (as the former tend to
    have better masks). 
    \item Recommendation level 3 (risk levels 4 and 5): Like level 2 plus practice stronger social distancing, which reduces the time duration of contacts by half. It also prevents people from visiting locations that haven't been visited in the past, making them more conservative.
    \item Recommendation level 4 (risk levels 6 to 15): this is the quarantine level, like level 3 plus recommendation to get tested, depending on the availability of a test.  The quarantined people work from home if they can, stay at home unless they are hospitalized (so there can still be household infections), they go out with a 10x reduced probability to stores or parks, but every time they go out they do not explore i.e. do not go to more than one location.
\end{enumerate}

A different simulation was then run to compare four different scenarios:
\begin{itemize}
    \item Unmitigated: agents are behaving according to pre-lockdown mobility statistics, leading to a value of $R_t$ slightly above 2 and rapid exponential growth of the infected population.
    \item Social distancing: all agents are following the same mobility policy, and a global parameter (corresponding to the strength of social distancing policies) controls the number of dangerous contacts. A maximal setting of this parameter would lead to full lockdown.
    \item Binary digital tracing (first order and second order): agents quarantine themselves if were in contact with someone who tested positive (first order), or were in contact with someone who was in contact with someone who tested positive (second order).
    \item AI-based app: agents use COVI's predictor to modulate their social distancing and self-isolation.
\end{itemize}
The global parameter controlling the strength of social distancing was modulated separately for the last three methods so as to equalize the global mobility (the number of contacts). Indeed, the AI-based app would otherwise tend to be favoured as more people would generally tend to get some form of recommendation to be prudent, whereas binary digital tracing only touches the immediate (or second-order) contacts of patients who tested positive. The results are shown in Figure~\ref{fig:results},
and suggest that ML-based risk prediction could very substantially reduce $R_t$, compared both to standard digital
tracing and to a uniform policy of social distancing. We see
that ML improves upon second-order tracing, which improves upon first-order
tracing, which improves upon no tracing at all (and only using uniform
social distancing for everyone).

This visualization focuses on the advantage of AI in
terms of reduced number of cases for some fixed general level of mobility, but it is possible to show how,
for a fixed choice of $R_t$ (obtained by globally decreasing mobility with more distancing and isolation), one can
obtain more mobility with an ML-based predictor. More details of this simulation, along with the code
used, will be provided in an upcoming technical report focusing on the simulation and ML aspects of this project.

\section{Empowering Citizens}
\label{sec:behaviour}
Decision making is difficult---doubly so in times of crisis when both the noise-to-signal ratio and the potential impact of decisions are orders of magnitude higher than in everyday life. Fields such as behavioural economics posit that this complexity leads people to make predictable mistakes \cite{heiner1983origin}. Policy-makers thus employ ``nudge'' strategies to ``fix'' those mistakes and align people's behaviour with certain normative standards. This strategy is certainly relevant in some contexts: for example a stay-at-home order or the instructions of washing your hands for 20 seconds \cite{Dreibelbis_2016}.

However, while fields such as behavioural economics \cite{leonard2008richard} are often used to achieve behaviour-change goals prescribed by organizations and governments, this approach also comes with a risk: a strategy based on libertarian paternalism (a common characterization of nudging) is only as powerful as the underlying motivations of the population to which it is applied. In the short-term, goals that are externally imposed can be promoted through clear, consistent, and prescriptive messaging; but sustained success relies on connecting to each user’s motivation and preferences.

Thus, over the period of a long crisis, it is necessary but not sufficient to ensure citizens can access and understand reliable information, internalizing what is most relevant to their own circumstance — rather, there must be feedback in the exchange of information. The public must have opportunities to articulate its goals and motivations — and indeed those goals and motivations must actually come to drive the information-focused components that are the initial focus of messaging. This approach provides an experience to users where they know that their voice is heard and the inputs they are provided are perceived as supporting their self-actualization rather than imposing external constraints.

So far in this crisis, citizens have been asked (or, in some cases, forced) to severely curtail their usual daily activities, with broad-based and stringent confinement measures enacted and enforced at various levels of government. To date, the stringency of these measures has actually made decision-making quite simple: the restrictions are so severe and so broad-based that citizens are left with minimal space to interpret what is being asked of them. 

As the crisis moves into its later phases, however, this situation will change substantially. Services will reopen progressively — initially in a limited capacity, and then more rapidly (perhaps with fluctuation to account for new outbreaks). The types of decisions the public will face during this phase of the crisis are likely to be more cognitively demanding, as the appropriate recommendations become more dependent on one’s specific situation (as opposed to broad-based). Similarly, the acceptance of measures to date will have worn down the resolve of the population to some degree. In combination, citizens will be both less clear on what to do and less motivated to do it. If an externally-motivated approach is still in place (as opposed to a self-actualization model), there could also be increasing tensions between the internal desire for personal freedom and the seemingly-external necessity for public health and safety. The implications of even small day-to-day decisions (e.g., ``Should I meet a friend at a coffee shop?'') can be profound.

We must therefore carefully craft tools that allow citizens to make these decisions in an evidence-based manner that considers (1) their potential risk-level to others, (2) their vulnerability in the case of an infection, and (3) their risk preferences. Rather than relying on coercive nudges, COVI promotes awareness, empowerment, and self-actualization. The application serves to create transparency so that each user can make the best possible decision for themselves and others around them. This transparency is driven by the preferences (notably risk preferences) that the users express through their use of the app. The sections below describe what this entails, in concrete terms, by introducing a set of foundational principles that drive core design decisions and ethical considerations.

\subsection{User preferences drive end-to-end experience}
While a strictly epidemiological approach to the Covid-19 crisis might prioritize minimizing the amount of risk taken on by citizens, at all costs — the need to also protect civil liberties and economic well-being requires any tool aiming to serve citizens in the long term to take account of their personal preferences. Designing tools based on an assumption that each user aims to solely minimize their risk of infection could lead to tone-deaf messaging that is ignored by users.

Given that one of the guiding principles of COVI is the commitment to providing actionable recommendations, it is important that this messaging be transparent, resonant, engaging, empowering, and supportive of the user's stated preferences. To achieve key public health goals, users must view COVI as a tool supporting their self-actualization rather than a tool for external actors to tell them what to do. This section discusses how various measurement mechanisms and frameworks must be leveraged to elicit personal preferences, which is critical to achieving public health goals.

COVI is built on an assumption of personal agency, such that recommendations are designed to inform choices rather than shift them (a subtle but critical distinction). We take an evidence-based approach and base the user experience on a foundation of clarity and collaboration. The best-practice frameworks and data gathering instruments listed below provide insight into how this principle applies to all aspects of the design and evolution of the application.

\subsubsection{Measuring in-application user behaviours}
Carefully measuring the kinds of information a user engages with allows us to create more meaningful interactions for them. In particular, past research in the cognitive science of mobile health-related applications has taught us that attention, memory, and reward processing are key in determining effective delivery models for health recommendations \cite{hilthychan2018}. Engagement data — gathered from in-application analytics tools and displayed on an internal engagement dashboard — allows us to leverage a combination of behavioural science frameworks established in the fields of public health \cite{mitchie2011refined}, behavioural product management \cite{eyal2013hooked} and ethical behaviour change \cite{jachimowicz2017community}. Doing so allows us to adjust the informational architecture of the application and personalize the user experience, prioritizing information that engages the user.

Importantly, certain kinds of information would be irresponsible to display, despite being highly engaging; for example, a map showing a real-time feed of infected users would surely draw a huge amount of attention, but the stigmatization of infected users and other negative impacts of this on society would be enormous. Information must be personalized to promote engagement, but not at all costs. Ethical frameworks described below must thus be used to elucidate how the pursuit of user engagement should be balanced against other crucial considerations, such as psychosocial well-being and inclusivity.

As noted above, the in-app analytics (of consenting users) contribute to machine learning algorithms that build epidemiological models. The combination of user interface tools and machine learning back-end allows us to provide tailored messaging for different user groups while remaining aligned with public health messaging across jurisdictions in Canada.

\subsubsection{Identifying unaddressed needs through population-level surveys}
When working with large data samples, it is easy to conflate the sample population with the overall population. The democratic principle of COVI’s guiding mission helps us to keep this difference in focus. To understand how these user preferences map onto the wider population, including any under-represented groups,  we periodically run a set of data-gathering exercises (such as quantitative surveys) with representative samples of Canadians. For example, while in-application data may show users are generally satisfied with the privacy options, survey data could reveal that the population at-large is more likely to interpret the application as autonomy-reducing — something that has been shown to reduce the effectiveness of health-related application messaging in Quebec and Alberta \cite{kongatsmcgetrickraine2019}.

The survey instruments we have designed focus on identifying key beliefs, attitudes, and unaddressed needs. They also focus on how Canadians perceive risk, what their intended courses of action are (based on this perception of risk) and how they consume information to guide the courses of action they will take.

Preliminary surveys have been conducted using MTurk, a platform frequently used in social science research. Though this platform is known to skew towards younger, more educated, and less affluent members of society, it nonetheless offers access to a population beyond our application users. By checking the demographics of our respondent population against census demographics available from Statistics Canada, the results of these surveys are then weighted to proportionately reflect the Canadian population (and subgroups thereof). Among those users that contribute their pseudonymized data to the COVI ML servers, similar adjustments are made (based on demographic information they input) to provide a representative image of Canadians generally.

To ensure the data gathering instruments sufficiently capture unaddressed needs and potential behavioural barriers, we refer to social network theory \cite{festinger1962theory,krause2007social}. These conceptual frameworks guide the design of data-gathering instruments such as surveys or interview questionnaires. The insights gathered from this research inform the prioritization of feature development and copy changes in the product roadmap. Summaries of these insights also allow us to connect with public-outreach teams to align their messaging with the preferences of Canadians (both to increase adoption of the application and to have a positive impact on Canadians looking to contribute to tackling the crisis without using the application).

\subsubsection{Applying usability checklists to improve experience}
A number of usability and user experience checklists are applied in the creation and testing of the application. These focus on both the cognitive and affective components of application usability \cite{zaharias2009developing}. Usability in this context focuses on 5 components described by J. Nielson \cite{nielsen1996usability}: 1) Learnability refers to “How easy is it for users to accomplish basic tasks the first time they encounter the design?,” 2) Efficiency refers to “Once have learned the design, how quickly can they perform tasks?,” 3) Memorability refers to “When users return to the design after a period of not using it, how easily can they reestablish proficiency?,” 4) Errors refers to “How many errors do users make, how severe are these errors, and how easily can they recover from the errors?,” and 5) Satisfaction refers to “How pleasant is it to use the design?.” Key principles as heuristic evaluation are used to detect and remove any potential design flaws that hinder usability.

In addition to ensuring that information and functionalities are understandable by users, the goal of these user experience audits is to ensure that user engagement is maximized. A key priority for us in improving the user experience is considering how we can empower users and tap into their intrinsic motivation so that they remain engaged in the long term. For this, we leverage research in human motivation, such as self-determination theory, self-efficacy, and self-regulation \cite{ryan2012willpower,bandura1978,kochnafziger2011}) to make sure that the user experience is designed for long-term and self-motivated behaviour change. 

Beyond applying best practices in UX design, we also work with a number of industry partners that are leaders in creating interfaces, and will provide guidance and validation on both new versions of the application as well as new features that are being rolled out. These engagements reflect the fact that the state of the art in UX design (and truly it is as much art as science) is not thoroughly documented in research articles or other written sources. Rather, the leading edge of UX is sometimes only visible in the work and the insights of its leading practitioners, as UX is a field of professional practice first and research second.

\subsection{User comprehension is prioritized and verified rather than assumed}
As noted throughout, privacy is a key consideration for the COVI project---as much because it is a social and democratic priority as because it plays a determining role in the adoption of the application. Research has shown that data sharing permissions are strongly affected by privacy experience, computer anxiety, and perceived control---factors that, if unaddressed, can have a profoundly negative effect on the privacy concerns of a significant segment of users \cite{degirmenci2020}. While there are examples of how cognitive biases such as over-choice and hyperbolic discounting can be exploited in platform design to coerce consent \cite{waldman2020}, we operate with the strong belief that this is an unethical and unsustainable strategy.

For this reason, it is critical that all sharing permissions be carefully crafted and validated throughout the application experience. To transparently demonstrate the privacy implications of using COVI, and to ensure that citizens who decide to use the application do so with confidence, the application cannot be designed or implemented in a way that assumes a user has infinite time, attention, or understanding to explore privacy implications.

To validate that the design is effectively supporting informed consent, evidence-based techniques and frameworks must be used to understand to what extent disclosure mechanisms are succeeding in informing users about key features---especially around the concept of privacy and data sharing. In particular, the usability testing features mentioned above are included to measure the choices made by users.

\subsubsection{Population-level testing about sharing preferences}
A substantial body of research shows that Canadian preferences on data sharing are informed by dubious beliefs about data collection and use policies \cite{ricebogdanov2018}. In order to reveal true population-level preferences, research must be conducted in a way that clearly outlines data usage policies, ensures understanding, and elicits candid responses. Thus, in addition to referencing the existing literature on the subject \cite{privacycanada2019}, our surveys seek to understand the privacy and data sharing preferences of Canadians. Importantly, the testing allows us to identify key gaps in knowledge that Canadians may have around what data sharing may entail in the context of a digital contact-tracing application. The application’s features and copy have been and continue to be informed by the general themes identified in this research regarding privacy and data sharing preferences of Canadians.

\subsubsection{Key terms and conditions are emphasized and progressively disclosed}
A wealth of cognitive ergonomics research supports the idea that people have limited cognitive bandwidth and fall back on heuristics to simplify their decision-making when interacting with complex systems \cite{markyoungkarel2015}. In fact, there is a line of evidence suggesting that cognitive scarcity correlates with more information disclosure behaviour \cite{veltriivchenko2017}. This means that consent could likely be coerced by overloading users (who are likely already emotionally strained) with information — but such an approach is antithetical to the ethical principles guiding COVI and is likely unsustainable.

Instead we must ensure that key components of the terms and conditions are well understood by users, and are not just agreed to haphazardly. This is done using a multi-layered, “progressive” disclosure approach, which has been shown to balance user experience and system transparency \cite{springerwhittaker2019}. For example, a graphics-heavy top layer illustrating privacy implications can link to a somewhat more textual second-layer — this can then link to the longer FAQ section on the website, which in turn sends users to the full privacy policy. 

Users are thus provided with disclosure information befitting their level of interest and literacy with the topic. Those who are satisfied with a top-level view are provided with that, while more interested users can continue digging into further and further details until their questions are answered.

\subsubsection{User comprehension is verified rather than assumed}
Consent is often gained by presenting users with a block of ``terms and conditions'' text and assuming they will read and understand it. Yet, this assumption has been invalidated by empirical evidence; as such, consent given under these conditions cannot be considered fully informed.
In order to make consent more meaningful, it is essential to know what information most affects user decisions, and to ensure this information is conveyed such that users are most likely to read and understand it.

We take a number of steps to achieve this. First, we apply in-app analytics to estimate users' comprehension — for example, by looking at the average user dropout at various layers of disclosure information. Second, we administer dynamic comprehension quizzes to a random sample of users, allowing us to understand what information has and has not been internalized. Finally, disclosure tools are iteratively revised based on the feedback from these measures to ensure they best cater to actual user behaviour.

\subsection{User empowerment to protect themselves and others is maximized}
As noted at the outset, a central challenge of easing lockdowns will be the loss of clear, uniform instructions for citizen behaviour. To provide continued clarity and to promote the empowerment of users, the COVI application leverages evidence-based methods. In particular, research on public health messaging suggests that two factors are particularly predictive of feelings of self-control: health consciousness (triggered by conservation and self-transcendence) and health knowledge (triggered by bonding and bridging social capital) \cite{chang2019}. By explicitly targeting these outcomes in its messaging, COVI can increase collaboration toward prosocial public health outcomes among its users.

In addition to leveraging empowerment-maximizing frameworks of public health messaging, a unique aspect of COVI is its ability to provide personalized information to users about what actions they can safely take based on their individual context. As noted above, this information is personalized to the preferences of the individual (to promote self-actualization, which is different from compliance) and aligned with the policy set by the public health authorities (who remain the legitimate decision-makers, even if COVI offers powerful tools to both inform those policies and deliver them to citizens).

Using a machine learning powered predictive algorithm, COVI works with scalar risk levels, rather than simple binaries of contact/no contact with an infected person. The predictive and scalar natures of the algorithm in turn facilitate a proactive, progressive approach for users to manage their risk — restricting their movements by degrees as their individual level of infection risk increases, even before they have certainty about whether they are infected or not (or have had direct contact with an infected person).

\subsubsection{Canadian beliefs and attitudes toward the crisis are closely monitored}
To ensure messaging is aligned with user preferences (to be empowering rather than perceived as imposed), evolving Canadian beliefs and attitudes toward the crisis (including health, social and economic dimensions) must be closely monitored. Various approaches are used in concert to accomplish this.

Large-scale surveys of the Canadian population have been and continue to inform our understanding of how Canadians perceive this crisis, what their preferences are, and so forth. This is complemented by our in-app data, which (as stated above) is census-weighted to reflect the Canadian population.

These instruments are informed by research across various fields, including health messaging, crisis communications, and revealed/stated preferences research from behavioural science. Previous research has demonstrated strong framing effects in health messaging, such as showing that gain-framed messages are more effective than loss-framed messages in promoting prevention behaviours \cite{gallagher2011perceived}. There is currently limited research on the framing effects of Covid-19 related messaging, a gap that research related to COVI seeks to address. Leading-edge research in crisis communications has highlighted the importance of prioritizing transparency, trust, and user empowerment \cite{chang2019}. Theories and frameworks pertaining to preferences are discussed in the section above---and also feed into the instruments described here, especially as the knowledge base increases and as the crisis evolves (along with the perceptions of the crisis).

These findings help us formulate messaging variants and feature design, which are then A/B tested in-app to determine the effects of the various approaches being considered. We achieve this by using our custom in-application analytics engine and questions (which are fed into the machine learning algorithm on the COVI Canada ML servers, as noted above), as well as with external surveys.

\subsubsection{In-application desensitization is closely monitored}
A well known effect in crisis communications is desensitization to messaging over time---otherwise known as ``alert fatigue.'' In fact, Baseman and colleagues \cite{baseman2013public} found that each additional public health message sent during the course of a week resulted in a statistically  significant 41.2\% decrease in the odds of recalling the message. Desensitization is also known to lower risk perception, which can affect user compliance with the recommendations provided. Because certain messages available through the application are critical (e.g.~a stay-at-home order based on your risk of infection, or vulnerability to Covid-19 due to a pre-existing condition), it is important that these messages are taken seriously by users. In brief, our in-app messaging must be carefully structured according to evidence-based practices \cite{rossman2017}, ensuring that low-priority alerts do not create noise that prevents high-priority alerts from being taken seriously and acted upon. There is a delicate balance to strike here, as the desire to drive application engagement (especially early on) could create pressure to escalate the urgency of the app’s messaging — but ultimately there needs to be space above the baseline messaging for urgent messages to stand out.

For this reason, the effect of application messaging on user desensitization must be assessed continuously. This is accomplished through user surveys (including single-question micro-surveys delivered in-application at carefully chosen moments), population surveys, and in-application analytics (examining rates and timeliness and compliance with low- or high-priority alerts). The design of the surveys is informed by frameworks for crisis communications, public messaging, and health messaging. 

\subsubsection{Credibility is well communicated}
Crisis and risk communications research \cite{cerc2019psychology} has shown that people look for simple, consistent, and credible information. Given these known preferences, users are likely to be extremely sensitive to any kind of indication that information presented in the application is biased, out of date, or false; the timeliness and credibility of information should be proactively identified and demonstrated to users. There is also general agreement among experts and researchers that communication by authorities to the public should include explicit information about uncertainties associated with events; therefore, the degree of certainty regarding information should be clearly communicated. A close collaboration with authorities to ensure that the information is valid and trustworthy is critical as a mitigation strategy for this, with the source of the data being made salient to the user. 

\subsubsection{Information is updated often and visibly}
Our research on existing digital applications relating to Covid-19 indicates that users place a lot of importance on receiving updated information. For this reason, special care must be taken to ensure that information available in the application is as up-to-date as possible and that the latest update timestamp is always available and salient to the user. The initial feature set of the application integrates APIs that are updated on a daily basis, with the goal of even more timely data being made available. As we shift toward proprietary sources of data for in-application visualizations, it is important to coordinate with the development teams to ensure that timestamps are (1) readily available and salient to users, and (2) made meaningful to users. For instance, we will communicate clearly whether timestamps refer to the date of infection, the date the test was performed, or the date the test result was available.

\subsection{User psycho-social well-being is promoted}
Due to the sensitive nature of the content being communicated through the application---as well as the general stress-level caused by the crisis---it is essential to approach both the application design and the copy through a lens of deep empathy. In particular, this means that undue stress on the user must at all times be minimized, and extra care should be provided to higher risk groups who are likely experiencing even higher stress levels.

Increased stress negatively impacts decision-making. Thus, given that the goal of COVI is to empower users through improved decision-making, adding stress runs directly counter to our objective. Furthermore, psychosocial well-being plays an important role in defining individual and collective narratives about our efforts to address the crisis, which has important democratic implications (e.g., government and market solutions are predominantly viewed as the main ways to solve collective-action problems; this project represents an experiment with a new decentralized form of coordination, one that is privacy-protecting and citizen-empowerment-first, and that can have impacts on future decisions about how to deal with large-scale societal challenges). There are also more tangible reasons to promote psychosocial well-being: increased stress and anxiety are linked to reduced immune functioning \cite{segerstrom2004psychological, morey2015current, glaser2000chronic}, and thus undue stress would decrease our biological resilience against Covid-19. 

Some of the strategies we employ to promote psychosocial well-being include the following.

\subsubsection{Creating features for users to assess risks to their psycho-social well-being}
During a global pandemic, people may be less attuned to their mental well-being, and many are in a position where they must make difficult trade-offs between health concerns and other issues (financial, professional, familial/social, cultural/religious). Even those who do not contract the virus may experience mental health consequences from additional anxiety and stress. Indeed, individuals with pre-existing mental health conditions are among the most vulnerable populations during periods of crisis and isolation \cite{yao2020patients}. As such, we include in-app features to help individuals assess their well-being, and provide mental health resources tailored to the user's profile (age, location, etc).

\subsubsection{Providing mental health resources}
As stated, we will provide users with resources (e.g.~psychological therapy exercises and mental health support lines) to help them make meaningful progress toward better psycho-social well-being. In cases where users report severe issues related to mental health, a simple triage system should be created that prompts them to contact a mental health practitioner or seek immediate support. Since mental health remains somewhat taboo in society, we have made it a priority to communicate the normality of mental health concerns, especially during a time like this (e.g.~by using social norms to indicate that X number of people in their neighbourhood used this resource).

\subsubsection{Creating a positive-skewed distribution of messaging valence}
Given that most messaging related to the Covid-19 crisis is overwhelmingly negative, care must be taken to reduce the stress imposed by messages framed in this way, particularly given that previous research has shown that message  framing results can lead to significant shifts in public response \cite{garfin2020novel}. This is done by using user-tests to measure the emotional valence of messaging and ensuring that message valence is framed positively and communicated effectively (while ensuring that they are not misinterpreted as being unserious). In fact, all messaging used in the application has been tailored to avoid eliciting negative sentiments and put an unnecessary burden on the user’s mental state. Each iteration of the application will continue to reflect the continuous tests that we carry out in this regard, to validate approaches and to improve the effect of the app on the user’s well-being.

\subsubsection{Proactively addressing risks of stigma and other social dynamics around privacy}

The COVI application was designed to protect privacy and empower its users. While a technical attack on infrastructure is an important vector to consider for privacy breaches, the user’s phone also needs to be considered as a point of vulnerability within the system. While we want to provide transparent and informative content to users, that same content viewed by someone else looking at their phone could be an untenable compromise of privacy. For example, a risk level displayed prominently could be of interest to users, but a shopkeeper or employer might also demand to see the risk level screen of users as a condition for being admitted onsite. As Bruns and colleagues suggest \cite{brunskraguljac2020}, these kinds of dynamics can create profound stigmatization risks that must be addressed through carefully planned risk mitigation strategies and appropriate social norms protecting the rights of individuals. Note that consent may not be a sufficient barrier for an employee who may have no choice but to accept the directives of their employer or lose their job. Further legal protection of privacy should be considered and put in place according to the collective preferences of citizens and the value of protecting the most vulnerable in society. Truly voluntary measures (e.g.~stay at home if at high risk) are preferable but should not come at a personal cost (e.g.~losing one's job).

To address these worries, we leverage “threat model” scenarios to ensure that we account for ways in which a user’s privacy might be compromised by someone who has access to their information through that user’s own phone. These threat models allow us to conduct privacy assessments structurally analogous to those created by the privacy infrastructure team.

\subsection{User inclusivity acknowledges the diversity of their needs}
Diversity and inclusion are important in the context of COVI for several reasons. First and foremost is the issue of social justice and fairness. Any application that brands itself as catalyzing a nationwide effort must support all members of the citizenry, providing opportunities for each person to step up and join the effort. An application that is less welcoming to or less effective for certain segments of the population undermines its claim to a truly nationwide effort. Furthermore, as the utility of the COVI application increases rapidly as the user base grows, promoting inclusivity is important for instrumental reasons insofar as it increases the power to serve the entire user base in addressing the crisis.

In the context of this aspiration, it is important to consider how various groups, especially those already marginalized, are likely to interact with the application. In order to do this, we are deploying strategies such as the following:

\subsubsection{Population-level inclusivity audits}
The application team must use data-gathering instruments to identify key demographic gaps in the use of the application. Dimensions highlighted as relevant here include: gender, race, age, language, income, education, sector of employment, family composition, region, rural/urban, Indigenous status, mental/physical health, mental/physical ability, housing status. This can be achieved by comparing user-base demographics to Statistics Canada census data. These insights would be shared with the public-outreach team to inform their strategies to reach a diverse user base. They would also be shared with public health authorities to help those authorities understand which population subgroups are or are not well represented in the inputs they receive from COVI Canada (including both aggregate data and epidemiological modeling).

One critical element for the success of this approach is the quality of data. Users are not required to input this level of demographic information. Rather, they are given the option to fill in as many or as few of these dimensions as they wish, knowing that the more they fill in the more personalized their experience is. Furthermore, for those users who opt in to contribute their data to the COVI ML server, their contribution of demographic data is critical for training the machine learning algorithms--- thus  providing those individual users and others like them with accurate recommendations.

\subsubsection{Integrating diversity dimensions into our other analyses}
Beyond simply understanding how the user base maps onto the diversity of sub-populations across Canada, it is important to identify and assess any meaningful differences in the way that sub-populations interact with and derive value from the COVI app. For instance, if younger, more educated, more affluent, urban residents are over-represented with the app’s user base (a plausible scenario given that this demographic group is also over-represented among smartphone owners), that could lead to the construction of an epidemiological model that is better adjusted to the reality of some users than others. If such a situation were not identified and addressed, it could lead to less accurate recommendations being supplied to users dissimilar to that group---providing less effective health protections to some users than others.

To identify and address these potential challenges, frameworks of bias assessment in clinical research are leveraged \cite{stone2019unified} and the analyses described throughout this document are disaggregated along these diversity dimensions. Additionally, the diversity dimensions are fed into the machine learning algorithm to identify epidemiological as well as behavioural differences across these subgroups. The training procedure for
risk prediction and epidemiological modeling can then
be modified to increase the weight of underrepresented groups, using a method of
importance sampling or importance weighting. For instance, if Indigenous people have different preferences than other Canadians in managing this crisis, and therefore respond differently to messaging, this could be identified by the analytical approach here, to ensure that Indigenous Canadians receive messaging that promotes their self-actualization rather than imposing messaging upon them that embeds the preferences of others.

\subsubsection{Engagement with at-risk populations}
The analyses above lay out the process for assessing demographic under-representation and tailoring user experience (including health recommendations) by demographics. However, the project teams assume that there are notable demographic groups that are so drastically underrepresented among the user base that it creates a risk that the entire subgroup is not well served by COVI. Furthermore, if COVI represents an effort to support and empower Canadians during this time of crisis, then at-risk groups (e.g., Canadians who are homeless or living in precarious housing situations) are the ones most likely to be systemically excluded from using COVI.

For these reasons, it is important for the application team to create an open discourse with entities representing groups within the population that are most at risk of being marginalized. Structural marginalization is foreseeable among the older adult population, people without access to smartphones or mobile data and people with disabilities, though a more thorough verification based on established frameworks for inclusive innovation is necessary \cite{georgemcgahan2012}. Since the beginning, we have been proactive in ideating alternative solutions to access groups of people who are at risk of being marginalized. In addition, our roadmap includes a very specific set of features that would allow the application to reach, albeit in a more limited way, members of at-risk groups---e.g.~custodial wallets for those who do not own a cellphone but have access to one. Finally, past research has shown that involving stakeholders from at-risk communities as domain-experts in the co-creation of solutions can drastically improve how ``data scientists approach the development of corpora and algorithms that affect people in marginalized communities and who to involve in that process'' \cite{freypatton2020}. Thus, direct engagement with marginalized communities is necessary throughout the project.

\section{Discussion}

Let us now briefly consider some of the critiques often made of digital tracing
and see how COVI fares in that light.
The biggest critiques of digital contact tracing generally center around privacy, trust, and adoption.

A high-level question
which comes up is whether or not it is worthwhile to take any risk with privacy
if we are not sure that such an app would help.
Studies suggest that at least about half of the population would have to
be on-board to beat the virus, a high threshold for adoption.
However, if there is any significant chance that an app like COVI
could succeed at being used at that level, we have to take it because the
consequences of not doing so are too great: the difference between a reproduction number of
1.5 and a reproduction number of 0.9 is huge in terms of human casualties, not
to mention the fact that targeted self-isolation offers the possibility that a
large fraction of the population be allowed to work while still keeping the
virus at bay. 

Of course, in a democratic country where we value freedom
and responsibility, we cannot make an app like this mandatory: the only option is trust.
For something like COVI to work, people have to trust
the organization managing it; hence the importance of privacy protection and of
a not-for-profit organization focused on the Covid-19 pandemic managing the effort.
Similarly, COVI Canada and governments have to trust that most citizens will act responsibly when they understand
what is at stake, i.e., the lives and health of fellow citizens.

However, even if the uptake of COVI was not sufficient for efficient automatic tracing of contacts to estimate contagion risks,
it would still reinforce the manual tracing efforts,
by allowing high-risk citizens (as estimated by COVI) to proactively get in touch with
public health autorities (ahead of when they would be called by public health, if they
would have been called at all), thus gaining
precious days during which contagions would otherwise likely occur. In addition, COVI could play a significant role in terms
of epidemiological modeling and forecasting. Indeed, it is enough that a small fraction
of the population consent to sharing their data for the ML models to already greatly enhance what is currently feasible in terms of epidemiological understanding and forecasting under different public health policies.

Focusing in on the privacy issue, the problem is multi-faceted, and COVI addresses the different aspects in different ways.
One major concern is about stigmatization resulting from the app.
We ensure that third parties or individuals with
whom a COVI contact is established will generally not be able to 
easily infer one's
risk level (unless it is obvious for other reasons). Spending time with a
diagnosed person would not mark one as a pariah because that information would
remain anonymous. Even the way that
infectiousness is communicated to the user, i.e., in the form of recommendations
which can depend on other factors, makes it more difficult for someone like a spouse
or an employer, simply by looking at your phone, to get an explicit readout of
the risk level (for example we avoid the kind of obvious colour scheme implemented
in other apps). COVI is not meant to be used as an immunity passport,
because of the concerns this would raise for human rights and dignity.

Another important privacy consideration is the concern that government
agencies would have access to one's detailed trajectory and network of contacts.
Again this is avoided to the extent possible, thanks to the decentralized approach to data management, the cryptographic mechanisms used to send risk messages, the privacy mechanisms on the machine learning side, and the creation of a strong data governance model (with COVI Canada, a not-for-profit organization)
to hold the pseudonymized medically relevant information (like questionnaire answers) for a period of three months and a single-minded mission to protect the health, privacy and dignity of citizens with
regard to the management of the collected data.
In addition, the pseudonymized nature of even the optional volunteered data makes it difficult to track people, as their phone number, IP address, name or other identifying 
information would not be collected and thus would not be available to anyone.

To promote trust, COVI Canada will have
open rules about its governance, open access to the code and aggregated
epidemiological models, and would be continuously monitored by its board,
internal experts committees, and external evaluations from independent
academic groups and governmental representatives, to make sure that it
stays faithful to its mission. COVI Canada's entire governance model is built around the core values of legitimacy, accountability, transparency, and efficiency.COVI complies with Canadian privacy laws and with the principles put forward in the Joint Statement by Federal, Provincial and Territorial Privacy Commissioners on May 7, 2020 \cite{privacystatement2020}.  These include consent, and trust, legal authority, necessity and proportionality, purpose limitation, de-identification, time-limitation, transparency and accountability and the deployment of safeguards.
This public white paper seeks to be an example of that transparency: we have tried to be explicit about the privacy risks that contact tracing does entail, and we hope that the end users will agree with our value proposition given the exception circumstances of the pandemic.

COVI Canada's single  mission of supporting Canadians in their fight against Covid-19 and not-for-profit nature ensure the data collected will never be used for commercial purposes, nor sold to private companies. It cannot be used for surveillance or to enforce quarantine by governments. The data is all stored in Canada and will be deleted as soon as the pandemic is over. COVI Canada’s Board of Directors will be chaired by a retired Canadian judge, and the governance model includes an Experts Advisory Council of recognized thought-leaders in relevant fields such as public health, ethics, human rights and privacy. COVI adheres to the Montreal Declaration for the Responsible Development of AI and was developed with the support of UNESCO. We recognize the unfortunate and unacceptable consequences that a technology can potentially have on marginalized groups and as such, COVI Canada will continue to work with human rights organizations, civil society groups, and legal and social science experts to prevent algorithmic bias, strengthen the technology’s accessibility and ensure inclusive representation at all levels of its governance model. COVI Canada will be dismantled at the end of the pandemic; only the science and technology will remain to help us in future similar situations.

\section{Conclusion}

As the pandemic progresses, leveraging digital strategies to minimize the spread of Covid-19 while preventing loss of privacy and intrusions on civil liberties remains a paramount objective. As healthcare and economic resources become significantly strained, the ability to efficiently and rapidly reduce the spread of Covid-19 and thereby reduce the morbidity and mortality associated with an infection is critical; however, in our view, this must not come at the expense of the civil liberties that lie at the core of democratic societies. `COVI' is a digital solution that combines digital contact tracing, user interface science, and machine learning with robust privacy protections while preserving independent agency and choice.

We cannot remove all of the privacy risks and trade-offs that are endemic to contact tracing, but combined with independent oversight, COVI aims to gain the public's trust and engage in responsible collective action against the pandemic. We view COVI as an opportunity to enhance a form of democracy where power truly rests in the hands of citizens: they decide whether to use this technology or not, balancing the risks and the benefits to themselves and their community according to their values. Of course, this requires a public discussion helping citizens understand what is at stake, and this democratic debate is a core component of COVI Canada's plan. Despite the many challenges associated with launching such a strategy, the balance achieved by COVI represents an important step in advancing the use of digital health and machine learning to combat a major world crisis. COVI aims to empower individuals by providing them with evidence-based and personalized information about their level of risk, thereby allowing them to act accordingly and responsibly to protect their loved ones and their community. COVI also empowers public health services with aggregated evidence which can be crucial to draw appropriate policies. Ultimately, we believe that COVI will empower Canadians to protect themselves, limit the spread of the virus and facilitate a smart and safe lifting of social distancing measures through collective and democratic action, as they go about their daily lives. 

\section*{Acknowledgment}
We would like to thank Sumukh Aithal, Behrouz Babaki, Henri Barbeau, Edmond Belliveau, Vincent Berenz, Olexa Bilaniuk, Am\'elie Bissonnette-Montminy, Pierre Boivin, Em\'elie Brunet, Jo\'e Bussi\`ere, Ga\'etan Marceau Caron, Ren\'e Cadieux, Pierre Luc Carrier, Hyunghoon Cho, Anthony Courchesne, Linda Dupuis, Justine Gauthier, Joumana Ghosn, Gauthier Gidel, Marc-Henri Gires, Simon Guist, Deborah Hinton, Bogdan Hlveca, Bernd Holznagel, Samuel Huberman, Shrey Jain, Jameson Jones-Doyle, Dilshan Kathriarachchi, Giancarlo Kerg, Soundarya Krishnan, David Lazar, Fr\'ed\'eric Laurin, Sacha Lepr\^etre, St\'ephane L\'etourneau, Libeo team, Alexandre Limoges, Danielle Langlois, Vincent Martineau, Lucas Mathieu, Philippe Matte, Rim Mohsen, Eilif Muller, Ermanno Napolitano,  David Noreau, Ivan Oreshnikov, Satya Ortiz-Gagn\'e, Jean-Claude Passy, Marie Pellat, Dan Popovici, Daniel Powell, Brad Rabin, Catherine Saine, Shanya Sharma, Kareem Shehata, Pierre-Luc St-Charles, Marie-Claude Surprenant, M\'elisande Teng, Julien Tremblay-Gravel, David Wu, and Lenka Zdeborova for their help. We would also like to thank NSERC, CIHR, CIFAR, FRQNT and Scale AI for their funding. A.S. is funded by the Fonds de la Recherche en Sante du Quebec Junior -1 Clinician Scientist award, the Lucien McGill Award, and the McGill Interdisciplinary Initiative in Infection and Immunity Research award. Y.W.Y. is funded by the University of Toronto Covid-19 Action Fund.

\bibliographystyle{./bibliography/IEEEtran}
\addcontentsline{toc}{section}{References}
\bibliography{./bibliography/main}

\begin{thebibliography}{100}
\providecommand{\url}[1]{#1}
\csname url@samestyle\endcsname
\providecommand{\newblock}{\relax}
\providecommand{\bibinfo}[2]{#2}
\providecommand{\BIBentrySTDinterwordspacing}{\spaceskip=0pt\relax}
\providecommand{\BIBentryALTinterwordstretchfactor}{4}
\providecommand{\BIBentryALTinterwordspacing}{\spaceskip=\fontdimen2\font plus
\BIBentryALTinterwordstretchfactor\fontdimen3\font minus
  \fontdimen4\font\relax}
\providecommand{\BIBforeignlanguage}[2]{{%
\expandafter\ifx\csname l@#1\endcsname\relax
\typeout{** WARNING: IEEEtran.bst: No hyphenation pattern has been}%
\typeout{** loaded for the language `#1'. Using the pattern for}%
\typeout{** the default language instead.}%
\else
\language=\csname l@#1\endcsname
\fi
#2}}
\providecommand{\BIBdecl}{\relax}
\BIBdecl

\bibitem{gates2020responding}
B.~Gates, ``Responding to covid-19—a once-in-a-century pandemic?'' \emph{New
  England Journal of Medicine}, 2020.

\bibitem{fernandes2020economic}
N.~Fernandes, ``Economic effects of coronavirus outbreak (covid-19) on the
  world economy,'' \emph{Available at SSRN 3557504}, 2020.

\bibitem{anderson2020will}
R.~M. Anderson, H.~Heesterbeek, D.~Klinkenberg, and T.~D. Hollingsworth, ``How
  will country-based mitigation measures influence the course of the covid-19
  epidemic?'' \emph{The Lancet}, vol. 395, no. 10228, pp. 931--934, 2020.

\bibitem{barometer2019january}
\BIBentryALTinterwordspacing
E.~T. Barometer, ``January 20, 2019,'' 2019. [Online]. Available:
  \url{https://www.edelman.com/sites/g/files/aatuss191/files/2019-02/2019_Edelman_Trust_Barometer_Global_Report_2.pdf}
\BIBentrySTDinterwordspacing

\bibitem{niehus2020quantifying}
R.~Niehus, P.~Martinez~de Salazar~Munoz, A.~Taylor, and M.~Lipsitch,
  ``Quantifying bias of covid-19 prevalence and severity estimates in {W}uhan,
  {C}hina that depend on reported cases in international travelers,'' 2020.

\bibitem{irfan2020case}
\BIBentryALTinterwordspacing
U.~Irfan, ``The case for ending the {COVID-19} pandemic with mass testing,''
  April 2020. [Online]. Available:
  \url{https://www.vox.com/2020/4/13/21215133/coronavirus-testing-covid-19-tests-screening}
\BIBentrySTDinterwordspacing

\bibitem{flint2003assessment}
J.~Flint, S.~Burton, J.~Macey, S.~Deeks, T.~Tam, A.~King, M.~Bodie-Collins,
  M.~Naus, D.~MacDonald, C.~McIntyre \emph{et~al.}, ``Assessment of in-flight
  transmission of sars--results of contact tracing, canada.'' \emph{Canada
  communicable disease report= Releve des maladies transmissibles au Canada},
  vol.~29, no.~12, p. 105, 2003.

\bibitem{ferretti2020quantifying}
L.~Ferretti, C.~Wymant, M.~Kendall, L.~Zhao, A.~Nurtay, L.~Abeler-D{\"o}rner,
  M.~Parker, D.~Bonsall, and C.~Fraser, ``Quantifying sars-cov-2 transmission
  suggests epidemic control with digital contact tracing,'' \emph{Science},
  2020.

\bibitem{tang2020contact}
D.~Tang, ``Contact-tracing strategies for sars-cov-2 eradication**** draft,''
  2020.

\bibitem{bay2020bluetrace}
\BIBentryALTinterwordspacing
J.~Bay, A.~Tan, C.~S. Hau, L.~Yongquan, J.~Tan, and T.~A. Quy, ``{BlueTrace}: A
  privacy-preserving protocol for community-driven contact tracing across
  borders,'' 2020. [Online]. Available:
  \url{https://bluetrace.io/static/bluetrace_whitepaper-938063656596c104632def383eb33b3c.pdf}
\BIBentrySTDinterwordspacing

\bibitem{chan2020pact}
J.~Chan, S.~Gollakota, E.~Horvitz, J.~Jaeger, S.~Kakade, T.~Kohno, J.~Langford,
  J.~Larson, S.~Singanamalla, J.~Sunshine \emph{et~al.}, ``Pact: Privacy
  sensitive protocols and mechanisms for mobile contact tracing,'' \emph{arXiv
  preprint arXiv:2004.03544}, 2020.

\bibitem{rivest2020pact}
\BIBentryALTinterwordspacing
R.~L. Rivest, J.~Callas, R.~Canetti, K.~Esvelt, D.~K. Gillmor, Y.~T. Kalai,
  A.~Lysyanskaya, A.~Norige, R.~Raskar, A.~Shamir, E.~Shen, I.~Soibelman,
  M.~Specter, V.~Teague, A.~Trachtenberg, M.~Varia, M.~Viera, D.~Weitzner,
  J.~Wilkinson, and M.~Zissman, ``The {PACT} protocol specification v0.1
  (4/8/2020),'' April 2020. [Online]. Available:
  \url{https://pact.mit.edu/wp-content/uploads/2020/04/The-PACT-protocol-specification-ver-0.1.pdf}
\BIBentrySTDinterwordspacing

\bibitem{applegoogle2020}
\BIBentryALTinterwordspacing
Apple and Google, ``Privacy-preserving contact tracing,'' April 2020. [Online].
  Available: \url{https://www.apple.com/covid19/contacttracing/}
\BIBentrySTDinterwordspacing

\bibitem{pepp}
\BIBentryALTinterwordspacing
``{Pan-European Privacy-Preserving Proximity Tracing},'' April 2020. [Online].
  Available: \url{https://pepp-pt.org/}
\BIBentrySTDinterwordspacing

\bibitem{trieu2020epione}
\BIBentryALTinterwordspacing
N.~Trieu, K.~Shehata, P.~Saxena, R.~Shokri, and D.~Song, ``Epione: Lightweight
  contact tracing with strong privacy,'' April 2020. [Online]. Available:
  \url{https://sunblaze-ucb.github.io/privacy/projects/epione.html}
\BIBentrySTDinterwordspacing

\bibitem{npr_contact_tracing}
\BIBentryALTinterwordspacing
F.~Ordonez, ``Ex-officials call for \$46 billion for tracing, isolating in next
  coronavirus package,'' \emph{NPR}, April 2020. [Online]. Available:
  \url{https://www.npr.org/2020/04/27/845165404/ex-officials-call-for-46-billion-for-tracing-isolating-in-next-coronavirus-packa}
\BIBentrySTDinterwordspacing

\bibitem{flanagan1988equality}
W.~F. Flanagan, ``Equality rights for people with aids: Mandatory reporting of
  hiv infection and contact tracing,'' \emph{McGill LJ}, vol.~34, p. 530, 1988.

\bibitem{levine1988contact}
M.~L. Levine, ``Contact tracing for hiv infection: a plea for privacy,''
  \emph{Colum. Hum. Rts. L. Rev.}, vol.~20, p. 157, 1988.

\bibitem{cho2020contact}
H.~Cho, D.~Ippolito, and Y.~W. Yu, ``Contact tracing mobile apps for covid-19:
  Privacy considerations and related trade-offs,'' \emph{arXiv preprint
  arXiv:2003.11511}, 2020.

\bibitem{raskar2020apps}
R.~Raskar, I.~Schunemann, R.~Barbar, K.~Vilcans, J.~Gray, P.~Vepakomma,
  S.~Kapa, A.~Nuzzo, R.~Gupta, A.~Berke \emph{et~al.}, ``Apps gone rogue:
  Maintaining personal privacy in an epidemic,'' \emph{arXiv preprint
  arXiv:2003.08567}, 2020.

\bibitem{burgess1963contact}
J.~Burgess, ``A contact-tracing procedure,'' \emph{British Journal of Venereal
  Diseases}, vol.~39, no.~2, p. 113, 1963.

\bibitem{millar2020well}
\BIBentryALTinterwordspacing
J.~Millar, ``A well-intentioned but unproven app could reinforce biases and
  create confusion and stress, something developers must take more time to
  consider,'' April 2020. [Online]. Available:
  \url{https://policyoptions.irpp.org/magazines/april-2020/five-ways-a-covid-19-contact-tracing-app-could-make-things-worse/}
\BIBentrySTDinterwordspacing

\bibitem{aclu2020principles}
\BIBentryALTinterwordspacing
D.~K. Gillmor, ``Principles for technology-assisted contact-tracing,'' April
  2020. [Online]. Available:
  \url{https://www.aclu.org/sites/default/files/field_document/aclu_white_paper_-_contact_tracing_principles.pdf}
\BIBentrySTDinterwordspacing

\bibitem{inria2020proximity}
\BIBentryALTinterwordspacing
Inria, ``Proximity tracing applications: The misleading debate about
  centralised versus decentralised approaches,'' April 2020. [Online].
  Available:
  \url{https://github.com/ROBERT-proximity-tracing/documents/blob/master/Proximity-tracing-discussion-EN.pdf}
\BIBentrySTDinterwordspacing

\bibitem{meyer2020controversy}
\BIBentryALTinterwordspacing
D.~Meyer, ``Controversy around privacy splits {E}urope’s push to build
  {COVID}-19 contact-tracing apps,'' \emph{Fortune}, April 2020. [Online].
  Available:
  \url{https://fortune.com/2020/04/20/coronavirus-contact-tracing-privacy-europe-pepp-pt-dp3t-covid-19-tracking/}
\BIBentrySTDinterwordspacing

\bibitem{greenberg2020clever}
\BIBentryALTinterwordspacing
A.~Greenberg, ``Clever cryptography could protect privacy in {Covid}-19
  contact-tracing apps,'' \emph{Wired}, April 2020. [Online]. Available:
  \url{https://www.wired.com/story/covid-19-contact-tracing-apps-cryptography/}
\BIBentrySTDinterwordspacing

\bibitem{sharma2018using}
A.~Sharma, R.~A. Harrington, M.~B. McClellan, M.~P. Turakhia, Z.~J. Eapen,
  S.~Steinhubl, J.~R. Mault, M.~D. Majmudar, L.~Roessig, K.~J. Chandross
  \emph{et~al.}, ``Using digital health technology to better generate evidence
  and deliver evidence-based care,'' \emph{Journal of the American College of
  Cardiology}, vol.~71, no.~23, pp. 2680--2690, 2018.

\bibitem{troncoso2020decentralized}
\BIBentryALTinterwordspacing
C.~Troncoso, M.~Payer, J.-P. Hubaux, M.~Salathe, J.~Larus, E.~Bugnion,
  W.~Lueks, T.~Stadler, A.~Pyreglis, D.~Antonioli, L.~Barman, S.~Chatel,
  K.~Paterson, S.~Capkun, D.~Basin, J.~Beutel, D.~Jackson, B.~Preneel,
  N.~Smart, D.~Singelee, A.~Abidin, S.~Guerses, M.~Veale, C.~Cremers, R.~Binns,
  and C.~Cattuto, ``Decentralized privacy-preserving proximity tracing,'' April
  2020. [Online]. Available:
  \url{https://github.com/DP-3T/documents/blob/master/DP3T%20White%20Paper.pdf}
\BIBentrySTDinterwordspacing

\bibitem{chen2017use}
J.~Chen, J.~Lieffers, A.~Bauman, R.~Hanning, and M.~Allman-Farinelli, ``The use
  of smartphone health apps and other mobile h ealth (mhealth) technologies in
  dietetic practice: a three country study,'' \emph{Journal of Human Nutrition
  and Dietetics}, vol.~30, no.~4, pp. 439--452, 2017.

\bibitem{dandekar2020safe}
R.~Dandekar, S.~G. Henderson, M.~Jansen, S.~Moka, Y.~Nazarathy, C.~Rackauckas,
  P.~G. Taylor, and A.~Vuorinen, ``Safe blues: A method for estimation and
  control in the fight against covid-19,'' 2020.

\bibitem{zikmund200428}
B.~J. Zikmund-Fisher, A.~Fagerlin, and P.~A. Ubel, ``“is 28\% good or bad?”
  evaluability and preference reversals in health care decisions,''
  \emph{Medical Decision Making}, vol.~24, no.~2, pp. 142--148, 2004.

\bibitem{kang2012development}
E.~Kang’ethe, V.~Kimani, D.~Grace, G.~Mitoko, B.~McDermott, J.~Ambia,
  C.~Nyongesa, G.~Mbugua, W.~Ogara, and P.~Obutu, ``Development and delivery of
  evidence-based messages to reduce the risk of zoonoses in nairobi, kenya,''
  \emph{Tropical animal health and production}, vol.~44, no.~1, pp. 41--46,
  2012.

\bibitem{gov_canada_covid_guidance}
\BIBentryALTinterwordspacing
G.~of~Canada, ``Coronavirus disease ({COVID}-19): Guidance documents,'' 2020.
  [Online]. Available:
  \url{https://www.canada.ca/en/public-health/services/diseases/2019-novel-coronavirus-infection/guidance-documents.html}
\BIBentrySTDinterwordspacing

\bibitem{hunger2013official}
M.~Hunger, L.~Schwarzkopf, M.~Heier, A.~Peters, R.~Holle, K.~S. Group
  \emph{et~al.}, ``Official statistics and claims data records indicate
  non-response and recall bias within survey-based estimates of health care
  utilization in the older population,'' \emph{BMC health services research},
  vol.~13, no.~1, p.~1, 2013.

\bibitem{he2020temporal}
X.~He, E.~H. Lau, P.~Wu, X.~Deng, J.~Wang, X.~Hao, Y.~C. Lau, J.~Y. Wong,
  Y.~Guan, X.~Tan, X.~Mo, Y.~Chen, B.~Liao, W.~Chen, F.~Hu, Q.~Zhang, M.~Zhong,
  Y.~Wu, L.~Zhao, F.~Zhang, B.~J. Cowling, F.~Li, and G.~M. Leung, ``Temporal
  dynamics in viral shedding and transmissibility of covid-19,'' \emph{Nature
  Medicine}, 2020.

\bibitem{lauer2020incubation}
S.~A. Lauer, K.~H. Grantz, Q.~Bi, F.~K. Jones, Q.~Zheng, H.~R. Meredith, A.~S.
  Azman, N.~G. Reich, and J.~Lessler, ``The incubation period of coronavirus
  disease 2019 (covid-19) from publicly reported confirmed cases: estimation
  and application,'' \emph{Annals of internal medicine}, 2020.

\bibitem{leung2020first}
K.~Leung, J.~T. Wu, D.~Liu, and G.~M. Leung, ``First-wave covid-19
  transmissibility and severity in china outside hubei after control measures,
  and second-wave scenario planning: a modelling impact assessment,'' \emph{The
  Lancet}, 2020.

\bibitem{wang2020response}
C.~J. Wang, C.~Y. Ng, and R.~H. Brook, ``{Response to COVID-19 in Taiwan: Big
  Data Analytics, New Technology, and Proactive Testing},'' \emph{JAMA}, 2020.

\bibitem{regan1995privacy}
P.~Regan, ``Legislating privacy,'' 1995.

\bibitem{solove2008understanding}
D.~J. Solove, ``Understanding privacy,'' 2008.

\bibitem{kundera1984being}
M.~Kundera, ``The unbearable lightness of being,'' 1984.

\bibitem{keith2010privacy}
M.~J. Keith, J.~S. Babb~Jr, C.~P. Furner, and A.~Abdullat, ``Privacy assurance
  and network effects in the adoption of location-based services: an iphone
  experiment.'' in \emph{ICIS}, 2010, p. 237.

\bibitem{whitman2003two}
J.~Q. Whitman, ``The two western cultures of privacy: Dignity versus liberty,''
  \emph{Yale LJ}, vol. 113, p. 1151, 2003.

\bibitem{bayer2000surveillance}
R.~Bayer and A.~L. Fairchild, ``Surveillance and privacy,'' 2000.

\bibitem{tcn}
\BIBentryALTinterwordspacing
T.~Coalition, ``{TCN} protocol,'' April 2020. [Online]. Available:
  \url{https://github.com/TCNCoalition/TCN}
\BIBentrySTDinterwordspacing

\bibitem{covidwatch}
``{COVID Watch},'' \url{https://covid-watch.org/}, 2020.

\bibitem{patil2014big}
H.~K. Patil and R.~Seshadri, ``Big data security and privacy issues in
  healthcare,'' in \emph{2014 IEEE international congress on big data}.\hskip
  1em plus 0.5em minus 0.4em\relax IEEE, 2014, pp. 762--765.

\bibitem{privacystatement2020}
\BIBentryALTinterwordspacing
O.~of~the Privacy Commissioner~of Canada, ``Joint statement by federal,
  provincial and territorial privacy commissioners: Supporting public health,
  building public trust: Privacy principles for contact tracing and similar
  apps.'' May 2020. [Online]. Available:
  \url{https://www.priv.gc.ca/en/opc-news/speeches/2020/s-d_20200507/}
\BIBentrySTDinterwordspacing

\bibitem{privacybydesign}
\BIBentryALTinterwordspacing
A.~Cavoukian, ``Privacy by design: the 7 foundational principles,'' January
  2011. [Online]. Available:
  \url{https://www.ipc.on.ca/wp-content/uploads/Resources/7foundationalprinciples.pdf}
\BIBentrySTDinterwordspacing

\bibitem{penney2018advancing}
\BIBentryALTinterwordspacing
J.~Penney, S.~McKune, L.~Gill, and R.~J. Diebert, ``{Advancing Human Rights by
  Design in the Dual Use Technology Industry},'' \emph{Columbia Journal of
  International Affairs}, December 2018. [Online]. Available:
  \url{https://jia.sipa.columbia.edu/advancing-human-rights-design-dual-use-technology-industry}
\BIBentrySTDinterwordspacing

\bibitem{nhs_app}
\BIBentryALTinterwordspacing
NHS, ``Nhs covid-19 app,'' April 2020. [Online]. Available:
  \url{https://www.nhsx.nhs.uk/covid-19-response/nhs-covid-19-app/}
\BIBentrySTDinterwordspacing

\bibitem{gregg2006easier}
A.~P. Gregg, B.~Seibt, and M.~R. Banaji, ``Easier done than undone: asymmetry
  in the malleability of implicit preferences.'' \emph{Journal of personality
  and social psychology}, vol.~90, no.~1, p.~1, 2006.

\bibitem{friese2006implicit}
M.~Friese, M.~W{\"a}nke, and H.~Plessner, ``Implicit consumer preferences and
  their influence on product choice,'' \emph{Psychology \& Marketing}, vol.~23,
  no.~9, pp. 727--740, 2006.

\bibitem{veltriivchenko2017}
\BIBentryALTinterwordspacing
G.~A. Veltri and A.~Ivchenko, ``The impact of different forms of cognitive
  scarcity on online privacy disclosure,'' \emph{Computers in Human Behavior},
  vol.~73, pp. 238 -- 246, 2017. [Online]. Available:
  \url{http://www.sciencedirect.com/science/article/pii/S0747563217301693}
\BIBentrySTDinterwordspacing

\bibitem{chang2019}
\BIBentryALTinterwordspacing
C.~Chang, ``Self-control-centered empowerment model: Health consciousness and
  health knowledge as drivers of empowerment-seeking through health
  communication,'' \emph{Health Communication}, vol.~0, no.~0, pp. 1--12, 2019,
  pMID: 31480856. [Online]. Available:
  \url{https://doi.org/10.1080/10410236.2019.1652385}
\BIBentrySTDinterwordspacing

\bibitem{gachter2011relationship}
M.~G{\"a}chter, D.~A. Savage, and B.~Torgler, ``The relationship between
  stress, strain and social capital,'' \emph{Policing: An International Journal
  of Police Strategies \& Management}, 2011.

\bibitem{devakumar2020racism}
D.~Devakumar, G.~Shannon, S.~S. Bhopal, and I.~Abubakar, ``Racism and
  discrimination in covid-19 responses,'' \emph{Lancet (London, England)}, vol.
  395, no. 10231, p. 1194, 2020.

\bibitem{gostin2020responding}
L.~O. Gostin, E.~A. Friedman, and S.~A. Wetter, ``Responding to covid-19: How
  to navigate a public health emergency legally and ethically,'' \emph{Hastings
  Center Report}, 2020.

\bibitem{wenham2020covid}
C.~Wenham, J.~Smith, and R.~Morgan, ``Covid-19: the gendered impacts of the
  outbreak,'' \emph{The Lancet}, vol. 395, no. 10227, pp. 846--848, 2020.

\bibitem{novid}
\BIBentryALTinterwordspacing
``{NOVID},'' April 2020. [Online]. Available: \url{https://novid.org/}
\BIBentrySTDinterwordspacing

\bibitem{white2020slowing}
\BIBentryALTinterwordspacing
T.~White, R.~Fenwick, I.~Becker-Mayer, J.~Petrie, Z.~Szabo, D.~Blank,
  J.~Colligan, M.~Hittle, M.~Ingle, O.~Nash, V.~Nguyen, J.~Schwaber,
  A.~Veeraghanta, M.~Voloshin, S.~V. Arx, and H.~Xue, ``Slowing the spread of
  infectious diseases using crowdsourced data,'' March 2020. [Online].
  Available: \url{https://www.covid-watch.org/article}
\BIBentrySTDinterwordspacing

\bibitem{bbcnews_nhs}
\BIBentryALTinterwordspacing
L.~Kelion, ``{Coronavirus: German contact-tracing app takes different path to
  NHS},'' \emph{BBC News}, May 2020. [Online]. Available:
  \url{https://www.bbc.com/news/technology-52650576}
\BIBentrySTDinterwordspacing

\bibitem{bbcnews_israel}
\BIBentryALTinterwordspacing
J.~Tidy, ``{Coronavirus: Israel enables emergency spy powers},'' \emph{BBC
  News}, March 2020. [Online]. Available:
  \url{https://www.bbc.com/news/technology-51930681}
\BIBentrySTDinterwordspacing

\bibitem{liu2015data}
V.~Liu, M.~A. Musen, and T.~Chou, ``Data breaches of protected health
  information in the {U}nited {S}tates,'' \emph{Jama}, vol. 313, no.~14, pp.
  1471--1473, 2015.

\bibitem{warrenbrandeis1890privacy}
S.~Warren and L.~Brandeis, ``The right to privacy,'' \emph{Harvard Law Review},
  vol.~4, pp. 193--220, 1890.

\bibitem{prosser1960privacy}
W.~Prosser, ``Privacy,'' \emph{California Law Review}, vol.~48, pp. 383--423,
  1960.

\bibitem{van15}
J.~Van Den~Hooff, D.~Lazar, M.~Zaharia, and N.~Zeldovich, ``{Vuvuzela: Scalable
  private messaging resistant to traffic analysis},'' in \emph{Proceedings of
  the 25th Symposium on Operating Systems Principles}, 2015, pp. 137--152.

\bibitem{tyagi2017stadium}
N.~Tyagi, Y.~Gilad, D.~Leung, M.~Zaharia, and N.~Zeldovich, ``Stadium: A
  distributed metadata-private messaging system,'' in \emph{Proceedings of the
  26th Symposium on Operating Systems Principles}, 2017, pp. 423--440.

\bibitem{corrigan15}
H.~Corrigan-Gibbs, D.~Boneh, and D.~Mazi{\`e}res, ``{Riposte: An anonymous
  messaging system handling millions of users},'' in \emph{2015 IEEE Symposium
  on Security and Privacy}.\hskip 1em plus 0.5em minus 0.4em\relax IEEE, 2015,
  pp. 321--338.

\bibitem{merkle1978secure}
R.~C. Merkle, ``Secure communications over insecure channels,''
  \emph{Communications of the ACM}, vol.~21, no.~4, pp. 294--299, 1978.

\bibitem{greschbach12}
B.~Greschbach, G.~Kreitz, and S.~Buchegger, ``{The devil is in the
  metadata—new privacy challenges in decentralised online social networks},''
  in \emph{2012 IEEE International Conference on Pervasive Computing and
  Communications Workshops}.\hskip 1em plus 0.5em minus 0.4em\relax IEEE, 2012,
  pp. 333--339.

\bibitem{berke2020assessing}
A.~Berke, M.~Bakker, P.~Vepakomma, R.~Raskar, K.~Larson, and A.~Pentland,
  ``Assessing disease exposure risk with location histories and protecting
  privacy: A cryptographic approach in response to a global pandemic,''
  \emph{arXiv preprint arXiv:2003.14412}, 2020.

\bibitem{chaum81}
D.~L. Chaum, ``{Untraceable electronic mail, return addresses, and digital
  pseudonyms},'' \emph{Communications of the ACM}, vol.~24, no.~2, pp. 84--90,
  1981.

\bibitem{reed1998anonymous}
M.~G. Reed, P.~F. Syverson, and D.~M. Goldschlag, ``Anonymous connections and
  onion routing,'' \emph{IEEE Journal on Selected areas in Communications},
  vol.~16, no.~4, pp. 482--494, 1998.

\bibitem{el2014concepts}
K.~El~Emam and B.~Malin, ``Concepts and methods for de-identifying clinical
  trial data,'' \emph{Paper commissioned by the Committee on Strategies for
  Responsible Sharing of Clinical Trial Data}, 2014.

\bibitem{census2016}
S.~Canada, ``Census division ({CD}),'' 2016.

\bibitem{sweeney2002k}
L.~Sweeney, ``k-anonymity: A model for protecting privacy,''
  \emph{International Journal of Uncertainty, Fuzziness and Knowledge-Based
  Systems}, vol.~10, no.~05, pp. 557--570, 2002.

\bibitem{fredrikson2015model}
M.~Fredrikson, S.~Jha, and T.~Ristenpart, ``Model inversion attacks that
  exploit confidence information and basic countermeasures,'' in
  \emph{Proceedings of the 22nd ACM SIGSAC Conference on Computer and
  Communications Security}, 2015, pp. 1322--1333.

\bibitem{garrett2017echo}
R.~K. Garrett, ``The “echo chamber” distraction: Disinformation campaigns
  are the problem, not audience fragmentation.'' 2017.

\bibitem{li1995global}
M.~Y. Li and J.~S. Muldowney, ``Global stability for the seir model in
  epidemiology,'' \emph{Mathematical biosciences}, vol. 125, no.~2, pp.
  155--164, 1995.

\bibitem{chang2020modelling}
S.~L. Chang, N.~Harding, C.~Zachreson, O.~M. Cliff, and M.~Prokopenko,
  ``Modelling transmission and control of the covid-19 pandemic in australia,''
  \emph{arXiv preprint arXiv:2003.10218v2}, 2020.

\bibitem{ferretti2020transmission}
L.~Ferretti, C.~Wymant, M.~Kendell, L.~Zhao, A.~Nurtay, L.~Abeler-Dörner,
  M.~Parker, D.~Bonsall, and C.~Fraser, ``Quantifying sars-cov-2 transmission
  suggests epidemic control with digital contact tracing,'' \emph{Science},
  2020.

\bibitem{jefferson2020mask}
T.~Jefferson, C.~B. Del~Mar, L.~Dooley, E.~Ferroni, L.~A. Al-Ansary, G.~A.
  Bawazeer, M.~L. van Driel, N.~S. Nair, M.~A. Jones, S.~Thoring, and J.~M.
  Conly, ``Physical interventions to interrupt or reduce the spread of
  respiratory viruses (review),'' \emph{Cochrane Database of Systematic
  Reviews}, 2020.

\bibitem{washpost_korea}
\BIBentryALTinterwordspacing
M.~J. Kim and S.~Denyer, ``{A ‘travel log’ of the times in South Korea:
  Mapping the movements of coronavirus carriers },'' \emph{The Washington
  Post}, March 2020. [Online]. Available:
  \url{https://www.washingtonpost.com/world/asia\_pacific/coronavirus-south-korea-tracking-apps/2020/03/13/2bed568e-5fac-11ea-ac50-18701e14e06d\_story.html}
\BIBentrySTDinterwordspacing

\bibitem{Kingma+Welling-ICLR2014}
D.~P. Kingma and M.~Welling, ``Auto-encoding variational bayes,'' in
  \emph{Proceedings of the International Conference on Learning Representations
  (ICLR)}, 2014.

\bibitem{hinton1995wake}
G.~E. Hinton, P.~Dayan, B.~J. Frey, and R.~M. Neal, ``The" wake-sleep"
  algorithm for unsupervised neural networks,'' \emph{Science}, vol. 268, no.
  5214, pp. 1158--1161, 1995.

\bibitem{bengio2013estimating}
Y.~Bengio, N.~L{\'e}onard, and A.~Courville, ``Estimating or propagating
  gradients through stochastic neurons for conditional computation,''
  \emph{arXiv preprint arXiv:1308.3432}, 2013.

\bibitem{williams1992simple}
R.~J. Williams, ``Simple statistical gradient-following algorithms for
  connectionist reinforcement learning,'' \emph{Machine learning}, vol.~8, no.
  3-4, pp. 229--256, 1992.

\bibitem{jang2016categorical}
E.~Jang, S.~Gu, and B.~Poole, ``Categorical reparameterization with
  gumbel-softmax,'' \emph{arXiv preprint arXiv:1611.01144}, 2016.

\bibitem{transformer}
A.~Vaswani, N.~Shazeer, N.~Parmar, J.~Uszkoreit, L.~Jones, A.~N. Gomez,
  L.~Kaiser, and I.~Polosukhin, ``Attention is all you need,'' \emph{NeurIPS},
  2017.

\bibitem{ke2018sparse}
N.~R. Ke, A.~G. A.~P. GOYAL, O.~Bilaniuk, J.~Binas, M.~C. Mozer, C.~Pal, and
  Y.~Bengio, ``Sparse attentive backtracking: Temporal credit assignment
  through reminding,'' in \emph{Advances in neural information processing
  systems}, 2018, pp. 7640--7651.

\bibitem{heiner1983origin}
R.~A. Heiner, ``The origin of predictable behavior,'' \emph{The American
  economic review}, vol.~73, no.~4, pp. 560--595, 1983.

\bibitem{Dreibelbis_2016}
\BIBentryALTinterwordspacing
R.~Dreibelbis, A.~Kroeger, K.~Hossain, M.~Venkatesh, and P.~Ram, ``Behavior
  change without behavior change communication: Nudging handwashing among
  primary school students in bangladesh,'' \emph{International Journal of
  Environmental Research and Public Health}, vol.~13, no.~1, p. 129, Jan 2016.
  [Online]. Available: \url{http://dx.doi.org/10.3390/ijerph13010129}
\BIBentrySTDinterwordspacing

\bibitem{leonard2008richard}
T.~C. Leonard, ``Richard h. thaler, cass r. sunstein, nudge: Improving
  decisions about health, wealth, and happiness,'' 2008.

\bibitem{hilthychan2018}
D.~Hilty and S.~Chan, ``Human behavior with mobile health: Smartphone/ devices,
  apps and cognition,'' \emph{Psychology and Cognitive Sciences - Open
  Journal}, vol.~4, pp. 36--47, 12 2018.

\bibitem{mitchie2011refined}
S.~Mitchie, S.~Ashford, F.~Sniehotta, S.~Dombrowski, A.~Bishop, and D.~French,
  ``A refined taxonomy of behaviour change techniques to help people change
  their physical activity and healthy eating behaviors: the calo-re taxonomy,''
  \emph{Psychol Health}, vol.~26, pp. 1479--1498, 2011.

\bibitem{eyal2013hooked}
N.~Eyal and R.~Hoover, ``Hooked: A guide to building habit-forming products,''
  2013.

\bibitem{jachimowicz2017community}
J.~M. Jachimowicz, S.~Chafik, S.~Munrat, J.~C. Prabhu, and E.~U. Weber,
  ``Community trust reduces myopic decisions of low-income individuals,''
  \emph{Proceedings of the National Academy of Sciences}, vol. 114, no.~21, pp.
  5401--5406, 2017.

\bibitem{kongatsmcgetrickraine2019}
K.~Kongats, J.~A. McGetrick, K.~D. Raine, C.~Voyer, and C.~I. Nykiforuk,
  ``Assessing general public and policy influencer support for healthy public
  policies to promote healthy eating at the population level in two canadian
  provinces,'' \emph{Public Health Nutrition}, vol.~22, no.~8, p. 1492–1502,
  2019.

\bibitem{festinger1962theory}
L.~Festinger, \emph{A theory of cognitive dissonance}.\hskip 1em plus 0.5em
  minus 0.4em\relax Stanford university press, 1962, vol.~2.

\bibitem{krause2007social}
J.~Krause, D.~P. Croft, and R.~James, ``Social network theory in the
  behavioural sciences: potential applications,'' \emph{Behavioral Ecology and
  Sociobiology}, vol.~62, no.~1, pp. 15--27, 2007.

\bibitem{zaharias2009developing}
P.~Zaharias and A.~Poylymenakou, ``Developing a usability evaluation method for
  e-learning applications: Beyond functional usability,'' \emph{Intl. Journal
  of Human--Computer Interaction}, vol.~25, no.~1, pp. 75--98, 2009.

\bibitem{nielsen1996usability}
J.~Nielsen, ``Usability metrics: Tracking interface improvements,'' \emph{Ieee
  Software}, vol.~13, no.~6, pp. 1--2, 1996.

\bibitem{ryan2012willpower}
N.~Ryan, ``Willpower: Rediscovering the greatest human strength, by roy f.
  baumeister and john tierney,'' 2012.

\bibitem{bandura1978}
\BIBentryALTinterwordspacing
A.~Bandura, ``Self-efficacy: Toward a unifying theory of behavioral change,''
  \emph{Advances in Behaviour Research and Therapy}, vol.~1, no.~4, pp. 139 --
  161, 1978, perceived Self-Efficacy: Analyses of Bandura's Theory of
  Behavioural Change. [Online]. Available:
  \url{http://www.sciencedirect.com/science/article/pii/0146640278900024}
\BIBentrySTDinterwordspacing

\bibitem{kochnafziger2011}
\BIBentryALTinterwordspacing
A.~K. Koch and J.~Nafziger, ``Self-regulation through goal setting*,''
  \emph{The Scandinavian Journal of Economics}, vol. 113, no.~1, pp. 212--227,
  2011. [Online]. Available:
  \url{https://onlinelibrary.wiley.com/doi/abs/10.1111/j.1467-9442.2010.01641.x}
\BIBentrySTDinterwordspacing

\bibitem{degirmenci2020}
\BIBentryALTinterwordspacing
K.~Degirmenci, ``Mobile users’ information privacy concerns and the role of
  app permission requests,'' \emph{International Journal of Information
  Management}, vol.~50, pp. 261 -- 272, 2020. [Online]. Available:
  \url{http://www.sciencedirect.com/science/article/pii/S0268401218307965}
\BIBentrySTDinterwordspacing

\bibitem{waldman2020}
\BIBentryALTinterwordspacing
A.~E. Waldman, ``Cognitive biases, dark patterns, and the ‘privacy
  paradox’,'' \emph{Current Opinion in Psychology}, vol.~31, pp. 105 -- 109,
  2020, privacy and Disclosure, Online and in Social Interactions. [Online].
  Available:
  \url{http://www.sciencedirect.com/science/article/pii/S2352250X19301484}
\BIBentrySTDinterwordspacing

\bibitem{ricebogdanov2018}
\BIBentryALTinterwordspacing
M.~D. Rice and E.~Bogdanov, ``Privacy in doubt: An empirical investigation of
  canadians' knowledge of corporate data collection and usage practices,''
  \emph{Canadian Journal of Administrative Sciences / Revue Canadienne des
  Sciences de l'Administration}, vol.~36, no.~2, pp. 163--176, 2019. [Online].
  Available: \url{https://onlinelibrary.wiley.com/doi/abs/10.1002/cjas.1494}
\BIBentrySTDinterwordspacing

\bibitem{privacycanada2019}
O.~of~the Privacy Commissioner~of Canada, ``2018-19 survey of canadians on
  privacy,'' 2019.

\bibitem{markyoungkarel2015}
\BIBentryALTinterwordspacing
M.~S. Young, K.~A. Brookhuis, C.~D. Wickens, and P.~A. Hancock, ``State of
  science: mental workload in ergonomics,'' \emph{Ergonomics}, vol.~58, no.~1,
  pp. 1--17, 2015, pMID: 25442818. [Online]. Available:
  \url{https://doi.org/10.1080/00140139.2014.956151}
\BIBentrySTDinterwordspacing

\bibitem{springerwhittaker2019}
\BIBentryALTinterwordspacing
A.~Springer and S.~Whittaker, ``Progressive disclosure: Empirically motivated
  approaches to designing effective transparency,'' in \emph{Proceedings of the
  24th International Conference on Intelligent User Interfaces}, ser. IUI
  ’19.\hskip 1em plus 0.5em minus 0.4em\relax New York, NY, USA: Association
  for Computing Machinery, 2019, p. 107–120. [Online]. Available:
  \url{https://doi.org/10.1145/3301275.3302322}
\BIBentrySTDinterwordspacing

\bibitem{gallagher2011perceived}
K.~M. Gallagher, J.~A. Updegraff, A.~J. Rothman, and L.~Sims, ``Perceived
  susceptibility to breast cancer moderates the effect of gain-and loss-framed
  messages on use of screening mammography.'' \emph{Health Psychology},
  vol.~30, no.~2, p. 145, 2011.

\bibitem{baseman2013public}
J.~G. Baseman, D.~Revere, I.~Painter, M.~Toyoji, H.~Thiede, and J.~Duchin,
  ``Public health communications and alert fatigue,'' \emph{BMC health services
  research}, vol.~13, no.~1, p. 295, 2013.

\bibitem{rossman2017}
\BIBentryALTinterwordspacing
C.~Rossmann, \emph{Content Effects: Health Campaign Communication}.\hskip 1em
  plus 0.5em minus 0.4em\relax American Cancer Society, 2017, pp. 1--11.
  [Online]. Available:
  \url{https://onlinelibrary.wiley.com/doi/abs/10.1002/9781118783764.wbieme0127}
\BIBentrySTDinterwordspacing

\bibitem{cerc2019psychology}
\BIBentryALTinterwordspacing
C.~for Disease~Control and Prevention, ``{CERC}: Psychology of a crisis,''
  2019. [Online]. Available:
  \url{https://emergency.cdc.gov/cerc/ppt/CERC_Psychology_of_a_Crisis.pdf}
\BIBentrySTDinterwordspacing

\bibitem{segerstrom2004psychological}
S.~C. Segerstrom and G.~E. Miller, ``Psychological stress and the human immune
  system: a meta-analytic study of 30 years of inquiry.'' \emph{Psychological
  bulletin}, vol. 130, no.~4, p. 601, 2004.

\bibitem{morey2015current}
J.~N. Morey, I.~A. Boggero, A.~B. Scott, and S.~C. Segerstrom, ``Current
  directions in stress and human immune function,'' \emph{Current opinion in
  psychology}, vol.~5, pp. 13--17, 2015.

\bibitem{glaser2000chronic}
R.~Glaser, J.~Sheridan, W.~B. Malarkey, R.~C. MacCallum, and J.~K.
  Kiecolt-Glaser, ``Chronic stress modulates the immune response to a
  pneumococcal pneumonia vaccine,'' \emph{Psychosomatic medicine}, vol.~62,
  no.~6, pp. 804--807, 2000.

\bibitem{yao2020patients}
H.~Yao, J.-H. Chen, and Y.-F. Xu, ``Patients with mental health disorders in
  the covid-19 epidemic,'' \emph{The Lancet Psychiatry}, vol.~7, no.~4, p. e21,
  2020.

\bibitem{garfin2020novel}
D.~R. Garfin, R.~C. Silver, and E.~A. Holman, ``The novel coronavirus
  (covid-2019) outbreak: Amplification of public health consequences by media
  exposure.'' \emph{Health Psychology}, 2020.

\bibitem{brunskraguljac2020}
\BIBentryALTinterwordspacing
D.~P. Bruns, N.~V. Kraguljac, and T.~R. Bruns, ``Covid-19: Facts, cultural
  considerations, and risk of stigmatization,'' \emph{Journal of Transcultural
  Nursing}, vol.~0, no.~0, p. 1043659620917724, 0, pMID: 32316872. [Online].
  Available: \url{https://doi.org/10.1177/1043659620917724}
\BIBentrySTDinterwordspacing

\bibitem{stone2019unified}
J.~C. Stone, K.~Glass, J.~Clark, Z.~Munn, P.~Tugwell, and S.~A. Doi, ``A
  unified framework for bias assessment in clinical research,''
  \emph{International Journal of Evidence-Based Healthcare}, vol.~17, no.~2,
  pp. 106--120, 2019.

\bibitem{georgemcgahan2012}
\BIBentryALTinterwordspacing
G.~George, A.~M. McGahan, and J.~Prabhu, ``Innovation for inclusive growth:
  Towards a theoretical framework and a research agenda,'' \emph{Journal of
  Management Studies}, vol.~49, no.~4, pp. 661--683, 2012. [Online]. Available:
  \url{https://onlinelibrary.wiley.com/doi/abs/10.1111/j.1467-6486.2012.01048.x}
\BIBentrySTDinterwordspacing

\bibitem{freypatton2020}
\BIBentryALTinterwordspacing
W.~R. Frey, D.~U. Patton, M.~B. Gaskell, and K.~A. McGregor, ``Artificial
  intelligence and inclusion: Formerly gang-involved youth as domain experts
  for analyzing unstructured twitter data,'' \emph{Social Science Computer
  Review}, vol.~38, no.~1, pp. 42--56, 2020. [Online]. Available:
  \url{https://doi.org/10.1177/0894439318788314}
\BIBentrySTDinterwordspacing

\end{thebibliography}

\addtocontents{toc}{\unexpanded{\unexpanded{{%
\vspace{2em}
\footnoterule
  \footnotesize
$^1$\textit{Inherent privacy limitations of decentralized automatic contact tracing} was published in revised form in JAMIA. Ref: \url{https://doi.org/10.1093/jamia/ocaa153}  \par\bigskip
}}}}

\end{document}